\documentstyle[prd,aps,psfig]{revtex}

\begin{document}

\draft

\title{Three Dimensional Numerical General Relativistic
Hydrodynamics I: Formulations, Methods,  and Code Tests}
\author{Jos\'e A. Font${}^{(1)}$, Mark Miller${}^{(*,2)}$,
Wai-Mo Suen${}^{(2,3)}$, Malcolm Tobias${}^{(2)}$}

\address{${}^{(1)}$Max-Planck-Institut f\"ur Gravitationsphysik \\
Albert-Einstein-Institut \\
Schlaatzweg 1, D-14473, Potsdam, Germany}

\address{${}^{(2)}$McDonnell Center for the Space Sciences \\
Department of Physics,
Washington University, St. Louis, Missouri 63130}

\address{${}^{(3)}$Physics Department \\
Chinese University of Hong Kong,
Hong Kong}

\date{\today}

\maketitle

\begin{abstract}

This is the first in a series of papers on the construction and validation of a
three-dimensional code for general relativistic hydrodynamics, and its 
application to general relativistic astrophysics.  This paper studies the 
consistency and convergence of our general relativistic hydrodynamic treatment 
and its coupling to the spacetime evolutions described by the full set of
Einstein equations with a perfect fluid source, complimenting a 
similar study of the (vacuum) spacetime part of the code~\cite{Bona98b}.

The numerical treatment of the general relativistic hydrodynamic equations is 
based on high resolution shock capturing schemes, specifically designed to 
solve non-linear hyperbolic systems of conservation laws.  These schemes rely 
on the characteristic information of the system.  A spectral decomposition for 
general relativistic hydrodynamics suitable for a general spacetime metric is
presented.  Evolutions based on different approximate Riemann solvers 
(flux-splitting, Roe, and Marquina) are studied and compared. The coupling 
between the hydrodynamics and the spacetime (the right and left hand side of 
the Einstein equations) is carried out in a treatment which is second order 
accurate in {\it both} space and time. The spacetime evolution allows for a 
choice of different formulations of the Einstein equations, and different 
numerical methods for each formulation. Together with the different 
hydrodynamical methods, there are twelve different combinations of spacetime 
and hydrodynamical evolutions. Convergence tests for all twelve combinations 
with a variety of test beds are studied, showing consistency with the 
differential equations and correct convergence properties. The test-beds 
examined include shocktubes, Friedmann-Robertson-Walker cosmology tests, 
evolutions of self-gravitating compact (TOV) stars, and evolutions of 
relativistically boosted TOV stars. Special attention is paid to the numerical 
evolution of strongly gravitating objects, e.g., neutron stars, in the full 
theory of general relativity, including a simple, yet effective treatment for
the surface region of the star (where the rest mass density is abruptly 
dropping to zero).

The code has been optimized for massively parallel computation, and
has demonstrated linear scaling up to 1024 nodes on a Cray T3E.

\end{abstract}


% Sec 1: Introduction.

\section{Introduction}

The field of computational astrophysics is entering an exciting and
challenging era.  The large amount of observational data involving
general relativistic phenomena requires the integration of numerical
relativity with the traditional tools of astrophysics, such as
hydrodynamics, magneto-hydrodynamics, nuclear astrophysics, and
radiation transport. General Relativistic
Astrophysics --- astrophysics involving gravitational fields so strong
and dynamical that the full Einstein field equations are required
for its accurate description, is quickly becoming a promising area
of research.

\medskip
--------------------------------------------------------------------
\newline
\noindent ${}^{(*)}$ {\small Correspondence should be 
addressed to Mark Miller.}

\pagebreak

As a first step in our study of ``Computational General Relativistic
Astrophysics'', our collaboration (the NCSA/Potsdam/Wash U numerical
relativity collaboration) is building a code called ``Cactus'' for
solving the full set of Einstein field equations coupled to a perfect
fluid source.  Such a code will have many applications for
astrophysical processes involving neutron stars and black holes.  In
this paper we present the formulation and methods of the 3D general
relativistic hydrodynamic part of the code, and its coupling to the
spacetime part of the code.  We also present various tests for the
validation of the code.  The complimentary presentation on the
(vacuum) spacetime evolution part of the code has been given in
~\cite{Bona98b}.  In the following we begin by discussing the
background of our code development effort.

\subsection{Motivation}

Two of the major directions of astronomy in the next century are high energy
astrophysics ($X$-ray and $\gamma$-ray astronomy)
and gravitational wave astronomy. The former is driven by
advanced X-ray and $\gamma$-ray satellite observations, e.g., CGRO, AXAF,
GLAST~\cite{nasa_missions}, XMM, INTEGRAL, that are either
current or planned in the next few
years.  High energy radiation is often emitted by highly relativistic
events in regions of strong gravitational fields, e.g.,  near black holes
(BHs) and neutron stars (NSs).  One of the biggest mysteries of modern
astronomy, $\gamma$-ray bursts, is likely related to processes
involving interactions of compact binaries (BH/NS or NS/NS) or highly
explosive collapse to a black hole (``hypernova'')
(see, e.g., \cite{Paczynski97} and references therein).
Such high energy astrophysical events often involve highly dynamical
gravitational fields, strong gravitational wave emissions, and ejecta
moving at ultrarelativistic speeds with relativistic Lorentz factors
up to $10^3-10^4$. The modeling of such events can only be achieved by
means of hydrodynamical simulations in the full theory of general
relativity.

The second major direction, gravitational wave astronomy, involves
the dynamical nature of spacetime in Einstein's theory of
gravity.  The tremendous recent interest in this frontier is driven by
the gravitational wave observatories presently being built or planned
in the US, Europe, and outer space, e.g., LIGO, VIRGO, GEO600, LISA, LAGOS
~\cite{nasa}, and the Lunar Outpost Astrophysics Program~\cite{nasa}.
The American LIGO and its European counterparts VIRGO and GEO600 are
scheduled to be on line in a few years~\cite{LIGO3}, making
gravitational wave astronomy a reality.  The space detector LISA has
been selected as one of the three ``cornerstone missions'' of
the European Space Agency~\cite{LISA}.  These observatories provide a
completely new window on the universe: existing observations are
mainly provided by the electromagnetic spectrum, emitted by individual
electrons, atoms, or molecules, and are easily absorbed, scattered, and
dispersed.  Gravitational waves are produced by the coherent bulk motion
of matter and travel nearly unscathed through space, coming to us
carrying the information of the strong field regions where they were
originally generated~\cite{Thorne95b}. This new window will provide very
different information about our universe that is either difficult or
impossible to obtain by traditional means.

The numerical (theoretical) determination of gravitational
waveforms is crucial for gravitational wave astronomy.  Physical
information in the data is to be extracted through template matching
techniques~\cite{Cutler93}, which {\em presupposes} that reliable
waveforms are known.  Accurate waveform detections are important both
as probes of the fundamental nature of gravity and for the unique
physical and astronomical information they carry, ranging from nuclear
physics (the equation of state of NSs~\cite{Cutler93}) to cosmology
(direct determination of the Hubble constant without going through the
``cosmic distance ladder''~\cite{schutz86b}). In most situations, the
waveform cannot be calculated without a numerical simulation based on
the full theory of general relativity. This need for waveform
templates is an important motivation of our effort.

In short, both of these frontiers of astronomy call for
Computational General Relativistic Astrophysics, i.e., the integration of
numerical relativity with traditional tools of computational
astrophysics, e.g., computational hydrodynamics, radiation transport,
nuclear astrophysics, and magneto-hydrodynamics.  If we are to fully
understand the observational data generated by the non-linear and
dynamical gravitational fields, detailed modeling taking dynamic
general relativity into full account must be carried out.

\subsection{Existing Work in General Relativistic Hydrodynamics}

We begin by briefly reviewing some of the significant existing
investigations in the field of numerical general relativistic
hydrodynamics (GR-Hydro in the following) to set the stage for the 
description of our own work.  While there has been much effort in the study of
relativistic hydrodynamics in {\it pre-determined} (fixed, or with its
time evolution specified) background
spacetimes, we focus on studies that are most relevant to {\it
dynamical} spacetimes with the matter flows acting as sources to the
Einstein equations.

The pioneering work dates back to the one-dimensional supernova
core-collapse code by May and White~\cite{May66}.  It was based on a
Lagrangian (i.e., coordinates co-moving with the fluid) finite difference scheme
with artificial viscosity terms included in the equations to damp the
spurious numerical oscillations caused by the presence of shock waves
in the flow solution. Numerous astrophysical simulations were based on
this approach. One drawback is that the Lagrangian character of the code
makes it difficult to be extended to the multidimensional case.

The pioneering Eulerian (i.e., coordinates not co-moving with the fluid) finite
difference GR-Hydro code was developed by Wilson~\cite{Wilson72} in
the early 70's. It used a combination of artificial viscosity (AV) and
upwind techniques.  It became the kernel of a large number of codes
developed in the 80's.  Many different astrophysical scenarios were
investigated with these codes, ranging from axisymmetric stellar
core-collapse~\cite{Nakamura81,Stark85,Evans86}, to accretion onto
compact objects~\cite{Hawley84a,Petrich89}, and to numerical
cosmology~\cite{Centrella84}.  In the following, we give a short
overview of this large body of work, paying more attention to the numerical
methods used than to the physical results obtained.

While there are a large number of numerical investigations in
pre-determined background spacetimes based on the AV approach
(e.g.,~\cite{Wilson72,Wilson79,Hawley84a,Petrich89,Hawley91}), we
focus on those using a fully self-consistent treatment evolving the
spacetime dynamically with the Einstein equations coupled to a
hydrodynamic source.  Although there is much recent interest in this
direction, only the spherically symmetric case (1D) can be considered
essentially
solved~\cite{Dykema80,Shapiro80,Evans84,Schinder88,Mezzacappa89}.  In
axisymmetry, i.e. 2D, only a few attempts have been made, with most of
them devoted to the study of the gravitational collapse and bounce of
rotating stellar cores and the subsequent emission of gravitational
radiation~\cite{Nakamura81b,Stark85,Evans86,Abrahams94a}.
\cite{Nakamura81b} was the first to calculate a general
relativistic stellar core collapse. The computation succeeded in
tracking the evolution of matter and the formation of a black hole but
the numerical scheme was not accurate enough to compute the emitted
gravitational radiation.  The code in~\cite{Stark85} used a radial
gauge and a mixture of polar and maximal slicing.  The GR-hydro
equations were solved with standard finite difference methods with AV
terms.  In~\cite{Evans86} the numerical scheme for the matter fields
was more sophisticated, using monotonic upwind reconstruction
procedures and flux limiters, with discontinuous solutions handled by
adding AV terms in the equations.  In~\cite{Abrahams94a}, a numerical
study of the stability of star clusters in axisymmetry was
performed. In this investigation, the source of the gravitational
field was assumed to be a configuration of collisionless (dust)
particles, which reduces the hydrodynamic computation to a
straightforward integration of the geodesic equations.

Three-dimensional extensions of these AV based GR-Hydro treatments have been
attempted over the last few years.  Wilson's original scheme has been
applied to the study of NS binary coalescence
in~\cite{Wilson95,Wilson96} under the assumption of a conformally flat
spacetime, which leads to a considerable simplification of the gravitational
field equations.  A code employing the {\it full} set of Einstein equations
and self-gravitating matter fields is currently being
developed~\cite{Oohara96}.  In this work the complete set of the
equations, spacetime and hydrodynamics, are finite differenced in a
uniform Cartesian grid using van Leer's scheme~\cite{vanLeer77} with
total variation diminishing (TVD) flux limiters 
(see, e.g.,~\cite{Leveque92} for definitions).  Shock waves
are spread out using a tensor AV algorithm.  With
this code they have studied the gravitational collapse of a rotating
polytrope to a black hole (comparing to the original axisymmetric computation
of Ref.~\cite{Stark85}) and the coalescence of a binary NS system.
Further work to achieve longer term stability is under way~\cite{Oohara96}.

The success of the artificial viscosity approach is well-known.
However, it has inherent difficulties in handling the
ultrarelativistic regime~\cite{Norman86}.  In Wilson's formulation of
the GR-hydro equations, there are explicit spacetime derivatives of
the pressure in the source terms.  This breaks the conservative
character of the system and introduces complications into the numerical
treatment.  This motivated, in recent years, the effort of extending to
relativistic hydrodynamics high-resolution shock-capturing schemes 
(HRSC) originally developed in classical (Newtonian) computational fluid 
dynamics.  Such schemes are based on the solution of local
Riemann problems, exploiting the hyperbolicity of the
hydrodynamic equations.  To use such numerical treatments, the hydrodynamic
equations are first cast into a first order (hyperbolic) system 
of conservation (or balance) laws. The characteristic fields of the system are
then determined which allows the construction of numerical schemes
which propagate the information along the fluid characteristics. We refer
the reader to~\cite{Leveque92} for a review of these methods for
general hyperbolic systems of conservation laws.

HRSC schemes were first introduced into GR-Hydro in~\cite{Marti91},
and applied in (spherical) dynamical spacetimes in~\cite{Bona93b}
and~\cite{Romero96}. The latter investigation focussed on, among other
problems, the study of supernova core collapse (including the
infall epoch, bounce, and shock propagation).  The numerical code was
based on the radial-gauge and polar-slicing coordinate
conditions~\cite{Bardeen83}.  
In~\cite{Banyuls97} the GR-Hydro
equations were analyzed in the ``3+1'' formalism and the theoretical
building blocks to construct a HRSC scheme in multidimensions were
presented.  Axisymmetric studies using HRSC schemes are currently
being carried out in~\cite{Brandt98}.  This investigation focussed on
the study of accretion phenomena onto (dynamic) rotating black holes
and the associated emission of gravitational radiation induced by the
presence of the matter fields.  
Axisymmetric studies will also provide useful ``test beds" in forthcoming
investigations with the present 3D code discussed in this paper.
As will be discussed in later sections
of this paper, our present code is based on the same HRSC algorithmic
machinery as in the aforementioned works.  We extend the treatment to
3D, and develop a code that makes no assumptions on the nature of the
spacetime, the form of the metric, or the slicing and spatial
coordinates.  We refer the interested reader to the above references
for a first understanding of the numerical schemes used in our work.

We also want to mention a completely
different approach for GR-Hydro based on pseudospectral
methods~\cite{Gourgoulhon91}.  These methods are well known
for having extraordinary accuracy in smooth regions of the
solution.  The numerical error is evanescent, i.e., it decreases
as $e^{-N}$ with $N$ being the number of coefficients in the spectral
expansion. The main drawback of pseudospectral methods has been,
traditionally, the inaccurate modeling of discontinuous solutions due
to the appearance of the so-called Gibbs phenomenon.  In the presence
of discontinuities, the numerical approximation of the solution does
not converge at the discontinuity and spurious oscillations
appear. Recently, however, an innovative pseudospectral method based
on a multidomain decomposition has been developed~\cite{Bonazzola98a}
which circumvents the Gibbs phenomenon.  This new approach has already
been shown to work remarkably well in the 3D numerical construction of
McLaurin and Roche equilibrium models.

\subsection{Issues of General Relativistic Hydrodynamics}

In this subsection we discuss the main issues we considered in
choosing our approach, building on the existing work discussed above.
The main aim of our program is to study violent and
highly-energetic astrophysical processes like NS/NS coalescence within
the framework of general relativity.  These scenarios involve strong
gravitational fields, matter motion with (ultra) relativistic speeds
and/or strong shock waves.  These features make the numerical
integration of the hydrodynamic equations a very demanding task.  The
difficulty is exacerbated by the intrinsic multidimensional character
of these astrophysical systems, and by the inherent complexities in
Einstein theory of gravity, e.g., coordinate degrees of freedom and
the possible formation of curvature singularities (e.g., collapse of
matter configurations to black holes).
These complications call for the use of advanced numerical
methodology, a flexible code construction which allows for
the use of different treatments, and a large amount of
careful testbed studies.  In the following we discuss these issues in
more detail.

Two major issues in GR-Hydro which are purely hydrodynamical in origin
are the numerical modeling of flows with large Lorentz factors and
strong shock waves. In ~\cite{Norman86} it was shown that the AV based
schemes have difficulties in handling ultrarelativistic velocity flows
with Lorentz factors $\geq 2$. As a result,~\cite{Norman86}
proposed using implicit finite difference schemes to handle the
GR-Hydro equations in the ultrarelativistic regime.  However,
investigations during the last decade have provided increasing
evidence that the most appropriate schemes to deal with
ultra-relativistic flow with strong shocks are those based on
(approximate or exact) Riemann solvers, i.e., HRSC schemes.  These methods
have high accuracy (second order or more) in regions where the flow
solution is smooth, and at the same time are able to resolve
discontinuities in the solution (e.g., shock waves) with little
smearing.  They have been extensively tested and found to be
applicable in the ultra-relativistic regime (see,
e.g.,~\cite{Ibanez98} for a recent review).

While we believe HRSC schemes may be capable of providing the technology
for treating the hydrodynamic part of the evolution, the field of
computational GR-Hydro still contains many issues that are as yet
unexplored, especially for cases where the relativistic fluid is coupled to
a dynamical spacetime.  For a fully dynamical spacetime, one major
issue is the handling of the gauge degrees of freedom.  This problem
is exacerbated in 3D simulations without symmetry assumptions.  In a
general 3D problem, there is no preferred choice of gauge to restrict
the metric functions as in lower dimension simulations (e.g., radial
gauge and polar slicing in spherically symmetric simulations).
Lagrangian coordinate systems are inappropriate for
complicated 3D flows.  The inevitable lower resolution in 3D
simulations also makes the problem more acute.  Even in vacuum
spacetime studies, the choice and implementation of appropriate gauge
conditions for a general dynamical evolution is a largely unexplored
territory~\cite{Balakrishna96a}.  How will the gauge choices be
affected by the presence of relativistic fluid flows or by the
existence of strong shocks which create sharp features in the sources
of the metric evolution?  For example, what will be a useful gauge
condition for a process like the inspiral of a NS/NS binary?  These
are completely open issues.  In order to provide the capability to
investigate these problems, the code we construct here is designed to
allow arbitrary gauge conditions, making no assumptions on the lapse
function or the shift vector.

Another class of problems involves the connection of the numerical
integration of the hydrodynamic equations to that of the spacetime
equations.  What is the best set of variables to use, locally
measured quantities, coordinate variables, densitized quantities or
some combination?  With the spacetime metric an evolved variable,
there are many choices.  What is the best way to
connect the hydrodynamics and the spacetime finite differencing steps
to achieve not only a second order accurate scheme in both space and
time, but also in a way that is suitable for long term evolutions?
Even in Newtonian strong field evolutions, coupling the hydrodynamic
integration to the gravitational potential calculation in different
ways can yield different long term behavior~\cite{Calder98}.  
As a consequence of the different character of the equations governing
the geometry of the spacetime and the evolution of the matter fields,
the numerical methods to handle them are drastically different.  What
are the effects of combining different methods, and is there a best
combination for a particular class of problems?  With the recent
development of hyperbolic formulations in GR, an interesting
possibility would be to consider all of the dynamical variables, both
spacetime and matter fields, to be members of one master state vector.
The entire system of equations could then be written as a single
(vector) conservation (or balance) equation.  One could then apply the
same HRSC schemes to the entire system.  What advantages would this
bring?  These are some of the issues that we have in mind in choosing
our approach in developing the code as will be discussed next.

\subsection{Outline of Our Approach}

Our overall goal is to develop an efficient, flexible, computational
tool for General Relativistic Astrophysics.  Specifically for this
paper, the aims are: (1) to establish the formulation, including the
spectral decomposition of the GR-Hydro equations, on which our code is
based, (2) to validate the numerical code we constructed for solving
the GR-Hydro equations, and (3) to compare the different numerical
schemes we used.

The set of differential equations we are attempting to solve consists of very
complicated, coupled partial differential equations involving
thousands of terms.  Considering the complexity and generality of the
code, along with the fact that the solution space of the differential
equations is largely unexplored, it is essential that any physical
result produced by a 3D GR-Hydro code be preceded by a series of tests
such as the ones we report here, in order to insure the fidelity of
the discretization to the original differential equations.  In fact,
we consider the tests presented here to be a minimal set: {\it any} 3D
GR-Hydro code should be able to reproduce these results.  Further
tests, especially those related to the long term stability of the code
and detailed comparisons of 3D and 1D results will be presented in a
forthcoming paper.

In exploring the very complex system of the GR-Hydro equations, it is also
essential to have the capability to compare results based on
different mathematical formulations and different numerical schemes.
Our code is currently set up to allow two different formulations of
the Einstein equations: the standard Arnowitt-Deser-Misner (ADM)
formulation~\cite{Arnowitt62} and the Bona-Mass\'o (BM) 
hyperbolic formulation~\cite{Bona98b} (other hyperbolic formulations
will be included and reported later).  The code allows for two
different choices for finite differencing the ADM equations: a
standard leapfrog scheme and an iterative Crank-Nicholson scheme.  The
BM equations are finite differenced using a Strang split to separate
the source and flux updates.  The latter are performed using a
MacCormack method.  As for the numerical treatment of the hydrodynamic
equations, the code has the capability of using three different HRSC
schemes: the first one is the flux split method, mainly chosen for its
simplicity. The second is Roe's method~\cite{Roe81} (note that contrary
to~\cite{Eulderink94}, we do not use Roe's averaging but
instead employ arithmetic averaging (see
section~\ref{sec:discrete} below for details)). The third scheme we use
is the recently developed Marquina's method~\cite{Donat96}. All three
schemes are coupled to the spacetime evolution solver in a way which
is second order accurate in both space and time.

In this code we also allow for arbitrary spacetime coordinate
conditions.  As mentioned previously, this enables the investigation 
of gauge choices in GR-Hydro and allows the use of different coordinate
systems for different astrophysical simulations.  This
capability is built into our development, and we
have carried out tests with non-trivial lapses and shifts in this
paper.  However, more investigation is needed
in this direction.

\subsection{Computational Issues}

As the aim of our program is to study {\it
realistic} astrophysical systems, which often require full 3D simulations and
involve many different time and length scales, it is important that the
computer code we develop be capable of carrying out large scale
simulations.  This requires the use of massively parallel
supercomputers.  
The ``Cactus'' code was built with this in mind.  Here we
give a brief overview of the computational infrastructure of the code
and its performance.  For a more extensive review, see
~\cite{Seidel98c}.

The Cactus code achieves parallelism through the MPI message passing
interface \cite{MPI}. This allows high performance portable
parallelism using a distributed memory model. All major high
performance parallel architectures, including the SGI Origin
2000, Cray T3E, HP/Convex Exemplar, and IBM SP-2 support this
programming model. The MPI layer of Cactus also allows computing on
clusters of networked workstations and PC's.  Parallelism 
in Cactus is based on a generic domain decomposition, distributing
uniform grid functions across multiple processors and providing ghost-zone
based communications for a variety of stencil widths and grid
staggerings. The code can also compile without MPI, allowing the same
source code to be run on a single processor workstation
and on massively parallel supercomputers.  The platforms currently
supported and tested include: 
the SGI Origin 2000 (up to 256 nodes), the Cray T3E
(up to 1024 nodes), SGI O2 clusters, NT clusters, DEC alphas, and SGI
workstations.  We have recently benchmarked a version of the code (the
``GR3D'' version, constructed for the NASA Neutron Star Grand
Challenge Project, see http://wugrav.wustl.edu/Relativ/nsgc.html) on a
1024 node T3E-1200, achieving over 140 GFlop/sec and a scaling efficiency
of over 95\% (for details of the benchmark, see
http://wugrav.wustl.edu/Codes/GR3D/).  Besides the floating point
and scaling efficiency, it is also noteworthy that a relatively
large grid size (644 x 644 x 1284 grid points for 32 bit
accuracy, and 500 x 500 x 996 grid points for 64 bit accuracy)
were used for the benchmarked run on the T3E-1200.
This is made possible by the efficient memory usage of the code.  With the
full set of the Einstein equations coupled to the relativistic hydrodynamics
equations, a large number of 3D arrays are required to evolve
the system.  In order to have reasonable resolutions for 
realistic simulations, it is essential that
the code make efficient use of available memory.  It is also essential
that the code be highly optimized in order for these large simulations
to be carried out in a reasonable time.

During the code development, 
special attention was also given to software engineering
problems, such as collaborative code development, maintenance, and management.
The code was developed to be shared by the entire community for the
investigation of general relativistic astrophysics.  To minimize
barriers associated with collaborative development, the code was
constructed to have: 1. A modular code structure that allows new code
modules to easily plug in as ``thorns'' to the core part of the 
code (the ``flesh'').  The ``Flesh'' contains the 
parallel domain decomposition
software, I/O, and various utilities.  2.  A consistency test suite
library to make sure that new thorns will not 
conflict with other parts of the code.  3. Various code development
tools, such as:  documentation, elliptic solvers, and
visualization tools, which
provide a complete environment for code development, and testing.
For detailed discussions of these and other features of
collaborative infrastructure of the code, see
\cite{Masso98c,Seidel98c}

These computational features of the code significantly
enhance our effort in constructing a multi-purpose code for 
general relativistic astrophysics.

\subsection{Organization of This Paper}

The organization of the paper is as follows: the formulation of the
differential equations are given in section~\ref{sec:formulation}.  A
spectral decomposition of the GR-Hydro equations suitable for a
general non-diagonal spatial metric is presented.
The details of the discretization of
the equations and of the coupling of the spacetime
and hydrodynamics are given in section~\ref{sec:discrete}.  Shock tube
tests are performed in section~\ref{sec:shocktube} for shocks along
the coordinate axes and along the diagonal.  These test the
hydrodynamic part of the code, with the background geometry held flat.
We then go on to test the coupling of the hydrodynamics to curved and
dynamical spacetimes.  Section~\ref{sec:frw} is on tests using
Friedmann-Robertson-Walker cosmologies with dust.
Section~\ref{sec:tov} contains tests using static spherical star
solutions with a polytropic equation of state.  We present a practical
procedure which gives stable evolution of the surface region of the
star.  Section~\ref{sec:boost} contains tests using the spherical star
solutions described in section~\ref{sec:tov} 
but now relativistically boosted along the
diagonal $\hat{x} + \hat{y} + \hat{z}$.  This is a strong test of the
fully coupled spacetime and hydrodynamics system, with all possible
terms in the equations activated and with a non-trivial lapse and
shift.  Finally, Section VIII contains a brief summary.

All tests presented in 
sections~\ref{sec:frw}-\ref{sec:boost} contain convergence studies 
performed in the following way: errors are obtained by subtracting
the exact solution at a specific time from the computed
solution for a number of dynamical variables.  These
errors are produced at three different resolutions, $\Delta x$,
$\Delta x /2$, and $\Delta x/4$.  To demonstrate
they have the correct convergence properties for a second order
accurate discretization we check that each error function 
decreases by a factor of four for each factor of two increase in 
resolution.  This is demonstrated by plotting the various
error functions along 1-D lines.  These convergence tests are an
essential part in validating the code.

\section{Formulation}
\label{sec:formulation}

\subsection{General Relativistic Hydrodynamic Equations}

In this subsection we present the 
hydrodynamic equations for a general curved spacetime in a form
suitable for advanced numerical treatment. The equations
for the evolution of the spacetime, including the
hydrodynamic source, will be presented in a later subsection.

The general relativistic hydrodynamic equations, written in the
standard covariant form, consist of the local conservation laws of the
stress-energy, $T^{\mu \nu}$, and the matter current density, $J^{\mu}$
\begin{equation}
\label{eq:stressenergycons}
{\nabla}_{\mu} T^{\mu \nu} = 0,
\end{equation}
\begin{equation}
\label{eq:masscons}
{\nabla}_{\mu} J^{\mu} = 0,
\end{equation}
where $J^{\mu} = \rho u^{\mu}$, $\rho$ is the rest mass density
and $u^{\mu}$ the 4-velocity of the fluid. ${\nabla}_{\mu}$
stands for the covariant derivative with respect to the 4-metric
of the underlying spacetime. Throughout this paper we
are using, unless otherwise stated, natural units ($G=c=1$). Greek
(Latin) indices run from 0 to 3 (1 to 3). In what follows we will
neglect viscous effects, assuming the stress-energy tensor to be that
of a perfect fluid
\begin{equation}
T^{\mu \nu} = \rho h u^{\mu} u^{\nu} + P g^{\mu \nu},
\label{perf_fluid}
\end{equation}
where $P$ is the fluid pressure 
and $g^{\mu \nu}$ is the 4-metric describing 
the spacetime. In addition, the relativistic specific enthalpy, 
$h$, is given by
\begin{equation}
h = 1 + \epsilon + P / \rho,
\end{equation}
where $\epsilon$ is the rest frame specific internal energy density
of the fluid.

The equations written in this covariant form are {\it not} suitable for
the use of advanced numerical schemes. 
In order to carry out numerical hydrodynamic
evolutions, and in particular to take advantage of the benefits of
HRSC methods, the hydrodynamic equations after the 3+1 split must be
written as a hyperbolic system of first order flux conservative
equations.  We introduce coordinates $(x^0 = t,x^1,x^2,x^3)$ and write
Eqs.~(\ref{eq:stressenergycons}) and (\ref{eq:masscons}) in terms of
coordinate derivatives. We project
Eq.(\ref{eq:stressenergycons}) and Eq.(\ref{eq:masscons}) onto the
basis $\{n^{\mu},{\left (
\frac {\partial}{\partial x^i} \right )}^{\mu} \}$, with $n^{\mu}$ being
a timelike vector normal to a given hypersurface.  A straightforward
calculation yields the set of equations in the desired
form
\begin{equation}
{\partial}_t \vec{\cal{U}} + {\partial}_i \vec{F^i} = \vec{S},
\label{balance}
\end{equation}
\noindent
where $\partial_t$ denotes a partial derivative with respect to time
and $\partial_i$ indicates a partial derivative with respect to the
spatial coordinate $x^i$.

The evolved state vector $\vec{\cal{U}}$ is given, in terms of the
primitive variables $(\rho,v^i,\epsilon)$, as
\begin{equation}
\vec{\cal{U}} = 
\left[ 
\begin{array}{c}
\tilde{D} \\
\tilde{S_j} \\
\tilde{\tau} \\
\end{array}  
\right]
= \left[ 
\begin{array}{c}
\sqrt{\gamma} W \rho \\
\sqrt{\gamma} \rho h W^2 v_j \\
\sqrt{\gamma} (\rho h W^2 - P - W \rho) \\
\end{array}  
\right],
\label{eq:evolvedvar}
\end{equation}
where $\gamma$ is the determinant of the 3-metric $\gamma_{ij}$,
$v_j$ is the fluid 3-velocity,  and
$W$ is the Lorentz factor, 
$W=\alpha {u^0} = {(1 - \gamma_{ij} v^i v^j)}^{-1/2}$. 
Notice that the spatial components of the 4-velocity $u^i$ are related
to the 3-velocity by the following formula:
\begin{equation}
u^i = W (v^i - {{\beta}^i}/\alpha),
\end{equation}
where $\alpha$ and $\beta^i$ are, respectively, the lapse function and the
shift vector of the spacetime.
Also notice that
we are using a slightly different set of variables as those used 
in~\cite{Banyuls97}. We are now ``densitizing" the evolved
quantities, $D$, $S_j$ and $\tau$, with the factor $\sqrt{\gamma}$.
The three flux vectors $\vec{F^i}$ are given by
\begin{equation}
\label{hydroflux}
\vec{F^i} = \left[ \begin{array}{c}
                    \alpha (v^i - \frac {1}{\alpha} {\beta}^i) \tilde{D} \\
                    \alpha ((v^i - \frac {1}{\alpha} {\beta}^i) 
                         \tilde{S_j} + \sqrt{\gamma} P {\delta}^i_j) \\
                    \alpha ( (v^i - \frac {1}{\alpha} {\beta}^i) 
                         \tilde{\tau} + \sqrt{\gamma} v^i P)
                 \end{array}  \right].
\end{equation}
Finally, the source vector $\vec{S}$ is given by
\begin{equation}
\vec{S} = \left[ \begin{array}{c}
                  0 \\
                  \alpha \sqrt{\gamma} T^{\mu \nu} g_{\nu \sigma}
                  { {\Gamma}^{\sigma} }_{\mu j} \\
                  \alpha \sqrt{\gamma} (T^{\mu 0} {\partial}_{\mu} \alpha -
                  \alpha T^{\mu \nu} { {\Gamma}^0}_{\mu \nu})
                 \end{array}  \right],
\end{equation}
where ${ {\Gamma}^{\alpha} }_{\mu \nu}$ is the 4-Christoffel symbol
\begin{equation}
{ {\Gamma}^{\alpha} }_{\mu \nu} = \frac {1}{2} g^{\alpha \beta}
({\partial}_{\mu} g_{\nu \beta} + {\partial}_{\nu} g_{\mu \beta} -
 {\partial}_{\beta} g_{\mu \nu}).
\end{equation}

A technical point must be included here. While the numerical code updates
the state vector $\vec{\cal{U}}$ forward in time it makes use, internally, of
the set of primitive variables defined above, $(\rho,v^i,\epsilon)$. 
Those are used throughout,
e.g., in the computation of the characteristic fields 
(see below). These variables cannot be 
obtained from the evolved ones in a closed functional form. 
Instead, they must be recovered through some appropriate root-finding 
procedure (an example of this can be found in~\cite{Marti96}).

\subsection{Spectral Decomposition and Characteristic Fields}
\label{sec:spectral}

The use of HRSC schemes, as will be presented in detail in the next
section, depends crucially on the knowledge of the
spectral decomposition of the Jacobian matrix 
of the system 
\begin{equation}
\frac {\partial \vec{F^i}} {\partial \vec{\cal{U}}}. 
\end{equation}
The characteristic speeds (eigenvalues) and
fields (eigenvectors) are the key ingredients of any HRSC
scheme. The spectral decomposition of the Jacobian matrices of
the general relativistic hydrodynamic equations with general equation
of state was first
reported in~\cite{Banyuls97} (for polytropic EOS see~\cite{Eulderink94}). 
However, we have found that the
eigenvectors reported in~\cite{Banyuls97} are correct only in
the case of a diagonal spatial metric. In this section
we display the full spectral decomposition valid for 
a generic spatial metric.
We focus on the $x$-direction, hence presenting the spectral decomposition
of $(\frac {\partial \vec{F^x}}{\partial \vec{\cal{U}}})$, as the other
two directions can be found by simple permutation of indices.

We start by considering an equation of state in which the pressure
$P$ is a function of $\rho$ and $\epsilon$,
$P = P(\rho,\epsilon)$. The relativistic speed of sound 
in the fluid $c_{s}$ is given by (see, e.g.,~\cite{Landau87})

\begin{equation}
c_{s}^2 =  { \left. {\frac {\partial P}{\partial E}} \right |}_{\cal{S}} =
\frac {\chi}{h} + \frac {P}{ {\rho}^2} \frac {\kappa}{h} ,
\end{equation}
where $\chi = {\left.\frac {\partial P}{\partial \rho} \right |}_\epsilon$,
$\kappa = {\left. \frac {\partial P}{\partial \epsilon} \right |}_\rho$, 
$\cal{S}$ is the entropy per particle, and
$E$ is the total rest energy density which in our case is
$E = \rho + \rho \epsilon$ .  We require
a complete set of eigenvectors $[\vec{r}_i]$ and corresponding
eigenvalues ${\lambda}_i$ along the $x$-direction, i.e.
\begin{equation}
\left [ \frac {\partial \vec{F^x}}{\partial \vec{\cal{U}}} \right ]
[\vec{r}_i] = {\lambda}_i [\vec{r}_i], \;\;\; i=1,\cdots,5.
\end{equation}
The solution contains a triply degenerate eigenvalue
\begin{equation}
{\lambda}_1 = {\lambda}_2 = {\lambda}_3 = \alpha v^x - {\beta}^x.
\end{equation}
A set of linearly independent vectors that
span this degenerate space is given by
\begin{equation}
\vec{r}_1 = { \left[\frac {\kappa}{h W (\kappa - \rho {c_s}^2)},
               v_x,v_y,v_z,
               1 - \frac {\kappa}{h W (\kappa - \rho {c_s}^2)}\right]}^T,
\end{equation}
\begin{equation}
\vec{r}_2 = { \left[W v_y, h (\gamma_{xy} + 2 W^2 v_x v_y),
                      h (\gamma_{yy} + 2 W^2 v_y v_y),
                      h (\gamma_{yz} + 2 W^2 v_y v_z),
                      v_y W (2 W h - 1)\right]}^T,
\end{equation}
\begin{equation}
\vec{r}_3 = { \left[W v_z, h (\gamma_{xz} + 2 W^2 v_x v_z),
                      h (\gamma_{yz} + 2 W^2 v_y v_z),
                      h (\gamma_{zz} + 2 W^2 v_z v_z),
                      v_z W (2 W h - 1)\right]}^T.
\end{equation}
The superscript $T$ denotes transpose.  The other two eigenvalues are given by
\begin{equation}
{\lambda}_{\pm} = \frac {\alpha}{1 - v^2 {c_s}^2}
\left \{ v^x (1 - {c_s}^2) \pm \sqrt{{c_s}^2 (1 - v^2)
   \left [ \gamma^{xx} (1 - v^2 {c_s}^2) - v^x v^x (1 - {c_s}^2) \right ]}
\right \} - {\beta}^x,
\end{equation}
with corresponding eigenvectors
\begin{equation}
\vec{r}_{\pm} = { \left[1,
   h W \left ( v_x - \frac {v^x - ({\lambda}_{\pm} + {\beta}^x)/{\alpha}}
           {\gamma^{xx} - v^x ({\lambda}_{\pm} + {\beta}^x)/{\alpha}} \right ),
   h W v_y, h W v_z,
   \frac {h W (\gamma^{xx} - v^x v^x)}{\gamma^{xx} - v^x
           ({\lambda}_{\pm} + {\beta}^x)/{\alpha}} - 1 \right] }^T.
\end{equation}

\subsection{Equations for a Dynamical Spacetime With a Hydrodynamic Source}

The dynamics of the gravitational field in general relativity theory is
described by Einstein's field equations
\begin{equation}
G_{\mu \nu} = 8 \pi T_{\mu \nu},
\label{eq:einstein}
\end{equation}
which relate the (ten) metric components $g_{\mu \nu}(=g_{\nu \mu})$
of the spacetime to the stress energy tensor $T_{\mu \nu}$. Here,
$G_{\mu \nu} $ is the Einstein tensor which involves second
derivatives, in both space and time, of the dependent variables
$g_{\mu \nu}$.  A formulation of the Einstein equations
suitable for numerical evolutions
has been known for more than three decades~\cite{Arnowitt62}.
In recent years, many new formulations have been proposed
(for a review, see~\cite{Bona96a,Reula98a} 
and references therein) seeking
to expose the hyperbolicity of the evolution components of the Einstein
field equations.

In the present paper, we discuss the mathematical and algorithmic issues 
related to the coupling of the hydrodynamic equations
to two different formulations
of the Einstein equations. We start with the more commonly used
ADM formulation~\cite{Arnowitt62}.  Then we discuss
the BM hyperbolic formulation of the Einstein equations~\cite{Bona98b}.  

\subsubsection{Arnowitt-Deser-Misner formulation}

In the ADM formulation~\cite{Arnowitt62}, spacetime is foliated into a set of
non-intersecting spacelike hypersurfaces.  There are two kinematic
variables which describe the evolution between these surfaces: the
lapse function $\alpha$, which describes the rate of advance of time
along a timelike unit vector $n^\mu$ normal to a surface, and the
spacelike shift vector $\beta^i$ that describes the motion of
coordinates. The line element is written as
\begin{equation}
 ds^2 = -(\alpha^{2} -\beta _{i}\beta ^{i}) dt^2 + 2 \beta _{i} dx^{i}
 dt +\gamma_{ij} dx^{i} dx^{j}.
\end{equation}

The ADM formulation 
casts the Einstein equations into a first order (in time) quasi-linear
\cite{Richtmyer67} system of equations.  The dependent variables are
the 3-metric $\gamma_{ij}$ and the extrinsic curvature $K_{ij}$. The
evolution equations read:

\begin{eqnarray}
\partial_t \gamma_{ij} &=& - 2 \alpha K_{ij}+\nabla_i \beta_j+
\nabla_j \beta_i,
\label{dtgij} \\
\partial_t K_{ij} &=& -\nabla_i \nabla_j \alpha + \alpha \left[
R_{ij}+K\ K_{ij} -2 K_{im} K^m_j 
- 8 \pi ( S_{ij} - \frac{1}{2}\gamma_{ij}S )
- 4 \pi {\rho}_{{}_{ADM}} \gamma_{ij}
\right] \nonumber \\ &\ & + \beta^m
\nabla_m K_{ij}+K_{im} \nabla_j \beta^m+K_{mj} \nabla_i \beta^m,
\label{dtkij}
\end{eqnarray}
where $\nabla_i$ denotes a covariant derivative with respect to the
3-metric $\gamma_{ij}$ and $R_{ij}$ is the Ricci curvature of the 3-metric.

In addition to the evolution equations, 
$\gamma_{ij},K_{ij},{\rho}_{{}_{ADM}},$ and $j^i$ must
satisfy the Hamiltonian constraint
\begin{equation}
{}^{(3)}R + K^2 - K_{ij} K^{ij} - 16 \pi {\rho}_{{}_{ADM}} = 0,
\end{equation}
and the momentum constraints
\begin{equation}
\nabla_j K^{ij} - \gamma^{ij} \nabla_j K - 8 \pi j^i = 0.
\end{equation}
Here, ${\rho}_{{}_{ADM}},j^i,S_{ij},S=\gamma^{ij} S_{ij}$ 
are the components of the stress
energy tensor projected onto the 3D hypersurface (for a more detailed 
discussion, see~\cite{York79}).  
In this paper we use the stress-energy tensor of a perfect fluid 
(Eq.~(\ref{perf_fluid})). Hence, 
explicitly in terms of the primitive hydrodynamic variables, 
\begin{eqnarray}
\label{adm_rho}
{\rho}_{{}_{ADM}} &=& \rho h W^2 - P, \\
j^i &=& \rho h {W^2}v^i, \\
S_{ij} &=& \rho h W^2 v_i v_j + \gamma_{ij} P, \\
\label{adm_s}
S &=& \rho h W^2 v_i v^i + 3 P. 
\end{eqnarray}

\subsubsection{Bona-Mass\'o hyperbolic formulation}

In the BM hyperbolic formulation~\cite{Bona97a,Bona98b},
the evolution equations are written as a first order balance law with,
formally, the same mathematical structure as the hydrodynamic equations,
Eq.~(\ref{balance}). Now, the state vector, containing the evolved quantities
for the spacetime, has the following $37$ components:
\begin{equation}
  \vec{\cal{U}}=\left( \gamma_{ij}, \alpha, K_{ij}, D_{kij}, A_k, V_k \right),
\end{equation}
where
\begin{eqnarray}
D_{kij} &=& \frac{1}{2} \partial_k \gamma_{ij}, \\
A_k &=& \partial_k \ln \alpha,  \\
V_i &=& {D_{ij}}^j - {D^j}_{ji}. 
\label{BM_v_def}
\end{eqnarray}
\noindent
Accordingly, the fluxes are given, using a self-explained notation, by:
\begin{eqnarray}
\label{fluxes}
  F^k_-\gamma_{ij} & = & 0 \;, \label{fluxgamma} \\
  F^k_-\alpha      & = & 0 \;, \label{fluxlapse} \\
  F^k_-K_{ij}      & = & -\beta^k\,K_{ij} + \alpha\;[\; D^k_{ij}
                         - n/2\;V^k\;\gamma_{ij}
                         \\ \nonumber
                   &   & + 1/2\;\delta^k_i\;(A_j+2\,V_j-D_{jr}^{\;\;r})
                         \label{fluxK} \\
                   &   & + 1/2\;\delta^k_j\;(A_i+2\,V_i-D_{ir}^{\;\;r})\;]
                         \;, \nonumber \\
  F^k_-D_{kij}     & = & -\beta^r D_{rij} + \alpha\;(K_{ij}-s_{ij})
                         \;, \label{fluxD} \\
  F^k_-A_k         & = & -\beta^r A_r + \alpha\;Q
                         \;, \label{fluxA} \\
  F^k_-V_i         & = & -\beta^k V_i + B^k_{\;i} - B_i^{\;k}
                         \;, \label{fluxV}
\end{eqnarray}
where
\begin{equation}
{B_k}^i = \frac{1}{2} \partial_k \beta^i,
\end{equation}
is calculated from the user supplied shift vector.
Finally, the source terms read:
\begin{eqnarray}
  S_-\gamma_{ij} &=& - 2\;\alpha\;(K_{ij}-s_{ij}) + 2\beta^r\,D_{rij}
                       \;,\label{sourcegamma} \\
  S_-\alpha      &=& - \alpha^2\;Q + \alpha\beta^r\,A_r
                       \;,\label{sourcelapse} \\
  S_-K_{ij}      &=&  2(K_{ir}B_j^{\;r}+K_{jr}B_i^{\;r}-K_{ij}B_r^{\;r})
                       \nonumber \\
      & & + \alpha\; [  \; 
                     - 2K_i^{\;k}K_{kj}+tr\,K\;K_{ij}
                       \nonumber  \\
                 & & \;\;\;\; - \Gamma^k_{\;ri}\Gamma^r_{\;kj}
                     + 2D_{ik}^{\;\;r}D_{rj}^{\;\;k}
                     + 2D_{jk}^{\;\;r}D_{ri}^{\;\;k}
                     + \Gamma^k_{\;kr}\Gamma^r_{\;ij}
                       \nonumber \\
            & & \;\;\;\;-(2\,D_{kr}^{\;\;k}-A_r)(D_{ij}^{\;\;r}+D_{ji}^{\;\;r})
                       \label{sourceK}  \\
  & & \;\;\;\; + A_i(V_j-1/2\;D_{jk}^{\;\;k}) + A_j(V_i-1/2\;D_{ik}^{\;\;k})
                       \nonumber \\
                 & & \;\;\;\; + A_j(V_i-1/2\;D_{ik}^{\;\;k})
                     - nV^kD_{kij} 
                       \nonumber  \\
  & & - 4 \pi (2 S_{ij} - \gamma_{ij} (S + p + (n-1) {\rho}_{{}_{ADM}})) \; ]
                       \nonumber  \\
  & & + n/4\;\alpha \gamma_{ij}\;[\; -D_k^{\;rs}\Gamma^k_{\;rs}
                     + D_{kr}^{\;\;r}D^{ks}_{\;\;s} -2\,V^kA_k \nonumber  \\
                 & &  \;\;\;\; + K^{rs}K_{rs}-(tr\,K)^2
                      \;]
                       \;, \nonumber \\
  S_-D_{kij}     &=& 0
                       \;, \label{sourceD} \\
  S_-A_{k}       &=& 0
                       \;, \label{sourceA} \\
  S_-V_i         &=&   \alpha\;[8 \pi j_i
                     + A_r\;(K^r_{\;i}-tr\,K\;\delta^r_i)
                       \nonumber \\
                 & & + K^r_{\;s}(D_{ir}^{\;\;s}-2D_{ri}^{\;\;s})
                     - K^r_{\;i}(D_{rs}^{\;\;s}-2D_{sr}^{\;\;s})]
                       \label{sourceV} \\
                 & & + 2(B_i^{\;r} - \delta_i^r\;tr\,B)\;V_r
                     + 2(D_{ri}^{\;\;s}-\delta^s_i\;D^j_{\;jr})B^r_{\;s}
                       \;, \nonumber
\end{eqnarray}
where
\begin{equation}
  s_{ij} = (B_{ij}+B_{ji})/\alpha,
\end{equation}
is used to simplify the equations.
The free parameter $n$ allows one to select a
specific evolution system (it is zero for the ``Ricci'' system and one for
the ``Einstein'' system) as discussed in~\cite{Bona97a}.

\section{Discretization of the equations}
\label{sec:discrete}

\subsection{Modern HRSC schemes for the GR-Hydro equations}

As stated previously, the main aim of this work is to
confirm the {\it consistency} of the coded finite difference equations with
the partial differential equations and the {\it convergence} of
three independent discretizations of the GR-Hydro equations.
All three approaches are based on finite-difference schemes 
employing HRSC schemes
to account, explicitly, for the characteristic
information of the equations. The methods considered are a flux split
method, Roe's approximate Riemann solver~\cite{Roe81},
and Marquina's recently developed scheme~\cite{Donat96}.

To simplify the discussion, let us examine the update for the 
state vector ${\vec{\cal{U}}}$ for a flux in the $x$-direction:
\begin{equation}
\frac {\partial \vec{\cal{U}}}{\partial t} +
\frac {\partial \vec{F}^x}{\partial x} = 0.
\end{equation}
The discretization of this equation takes the form
\begin{equation}
\frac {\partial {\vec{\cal{U}}}_{i}} {\partial t} + 
\frac {{(\vec{f}^{*})}_{i + 1/2} - 
{(\vec{f}^{*})}_{i - 1/2}}{\Delta x} = 0,
\end{equation}
where ${(\vec{f}^{*})}_{i \pm 1/2}$ is the ``numerical flux"
function calculated at the interfaces $i \pm 1/2$ of the spatial cell $i$.
The different methods we are using simply differ in the way
the numerical fluxes are calculated.
The way the source terms are
integrated is explained later in this section.

\subsubsection{Flux Split Method}
The first scheme
is a flux split method, where the flux is decomposed into
the part contributing to the eigenfields with positive eigenvalues
(fields moving to the right) and the part with
negative eigenvalues (fields moving to the left).  These fluxes
are then discretized with one sided derivatives whose side depends on the
sign of the particular eigenvalue.

For the flux split method, one makes the assumption that
\begin{equation}
\vec{F}^x(\sigma \vec{\cal{U}}) = \sigma \vec{F}^x(\vec{\cal{U}}),
\end{equation}
for any constant $\sigma$.  This is only true for the
fluxes of Eq.~(\ref{hydroflux}) if the equation of state
has the following functional form
\begin{equation}
P = P(\rho,\epsilon) = \rho f(\epsilon),
\end{equation}
for some function $f(\epsilon)$.
For the flux split method we therefore assume the equation of
state to be in the following form
\begin{equation}
P = (\Gamma - 1) \rho \epsilon,
\end{equation}
with $\Gamma$ being the (constant) adiabatic index of the fluid.
It is easy to show that, under the above assumptions, the flux
vector $\vec{F}^x$ can be written
\begin{equation}
\vec{F}^x = \left(\frac {\partial \vec{F}^x}
{\partial \vec{\cal{U}}}\right) \vec{\cal{U}}.
\end{equation}
Using the spectral decomposition of Section~\ref{sec:spectral} one can
express the Jacobian matrix 
$\frac {\partial \vec{F}^x}{\partial \vec{\cal{U}}}$ as
\begin{equation}
\frac {\partial \vec{F}^x}{\partial \vec{\cal{U}}} = M \Lambda M^{-1},
\end{equation}
where M is the matrix whose columns are the right-eigenvectors
of the system and $\Lambda$ is a diagonal matrix constructed from
the corresponding eigenvalues (see Section~\ref{sec:spectral}).

Given the characteristic information, we can now split the flux into
the part that is moving to the right and the part that is
moving to the left:
\begin{equation}
\vec{F}^x = {(\vec{F}^x)}^{+} + {(\vec{F}^x)}^{-} =
(M {\Lambda}^{+} M^{-1}) \vec{\cal{U}} + 
(M {\Lambda}^{-} M^{-1}) \vec{\cal{U}},
\end{equation}
where ${\Lambda}^{+} = \frac {1}{2} (\Lambda + |\Lambda|)$, and
${\Lambda}^{-} = \frac {1}{2} (\Lambda - |\Lambda|)$.
The numerical flux which corresponds to an upwind method
(first order in space) is then simply
\begin{equation}
{({{\vec{f}^{*}}_{i+1/2}})}_{\verb+first order+} = 
{(\vec{F}^x)}^{+}_{i} + {(\vec{F}^x)}^{-}_{i+1}.
\end{equation}
One could attempt to construct a numerical flux based on
one-sided derivatives that were second order accurate in space:
\begin{eqnarray}
{({{\vec{f}^{*}}_{i+1/2}})}_{\verb+second order -- non TVD+}  & = &
{({{\vec{f}^{*}}_{i+1/2}})}_{\verb+first order+} +
\frac {1}{2} (\vec{F}_{i} - 
   {({{\vec{f}^{*}}_{i-1/2}})}_{\verb+first order+}) + \\
 & & \frac {1}{2} (\vec{F}_{i+1} - 
         {({{\vec{f}^{*}}_{i+3/2}})}_{\verb+first order+}).
\end{eqnarray}
However, the method would not have the (numerically desirable)
Total Variation Diminishing (TVD) property (see, e.g.~\cite{Leveque92} 
for definition) unless flux
limiters are used in front of the second order correction terms:
\begin{eqnarray}
{({{\vec{f}^{*}}_{i+1/2}})}_{\verb+second order+}  & = &
{({{\vec{f}^{*}}_{i+1/2}})}_{\verb+first order+} +
\frac {1}{2} {\psi}^{+}_{i-1/2} (\vec{F}_{i} - 
   {({{\vec{f}^{*}}_{i-1/2}})}_{\verb+first order+}) + \\
 & & \frac {1}{2} {\psi}^{-}_{i+3/2} (\vec{F}_{i+1} - 
         {({{\vec{f}^{*}}_{i+3/2}})}_{\verb+first order+}),
\end{eqnarray}
with
\begin{eqnarray}
{\psi}^{+}_{i+1/2} & = & \psi 
\left(\frac {\vec{F}^x_{i+2} - {({{\vec{f}^{*}}_{i+3/2}})}}
       {\vec{F}^x_{i+1} - {({{\vec{f}^{*}}_{i+1/2}})}}\right), \\
{\psi}^{-}_{i+1/2} & = & \psi 
\left(\frac {\vec{F}^x_{i-1} - {({{\vec{f}^{*}}_{i-1/2}})}}
       {\vec{F}^x_{i} - {({{\vec{f}^{*}}_{i+1/2}})}}\right), 
\end{eqnarray}
where we are using van Leer flux limiters~\cite{vanLeer79}
\begin{equation}
\label{eq:vanleer}
\psi (\sigma) = \frac { \sigma + |\sigma|}{1 + \sigma}.
\label{vanleer}
\end{equation}

\subsubsection{Roe's Method}

The second scheme we use to integrate the hydrodynamic equations
makes use of Roe's approximate Riemann solver~\cite{Roe81}. This is
undoubtedly the most established method for the accurate 
integration of non-linear hyperbolic systems of conservation
laws. The suitability of Roe's method for the relativistic hydrodynamic
equations have been shown in~\cite{Font94,Eulderink94,Banyuls97}. 
This method makes no assumption on the equation of state,
and, in this respect, is more flexible than the flux split methods.
As the method is well documented in the literature
(see, e.g.,~\cite{Hirsch92}) it will only be briefly 
outlined here. As mentioned in the introduction, all simulations
reported in this paper using Roe's scheme are performed employing
arithmetic averages. For the use of Roe's averaging in GR-Hydro 
see~\cite{Eulderink94}.

A monotonic piecewise-constant 
(piecewise-linear) reconstruction of
the cell centered values of the primitive variables 
to the cell interfaces provides first-order (second-order)
accuracy in space~\cite{vanLeer79}.
In order to get second-order convergence we have implemented
a standard minmod piecewise-linear reconstruction 
algorithm~\cite{vanLeer79}.
The numerical fluxes across interfaces are
calculated according to
\begin{equation}
{(\vec{f}^{*})}_{i + 1/2} =
\frac {1}{2}  [\vec{F}^x_{R} + \vec{F}^x_{L}
- \sum_{m=1}^5 |\widetilde{\lambda}_m| \; \Delta \widetilde{\omega}_m \; 
\widetilde{\vec{r}}_m ],
\label{roeflux}
\end{equation}
\noindent
where ${R}$ and ${L}$ indicate the right and left sides of a cell
interface.
In addition, $\{\widetilde{\lambda}_m, \widetilde{\vec{r}}_m\}_{m=1,..,5}$ 
are, respectively, the eigenvalues and right-eigenvectors of the Jacobian
matrix of the system calculated at the cell interfaces as the arithmetic
mean of the left and right reconstructed (interpolated) primitive
variables. Averaged quantities in Eq.~(\ref{roeflux}) 
are identified by a ``tilde".
Finally, the quantities $\{\Delta \widetilde{\omega}_n\}_{n=1,..,5}$, 
the jumps of the characteristic variables across 
each characteristic field, are obtained from
\begin{equation}
\vec{\cal{U}}_{R} - \vec{\cal{U}}_{L} = \sum_m \Delta \widetilde{\omega}_m 
\widetilde{\vec{r}}_m.
\end{equation}
\noindent

\subsubsection{Marquina's Method}

In \cite{Donat96} Donat and Marquina proposed a new flux formula
to compute the numerical flux at a cell interface.
The new flux formula has a clear flux splitting
structure, and leads to an upstream scheme.
The novelty of Marquina's approach lies in the
extension of Shu and Osher's 
entropy-satisfying numerical flux~\cite{Shu89}
to systems of hyperbolic conservation laws.
In this scheme there are no artificial intermediate
states constructed at each cell interface. This implies that
there are no Riemann
solutions involved (either exact or approximate); moreover,
the scheme has been proven to alleviate
several numerical pathologies associated to the introduction
of an averaged state (as Roe's method does)
in the local diagonalization procedure (see~\cite{Donat96,Donat98}).

To compute the numerical flux at a given
interface, separating the states $\vec{\cal{U}}_L$ and $\vec{\cal{U}}_R$,
we compute first the sided local characteristic
variables and fluxes:
\begin{eqnarray}
\begin{array}{rclcrcl}
\omega_{l}^p &=& {\vec{l}}^p({\vec{\cal{U}}}_l) \cdot 
{\vec{\cal{U}}}_l , &\quad&
\phi_{l}^p &=& {\vec{l}}^p({\vec{\cal{U}}}_l) \cdot 
{\vec{f}}({\vec{\cal{U}}}_l), \\
\omega_{r}^p &=& {\vec{l}}^p({\vec{\cal{U}}}_r) \cdot 
{\vec{\cal{U}}}_r , &\quad&
\phi_{r}^p &=& {\vec{l}}^p({\vec{\cal{U}}}_r) \cdot 
{\vec{f}}({\vec{\cal{U}}}_r),
\end{array}
\end{eqnarray}
for $p=1,2\ldots,5$.
Here ${\vec{l}}^p({\vec{\cal{U}}}_l)$, ${\vec{l}}^p({\vec{\cal{U}}}_r)$,
are the (normalized) left eigenvectors 
of the Jacobian matrices of the system.
Let $\lambda_1({\vec{\cal{U}}}_l),\ldots,\lambda_5({\vec{\cal{U}}}_l)$ and
$\lambda_1({\vec{\cal{U}}}_r),\ldots,\lambda_5({\vec{\cal{U}}}_r)$
be their corresponding eigenvalues. For $k=1,\ldots, 5$
the procedure is the following:

\begin{itemize}

\item
If    $ \displaystyle{\lambda_k({\vec{\cal{U}}})}$ does not change sign in
  $[{\vec{\cal{U}}}_l,{\vec{\cal{U}}}_r]$, then the
  scheme is upwind  \newline
    \newline
    \indent  \ \ \ \ \ \ \ \ \ \ \ \ \ If
    $\displaystyle{ \lambda_k({\vec{\cal{U}}}_l)> 0}$ then 
    \begin{eqnarray}
    \begin{array}{c}
    \phi_+^k= \phi_l^k, \\
    \phi_-^k= 0,
    \end{array}
    \end{eqnarray}
    \indent  \ \ \ \ \ \ \ \ \ \ \ \ \ else 
    \begin{eqnarray}
    \begin{array}{c}    
    \phi_+^k= 0, \\
    \phi_-^k= \phi_r^k,
    \end{array}
    \end{eqnarray}
    \indent  \ \ \ \ \ \ \ \ \ \ \ \ \ endif \newline
    \indent  
\item 
Otherwise, the scheme is switched to the more viscous,
 entropy-satisfying, local-Lax-Friedrichs scheme \newline
 \newline
    \begin{eqnarray}
    \begin{array}{l}
    \alpha_k  = \max |\lambda_k({\vec{\cal{U}}})|, {{\vec{\cal{U}}} \in \Gamma
    ( {\vec{\cal{U}}}_l ,{\vec{\cal{U}}}_r)}, \\
     \phi_+^k = .5(\phi_l^k+\alpha_k \omega_l^k), \\
     \phi_-^k = .5(\phi_r^k-\alpha_k \omega_r^k),
    \end{array}
    \end{eqnarray}
     \indent 
\end{itemize}
\noindent
$\Gamma( {\vec{\cal{U}}}_l,{\vec{\cal{U}}}_r)$ is a curve in phase space connecting
${\vec{\cal{U}}}_l$ and ${\vec{\cal{U}}}_r$. 
In addition, $\alpha_k$ can be determined as
\begin{equation}
\alpha_k= \max\{|\lambda_k({\vec{\cal{U}}}_l)|,
|\lambda_k({\vec{\cal{U}}}_r)|\}.
\end{equation}

Marquina's flux formula is then:
\begin{equation} 
\label{mar}
{(\vec{f}^{*})}^n_{i + 1/2} =
\sum_{p=1}^m \left ( \phi^p_+ {\vec{r}}^p({\vec{\cal{U}}}_l) +
          \phi^p_- {\vec{r}}^p({\vec{\cal{U}}}_r) \right ),
\end{equation}
where, 
${\vec{r}}^p({\vec{\cal{U}}}_l)$, ${\vec{r}}^p({\vec{\cal{U}}}_r)$,
are the right (normalized) eigenvectors of the system. For further 
technical information about this solver we refer the reader to~\cite{Donat96}.
The suitability of this scheme for the accurate integration of
the hydrodynamic equations and 
many of its desirable properties
can be found in~\cite{Donat96} (Newtonian hydrodynamics)
and~\cite{Donat98,Marti97} (relativistic hydrodynamics).

\subsection{Discretization Techniques for 
the Spacetime and Spacetime-Hydrodynamics Coupling}

\subsubsection{Spacetime Discretization}

In this section we outline the discretization techniques used in the
vacuum spacetime part of the code.  For a more detailed discussion
we refer the reader to~\cite{Bona98b}.  Here we give the essential 
formulae for completeness and discuss in detail only the issues 
relevant to its coupling to hydrodynamics described in the next subsection.

The BM system uses the so-called Strang splitting~\cite{Strang68} 
to separate Eq.~(\ref{balance}) into two evolution steps. In the first step,
only the source terms are used to update the variables
\begin{equation}
\partial_t \vec{\cal{U}} = \vec{S},
\end{equation}
while in the second step, only the flux terms are used for the update
\begin{equation}
 \partial_t \vec{\cal{U}} + \partial_i \vec{F}^i = 0.
\end{equation}
To ensure second order accuracy in both space and time, this is done
by first evolving the source terms forward in time half a time
step, then evolving with only the flux terms a full time step, and
finally evolving with only the source terms another half time step.
The source terms are evolved forward using a second order accurate
predictor-corrector method, while the flux terms are evolved using a
second order accurate MacCormack scheme. 
Specific details of these methods are discussed in \cite{Bona98b}.  

The ADM system supports the use of several different numerical
schemes.  Currently, a leapfrog (non-staggered in time)
and iterative Crank-Nicholson scheme have been
coupled to the hydrodynamic solver.

The leapfrog method assumes that all variables exist on both the
current time step $n$ and the previous time step $n-1$.  
Variables are updated from $n-1$ to $n+1$ (future time)
evaluating all terms in the evolution equations on the current time 
step $n$.  

The iterative Crank-Nicholson solver first evolves the data from the
current time step $n$ to the future time step $n+1$ using a
forward in time, centered in space (FTCS) first order method.  
The solution at steps $n$ and $n+1$ 
are then averaged to obtain the solution on the half time step 
$n+\frac{1}{2}$.  This solution at the half time step $n+\frac{1}{2}$
is then used in a leapfrog step to 
re-update the solution at the final time step $n+1$. This
process is then iterated.
The error is defined as the difference
between the current and previous solutions on the half time 
step $n+\frac{1}{2}$. 
This error is summed over all grid points and all evolved variables.
This process is repeated until some desired tolerance is reached.
Care is taken to make sure that at least two iterations are taken to
make the process second order accurate.  

\subsubsection{Spacetime-Hydrodynamics Coupling}

Our code evolves the spacetime geometry and the matter fields
separately.  This allows different methods to be used for each
system (spacetime and hydrodynamics).  
The coupling of those different evolution algorithms
in a way that is second order accurate in both
space and time is highly method dependent.  We will therefore discuss
the coupling of each system, ADM or BM, with hydrodynamics, separately.
A summary of the different combined schemes appears
in Table~\ref{table:discrete_names}.

\begin{table}  [tb]
\begin{tabular}{|c|c|c|} \hline
abbreviation & spacetime formulation / evolution scheme &
hydrodynamics update method \\ \hline
ADMLEAP\_ROE & ADM / leapfrog & Roe \\ \hline
ADMLEAP\_FLUX & ADM / leapfrog & flux split \\ \hline
ADMLEAP\_MAR  & ADM / leapfrog & Marquina \\ \hline
ADMICN\_ROE & ADM / iterative Crank-Nicholson & Roe \\ \hline
ADMICN\_FLUX & ADM / iterative Crank-Nicholson & flux split \\ \hline
ADMICN\_MAR  & ADM / iterative Crank-Nicholson & Marquina \\ \hline
BMEIN\_ROE & BM (Einstein) / MacCormack & Roe \\ \hline
BMEIN\_FLUX & BM (Einstein) / MacCormack & flux split \\ \hline
BMEIN\_MAR  & BM (Einstein) / MacCormack & Marquina \\ \hline
BMRIC\_ROE & BM (Ricci) / MacCormack & Roe \\ \hline
BMRIC\_FLUX & BM (Ricci) / MacCormack & flux split \\ \hline
BMRIC\_MAR  & BM (Ricci) / MacCormack & Marquina \\
\end{tabular}
\caption{This table summarizes the abbreviations used for the various
methods used for the spacetime and hydrodynamical evolutions.}
\label{table:discrete_names}
\end{table}

\begin{figure}
\centerline{\psfig{figure=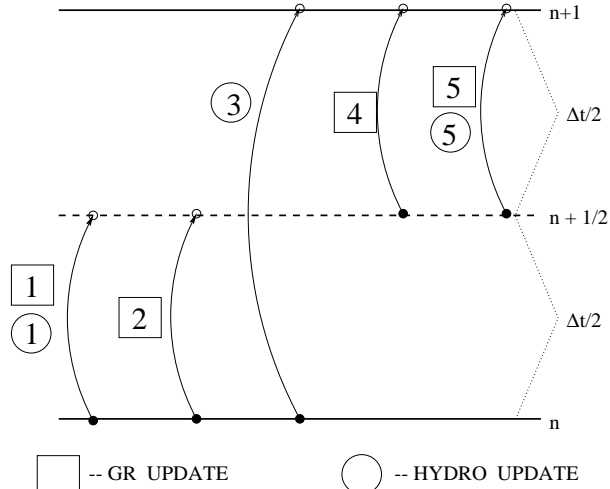,width=8.0cm}}
\caption{This figure represents the coupling between the
hydrodynamics and the MacCormack evolution scheme with the BM 
formulation of the field equations (either BMEIN or BMRIC). 
{\bf STEP 1:} Simultaneous update of the spacetime and 
hydrodynamic variables with the source terms via a two-step
predictor-corrector scheme to the half-timestep $n+1/2$.
{\bf STEP 2: } Update of the spacetime variables with
the flux terms via a MacCormack scheme to the half-timestep $n+1/2$.
{\bf STEP 3: } Update of the hydrodynamic variables with the
flux terms via a two-step predictor-corrector scheme to the $n+1$ step.
{\bf STEP 4: } Same as step 2 but the update is to the final time
$n+1$. {\bf STEP 5:} Same as step 1 but updating both sets of
variables to the final $n+1$ step.}
\label{fig:bm_couple}
\end{figure}

The coupling between the BM system (for both the ``Einstein" (BMEIN) and
``Ricci" (BMRIC) systems) with the
hydrodynamic solver is fairly straightforward as both systems of
equations take a similar (first-order flux-conservative) form.
The steps involved in the coupling are outlined in Fig.~\ref{fig:bm_couple}.
In step 1 we simultaneously update the spacetime and
hydrodynamic variables with the source terms via a two-step
predictor-corrector scheme (second order accurate in time) to
the half-timestep $n+1/2$. In step 2, the spacetime variables
are updated with the flux terms using a
MacCormack scheme (second order accurate
in time) again to the half-timestep $n+1/2$. In
step 3, we update the hydrodynamic variables with the
flux terms via a two-step predictor-corrector scheme
to the $n+1$ step. In step 4 the spacetime variables
are updated with the flux terms
via a MacCormack scheme to the $n+1$ step. Finally, in step 5,
the spacetime and hydrodynamic variables are simultaneously updated
with the source terms via a two-step
predictor-corrector scheme to the final $n+1$ step.

% Figure

\begin{figure}
\centerline{\psfig{figure=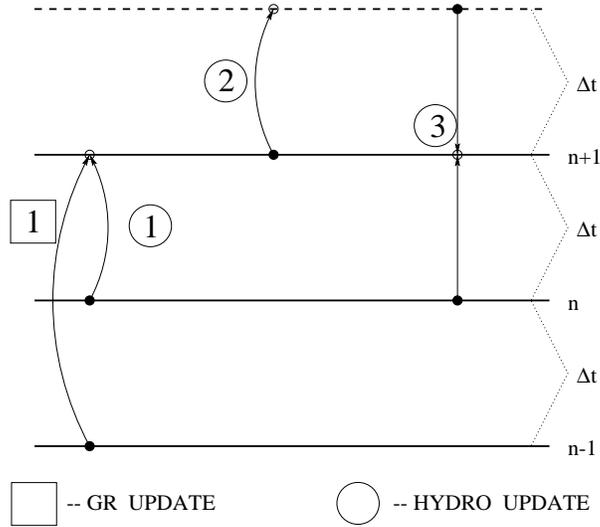,width=8.0cm}}
\caption{This figure represents the coupling between the
hydrodynamics and the leapfrog evolution scheme used for the
ADM spacetime equations (ADMLEAP).  {\bf STEP 1:} Simultaneous update 
of the ADM equations via a leapfrog step (second order accurate in time)
and of the hydrodynamic equations via a (first order accurate in time) 
Euler-predictor step using the method
of lines.  {\bf STEP 2:} Update of the hydrodynamic equations to a virtual
$n+2$ timestep via a first order in time Euler-corrector step.
{\bf STEP 3:} A second order accurate in time
update of the hydrodynamic variables is obtained by averaging the corrected
quantities obtained in step 2 with the original state of step $n$.}
\label{fig:leap_couple}
\end{figure}

In Fig.~\ref{fig:leap_couple} we display the coupling between the
ADM leapfrog evolution (ADMLEAP) and the hydrodynamical evolution.
In step 1 we simultaneously update the ADM equations
via a leapfrog step (second order accurate in time) and update the 
hydrodynamic equations with an Euler-predictor step (first order
in time) using the method of lines. In step 2, we update the
fluid variables to a virtual $n+2$ timestep with a first order in 
time Euler-corrector step using the method of lines. Finally, in step 3, 
we obtain a second order accurate in time update of the hydrodynamic
variables by averaging the corrected quantities obtained in step 2
with the original state of step $n$.

% Figure

\begin{figure}
\centerline{\psfig{figure=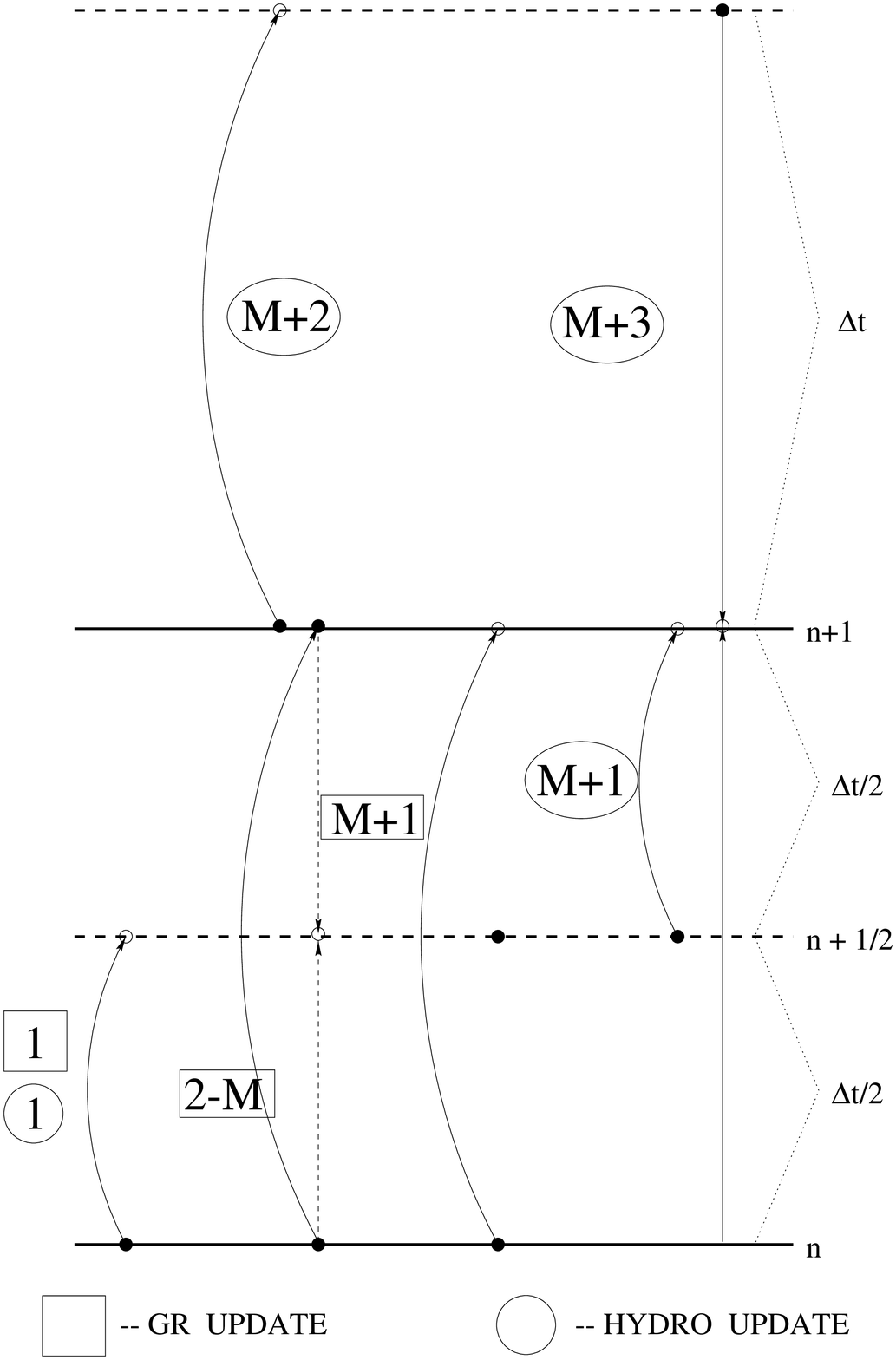,width=8.0cm}}
\caption{This figure represents the coupling between the
hydrodynamic evolution scheme and the iterative Crank-Nicholson method 
used for the integration of the ADM equations (ADMICN).
{\bf STEP 1:} Simultaneous update of the ADM and hydrodynamic equations
via a Euler-predictor step (first order in time) to the half timestep $n+1/2$.  
{\bf STEP 2 through M:} Update of the ADM equations via an iterative
Crank-Nicholson scheme (second order accurate in time) to the 
$n+1$ timestep, then compute a corrected $n+1/2$ state by averaging
the $n+1$ and $n$ states. 
{\bf STEP M+1: } Simultaneous update of the ADM equations via 
a leapfrog step (second order in time) based on the $n$ and 
$n+1/2$ states and the hydrodynamics equations via the second half of
the Euler-predictor step (first half applied in step 1) 
using a method of lines.  
{\bf STEP M+2: } Update of the hydrodynamic equations to a virtual
$n+2$ timestep via a (first order in time) Euler-corrector step
using method of lines.  {\bf STEP M+3: } A second order (in time)
hydrodynamics update is obtained by averaging the corrected quantities 
of step $M+2$ and the original variables of step $n$.}
\label{fig:icn_couple}
\end{figure}

Our last combination appears plotted in 
Fig.~\ref{fig:icn_couple}. Here we display the coupling between the
iterative Crank-Nicholson evolution scheme for the ADM equations (ADMICN)
and the hydrodynamical evolution. First, in step 1,
we simultaneously update the ADM and hydrodynamic equations
using an Euler-predictor step, which is first-order order accurate in time,
to the half timestep $n+1/2$. In step 2 through M, we update the ADM 
equations via an iterative Crank-Nicholson scheme (second order in time) 
to the $n+1$ timestep and average the $n+1$ and $n$ states to produce
a corrected $n+1/2$ state.
The solution is guaranteed to be second 
order accurate (in time) for $M \geq 2$. In step M+1 we simultaneously
update the ADM equations via a leapfrog step (second order in time) based 
on the $n$ and $n+1/2$ states and the hydrodynamic equations via the second 
half of the Euler-predictor step (first half applied in step 1) using a 
method of lines. In step M+2 the hydrodynamic equations are updated to a
virtual $n+2$ timestep via an Euler-corrector step using a method of lines.
Finally, in step M+3 we obtain a (second order in time)
hydrodynamics update by averaging the corrected variables 
obtained in step $M+2$ and the original state of step $n$.

% Figure

\section{Shock tube tests}
\label{sec:shocktube}

We start testing the code with one
of the standard tests in fluid dynamics, 
the shock tube (see 
references~\cite{Marti96,Norman86,Schneider90,Donat98} for a sample of 
previous relativistic simulations). In this test, the fluid initially
has two different thermodynamical states on either side
of an interface. When this interface is removed, 
the fluid evolves in such a way that
four states appear.
Each state is separated by one of three elementary waves: a shock 
wave, a contact discontinuity, and a rarefaction wave. This time-dependent
problem has an exact solution to which our numerical integration 
can be compared. This problem only checks the hydrodynamical
part of the code, as it assumes a flat background metric. However,
it provides a good test of the shock capturing properties of
any HRSC scheme.  The integration of the
hydrodynamic equations in each of the three spatial directions 
can be tested  independently by placing the initial discontinuity along each
of the coordinate axes or in the fully multidimensional
case when the interface is placed along the main diagonal
of the computational domain.

The initial state of the fluid is specified by
$P_L=13.3$, $\rho_L=10$ on the left side of the interface and
$P_R=0.66\cdot 10^{-6}$, $\rho_R=1$ on the right side. 
The fluid is assumed to
be initially at rest on both sides of the interface. 
We have tested the code for each direction, $x$, $y$, and $z$, 
separately. The integration domain extends from $-0.5$ to $0.5$ and
at $t=0$ the interface is placed at $x=0$ (similarly 
when testing in the $y$ and
$z$ directions). We use a perfect fluid equation of state,
$P=(\Gamma-1)\rho\epsilon$ with $\Gamma=5/3$.

\begin{center}
\begin{figure}
\centerline{\psfig{figure=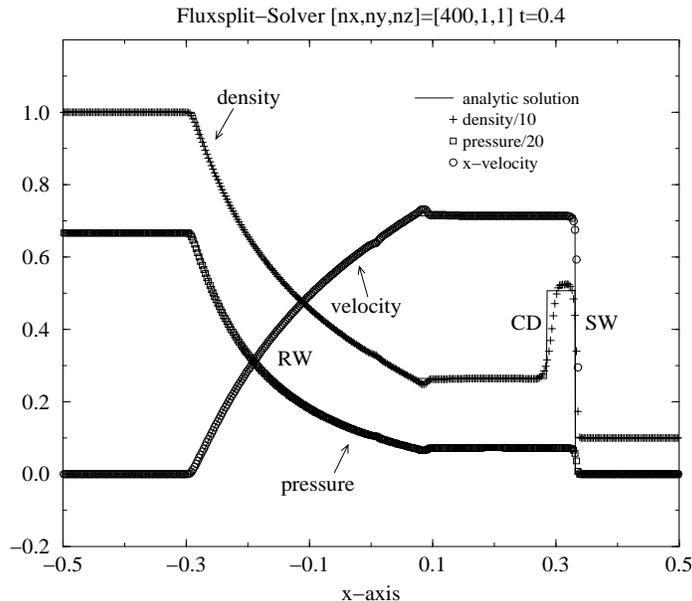,width=11.0cm,height=9.0cm}}
\caption{Numerical (symbols) versus analytic (solid lines) solution for 
the shock tube test for time $t=0.4$. The initial discontinuity is
placed along the $x$-axis. The numerical results are
obtained with the flux split method. We show normalized profiles
of density, pressure, and velocity as a function of the $x$ coordinate.
We use a uniform grid of $400\times 1\times 1$ zones for the
$x$, $y$, and $z$ directions respectively. The main features of this
time-dependent problem are the leading shock wave (`SW'), the trailing
rarefaction wave (`RW'), and the contact discontinuity (`CD'). The latter
discontinuity is only present in the density.  Note the ``overshooting" in the
velocity and the ``undershooting" in density and pressure, found at the
leading part of the rarefaction wave ($x\approx 0.1$). Also note the
higher density between the shock and the
contact discontinuity as well as the deviation from the exact solution 
in the transonic rarefaction region around $x=0$, typical of non-entropy
satisfying schemes.}
\label{fig:shocktube1}
\end{figure}
\end{center}

\begin{center}
\begin{figure}
\centerline{\psfig{figure=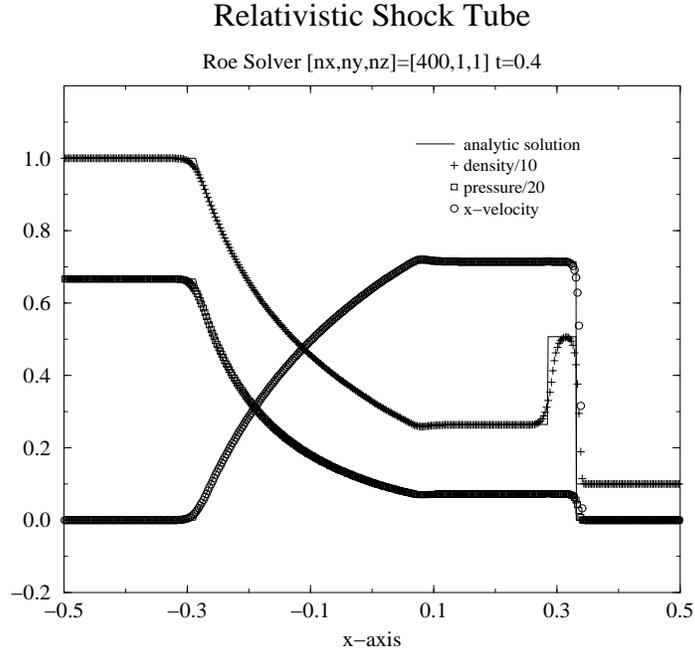,width=11.0cm,height=9.0cm}}
\caption{
Numerical (symbols) versus analytic (solid lines) results for the 
one-dimensional shock tube tests at $t=0.4$ using Roe's method. Shown are 
normalized profiles of density, pressure, and velocity as functions of
the $x$ coordinate. A uniform grid of $400\times 1\times 1$ zones for the
$x$, $y$, and $z$ directions, respectively, was used.
}
\label{fig:shocktube2}
\end{figure}
\end{center}

\begin{center}
\begin{figure}
\centerline{\psfig{figure=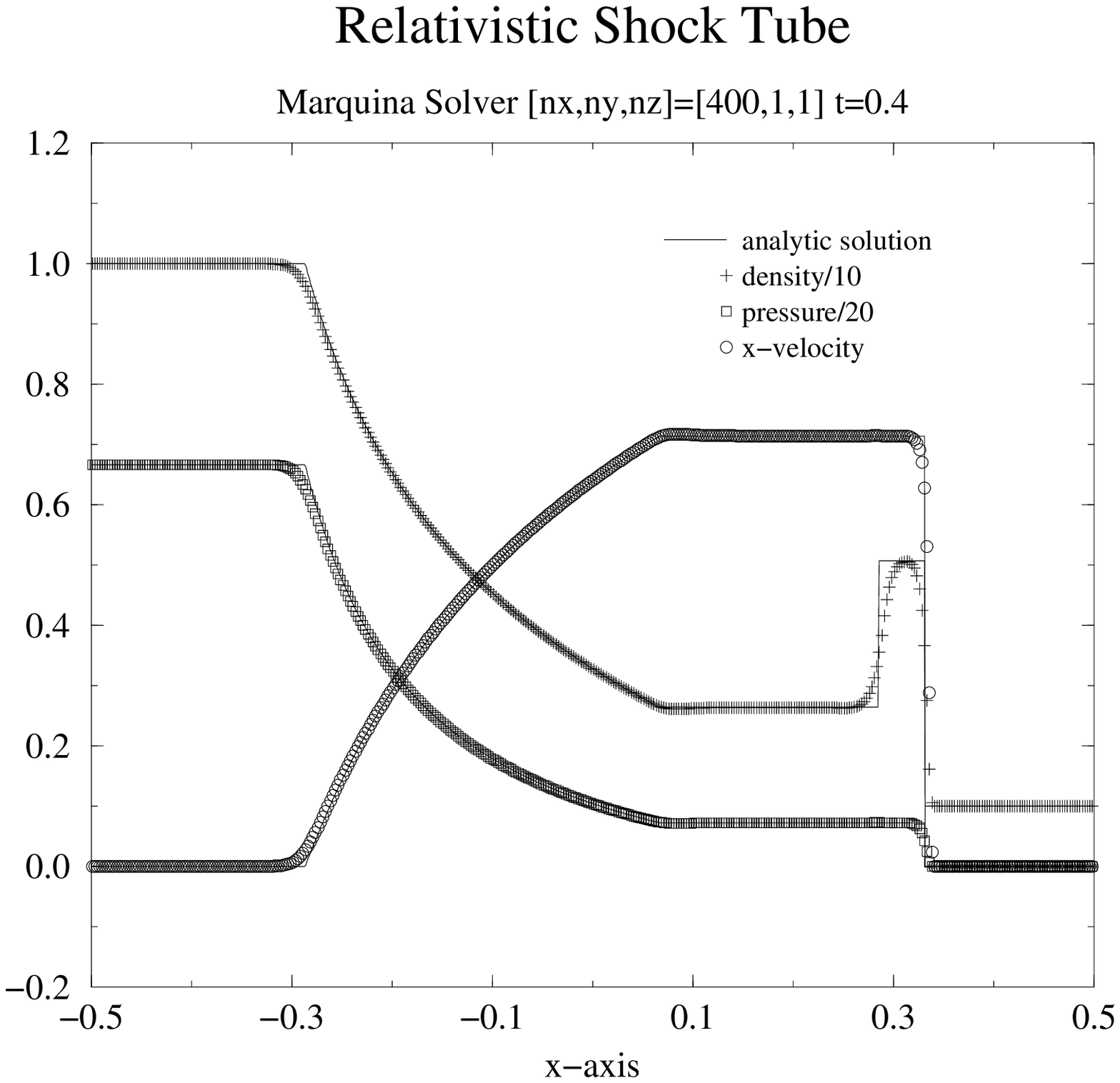,width=11.0cm,height=9.0cm}}
\caption{
Numerical (symbols) versus analytic (solid lines) results for the 
one-dimensional shock tube tests at $t=0.4$ using Marquina's method. 
Shown are normalized profiles of density, pressure, 
and velocity as functions of
the $x$ coordinate. A uniform grid of $400\times 1\times 1$ zones for the
$x$, $y$, and $z$ directions, respectively, was used.
}
\label{fig:shocktube3}
\end{figure}
\end{center}

We present the results of the evolution in Figs.~\ref{fig:shocktube1},
\ref{fig:shocktube2}, and~\ref{fig:shocktube3} . 
These figures correspond to, effectively, 
one-dimensional evolutions, as the discontinuity is located along the
$x$ axis. We do not show results for the $y$ and $z$ directions as they
exhibit no differences with respect to the $x$ direction.
Fig.~\ref{fig:shocktube1} shows the results obtained with the flux split
method. Fig.~\ref{fig:shocktube2} corresponds to the Roe method
and Fig.~\ref{fig:shocktube3} to the 
Marquina solver.  The solid lines
represent the exact solution while the different symbols indicate the
numerical results for the
density (plus signs), pressure (squares), both scaled to 
fit on the same graph, and
velocity (circles). We use a grid of $400$ zones along the
relevant direction ($x$ in this case) and one zone in the two other 
directions. The final time of the evolution is $t=0.4$ which corresponds
to $320$ iterations with a CFL number of 0.5.
From these figures we conclude that our numerical
evolutions show good agreement with the analytic solution. All
features of the solution, the trailing rarefaction wave (labeled `RW'
in Fig.~\ref{fig:shocktube1}), the leading
shock wave (`SW' in Fig.~\ref{fig:shocktube1}), and
the contact discontinuity which is only present in the density
(`CD' in Fig.~\ref{fig:shocktube1})
are well resolved. 
The constant state in the density between
the shock and the contact discontinuity is properly captured with $400$
zones. In order to quantify the quality of the simulations we
indicate in Table~\ref{table:shocktube} the L1-norm errors of the different 
hydrodynamic quantities.

By direct inspection of the figures and the table we see that
the Marquina method gives better results than the other two schemes.
The improvement with respect to the Roe solver is not too sensitive
for high resolutions ($\Delta x=1/400$). It is however quite relevant for the
coarse grid ($\Delta x=1/200$) evolutions.
Also noticeable are the large errors found with the flux split
method for the coarse grid although they are
drastically reduced when the resolution is doubled.
All errors reported in Table II are measured taking into account the whole
domain of integration, i.e., including the discontinuities.
Obviously, the smooth parts of the solution, e.g., the
rarefaction wave, have less numerical error.
It is also of interest
to mention the numerical ``kinks" (under or overestimations of the
solution at the leading edge of the rarefaction wave)
found with the flux split method.
In addition, this method shows some remnant
of the initial discontinuity at $x=0$. This is
a common feature of all non-entropy satisfying schemes.
These features are absent in the other two schemes.
Finally, we note that
the flux split method leads to a slightly larger density between
the leading shock wave and the contact discontinuity.

\vspace{0.5cm}
\begin{table}
\begin{tabular}{|c|c|c|c|c|c|} 
   Dimension & $\Delta x=\Delta y =\Delta z=$ & Solver & $||E(\rho)||_1$ &
$||E(v)||_1$ & $||E(p)||_1$ \\
\hline
Along axis & $\frac{1}{200}$          & Flux-Split & 
$1.99\cdot 10^{0}$ & $3.46\cdot 10^{-1}$ & $2.88\cdot 10^{0}$  \\
   &                          & Roe        & 
$1.19\cdot 10^{-1}$ & $1.36\cdot 10^{-2}$ & $8.15\cdot 10^{-2}$ \\
   &                          & Marquina   & 
$7.65\cdot 10^{-2}$ & $8.13\cdot 10^{-3}$ & $4.60\cdot 10^{-2}$ \\ \hline
   & $\frac{1}{400}$          & Flux-Split & 
$6.61\cdot 10^{-2}$ & $6.67\cdot 10^{-3}$ & $4.25\cdot 10^{-2}$  \\ 
   &                          & Roe        & 
$6.90\cdot 10^{-2}$ & $7.72\cdot 10^{-3}$ & $4.33\cdot 10^{-2}$  \\
   &                          & Marquina   & 
$4.65\cdot 10^{-2}$ & $4.84\cdot 10^{-3}$ & $2.41\cdot 10^{-2}$  \\ \hline
diagonal & $\frac{1/\sqrt{3}}{128}$ & Flux-Split        &
$7.95\cdot 10^{-2}$ & $8.35\cdot 10^{-3}$ & $6.62\cdot 10^{-2}$  \\ 
   &                          & Roe        & 
$9.12\cdot 10^{-2}$ & $9.39\cdot 10^{-3}$ & $7.53\cdot 10^{-2}$ \\
   &                          & Marquina        & 
$9.23\cdot 10^{-2}$ & $9.66\cdot 10^{-3}$ & $7.98\cdot 10^{-2}$ \\
\end{tabular}
\caption{
L1-norm errors of different hydrodynamical quantities, density,
velocity, and pressure for the shock tube tests. The results correspond
to the three different schemes we employ to integrate the hydrodynamic
equations.  The waves along the axis use either $200$ of $400$ 
grid zones in the direction of propagation, and one zone in the
remaining two directions.  This allows our 3D code to be effectively
run as a 1D code.
The diagonal
shock tube test is run with $128$ grid zones in each direction. All three
solvers are found to perform nicely in the multidimensional case.
}
\label{table:shocktube}
\end{table}
\vspace{0.5cm}

% Figure

\begin{center}
\begin{figure}
\centerline{\psfig{figure=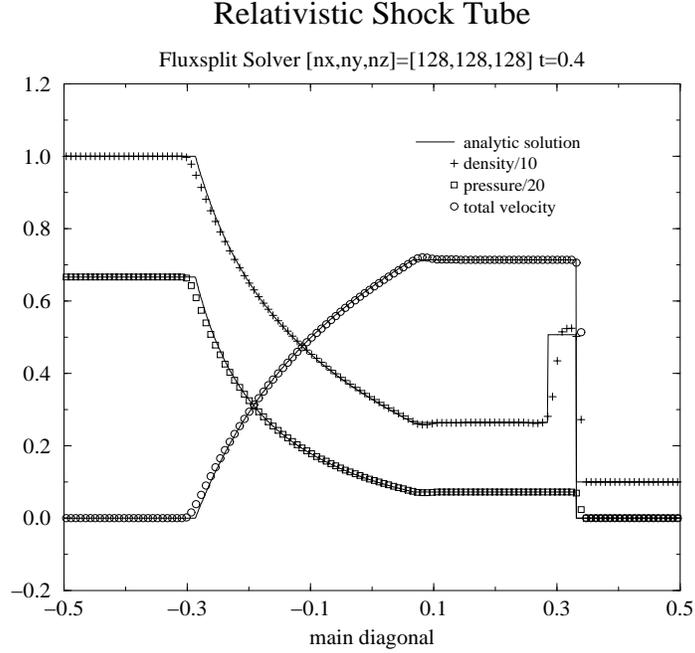,width=11.0cm,height=9.0cm}}
\caption{
Numerical (symbols) versus analytic (solid lines) solution for the 
three-dimensional shock tube test at time $t=0.4$. The initial 
discontinuity is
placed along the main diagonal of a $(1/3)^{3/2}$ volume cube. 
The numerical results are obtained with the flux split method. We show 
the normalized profiles of density, pressure, and velocity as functions of 
the coordinate distance along the main diagonal. 
A uniform Cartesian grid of $128^3$ zones was used.
Note the small deviations in the leading part of the rarefaction wave and
that the density plateau between the
shock front and the contact discontinuity is higher than the analytic solution.
}
\label{fig:shocktube4}
\end{figure}
\end{center}

\begin{center}
\begin{figure}
\centerline{\psfig{figure=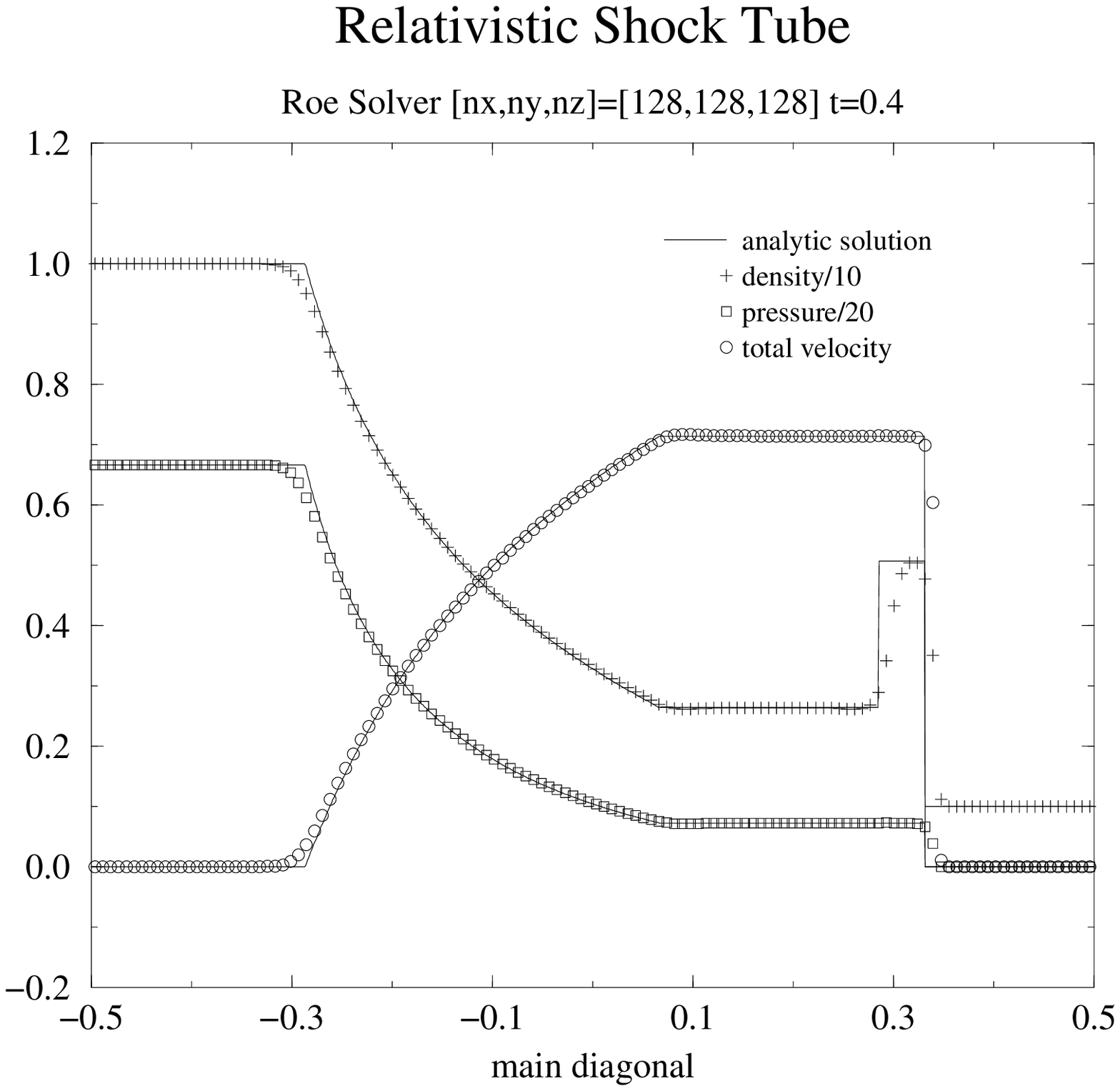,width=11.0cm,height=9.0cm}}
\caption{
Numerical (symbols) versus analytic (solid lines) results for the 
three-dimensional shock tube tests at $t=0.4$ using Roe's method.
Shown are 
normalized profiles of density, pressure, and velocity as functions of
the coordinate distance along the main diagonal. A uniform Cartesian grid
of $128^3$ zones was used.
}
\label{fig:shocktube5}
\end{figure}
\end{center}

\begin{center}
\begin{figure}
\centerline{\psfig{figure=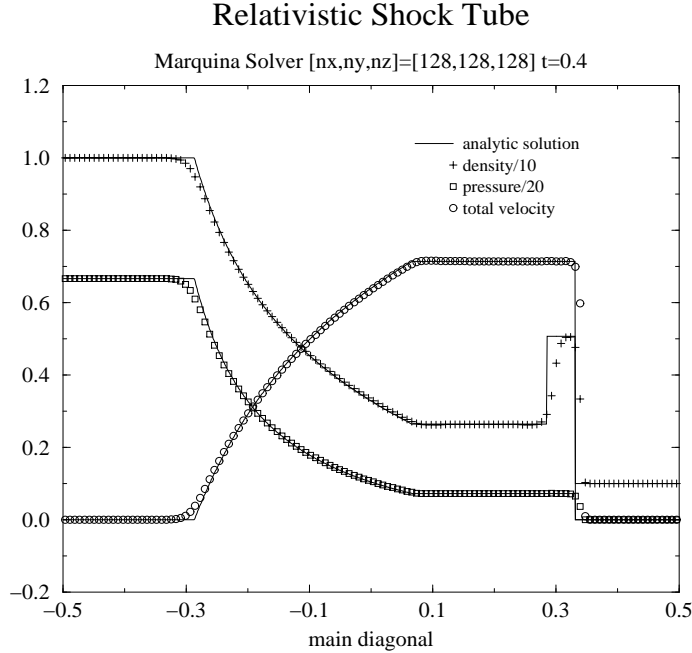,width=11.0cm,height=9.0cm}}
\caption{
Numerical (symbols) versus analytic (solid lines) results for the 
three-dimensional shock tube tests at $t=0.4$ using Marquina's method.
Shown are normalized profiles of density, pressure, and velocity 
as functions of the coordinate distance along the main diagonal. 
A uniform Cartesian grid of $128^3$ zones was used.
}
\label{fig:shocktube6}
\end{figure}
\end{center}

We next test the code by placing the initial discontinuity along the
diagonal of the computational domain. With this setup we are checking the
finite-differencing of all three directions simultaneously. We
consider a grid of $128^3$ zones, spanning an interval of length
$1/\sqrt{3}$ in every direction. The diagonal of the cube therefore has
unit length. We evolve to the same time as in the 1D tests,
$t=0.4$, but discovered we need a lower CFL factor in order to
get stable evolutions. We found that a CFL number of $0.25$ sufficed for
this purpose.
This corresponds to $640$ update iterations.
The results of the evolution for flux split, Roe, and
Marquina's method are depicted in 
Figs.~\ref{fig:shocktube4}-~\ref{fig:shocktube6}.
We find good agreement
between the numerical and analytic results. Notice that, due to the lack
of resolution, some features such as the constant intermediate state
in density, are less resolved than in the one dimensional case. 
The errors of the hydrodynamic quantities for this run are also contained in
Table~\ref{table:shocktube}.  By inspection of this table, we observe 
that, although the errors are very similar for all schemes,
the flux split method is slightly more accurate than both Roe and 
Marquina's method.  However,
the existence of the aforementioned ``kink" in the
leading part of the rarefaction wave (which was more
clear Fig.~\ref{fig:shocktube1}) is still present in the flux
split method.

% Figure

% Figure

% Figure

% Figure

% Figure

We summarize this section by stressing the shock-capturing 
capabilities of the different numerical schemes we use to
integrate the hydrodynamic equations. Such capabilities 
are essential to
our final goal of performing accurate simulations of interesting
astrophysical scenarios, such as coalescing NS binaries.

\section{Friedmann-Robertson-Walker cosmology tests}
\label{sec:frw}

For the first testbed of the coupled GR-Hydro code with dynamical
spacetimes, we use the Friedmann-Robertson-Walker (FRW) model of an expanding
cosmology.  We use the standard form of the FRW metric
\begin{equation}
ds^2 = -dt^2 + R^2(t) \left[ \frac{dr^2}{1-kr^2} + r^2 ( d\theta^2 
+ sin^2\theta d\phi^2 ) \right],
\label{frw_metric}
\end{equation}
corresponding to an open ($k=-1$), flat ($k=0$), or closed ($k=1$)
universe, with scale factor $R(t)$.  For the special case $k=0$,
each constant time slice is spatially flat.  For this case, all
terms involving spatial derivatives drop out of both the spacetime and
hydrodynamic evolution equations.  
Although we still use the fully general form of the evolution
equations to evolve the initial data,
this allows us to concentrate on
the coupling of the two codes in time.  We note
that the case $k=0$ has been extensively used in the literature for
calibration of cosmological codes.  We also study the non-spatially flat 
case $k=-1$, which involves non-trivial spatial derivatives.

We take the matter to be collisionless dust, $P=\epsilon=0$. 
Under these assumptions, the hydrodynamic evolution equations reduce to
\begin{equation}
\partial_t ( \rho R^3) = 0,
\label{eq:frw_hydro}
\end{equation}
while the Einstein equations take the form
\begin{equation}
(\partial_t R)^2 + k = \frac {8 \pi}{3} \rho R^2,
\label{eq:frw_gr}
\end{equation}
(see, for example, \cite{Schutz85}).

It is important to stress that we are using the {\it full} set 
of evolution equations to integrate the initial data.  
The only assumption we make is that the stress-energy tensor
take the form of a perfect fluid with zero pressure and internal energy.
We only use the simplified solutions given by Eqs.~(\ref{eq:frw_hydro},
\ref{eq:frw_gr}) to check the validity of our numerical evolutions.

\subsection{k=0 Convergence Tests}

For the $k=0$ case we have a flat spatial geometry.  Notice
that Eq.~(\ref{eq:frw_hydro}) implies
that the evolved ``densitized" variable $\tilde{D}$ is a constant of motion
\begin{equation}
\tilde{D} = \sqrt{\gamma} W \rho = \rho R^3 = constant.
\end{equation}
This, combined with the fact that
no fields have any spatial dependence, results in the hydrodynamical
finite difference equations becoming exact for our choice of variables.
Therefore, we see no difference between the three methods which we
use to integrate the hydrodynamical variables.  
Because of this, we only
present results for the flux split hydrodynamics method, and note
that the results for Roe and Marquina methods are equivalent.
We concentrate on the spacetime
evolution as driven by the hydrodynamic source.  This is the first
direct test of the spacetime-hydrodynamics coupling aspects of our code.

Since numerical error only comes from finite differencing
errors in time, we could
run the convergence tests by simply decreasing the CFL number.
However, to minimize boundary effects, we run the convergence
tests at three different resolutions: $40^3, 20^3, 10^3$ with a fixed
CFL factor as summarized in Table \ref{table:frwk0grid}.
For initial conditions we choose $R(t=0)=1$, and $\rho(t=0)=0.01$.

\begin{figure}
\psfig{figure=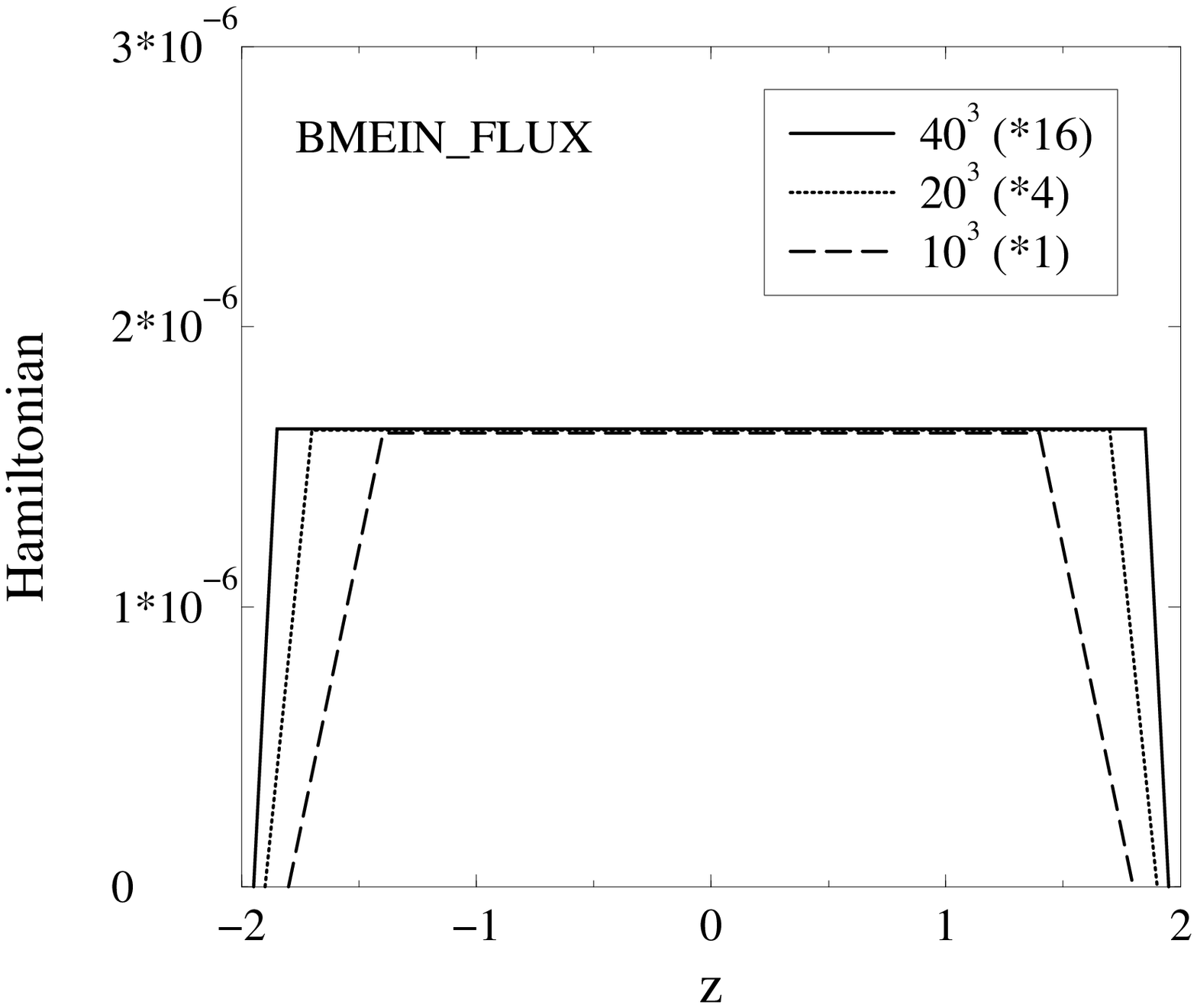,width=7.5cm}
\vspace{-6.4cm}
\hspace{7.5cm}
\psfig{figure=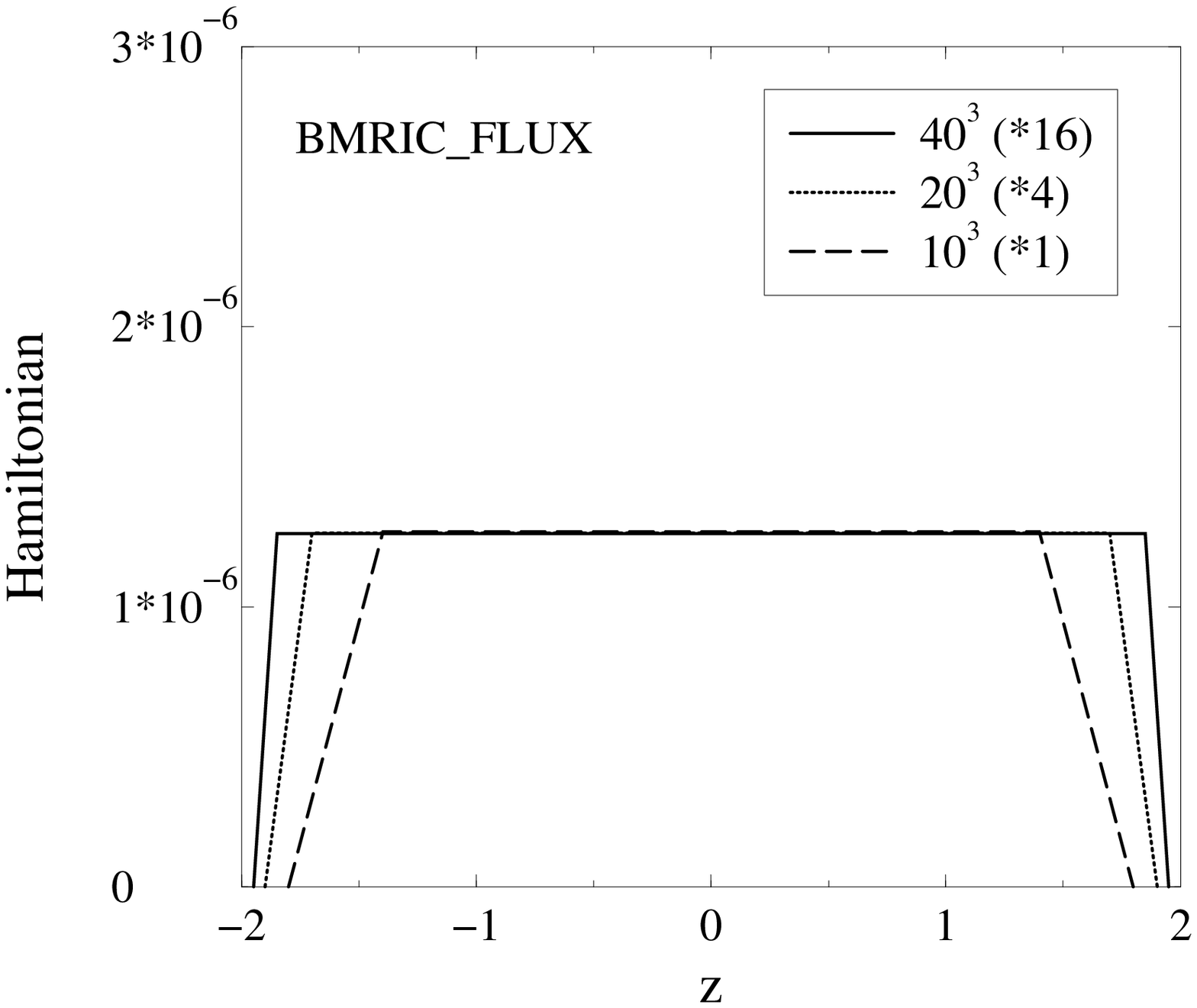,width=7.5cm}
\caption{The Hamiltonian constraint violation
for the BMEIN\_FLUX and BMRIC\_FLUX systems is compared.  Since the
hydrodynamic code gives the exact solution for the special case of
$k=0$ initial data, there is no need to monitor the errors in the
matter field evolutions, or to compare different hydrodynamic methods.  We
find that both the BMEIN\_FLUX and BMRIC\_FLUX systems converge to
second order, with BMRIC giving a slightly smaller value for
the Hamiltonian constraint violation.}
\label{fig:k0_bm_flux}
\end{figure}

In Fig.~\ref{fig:k0_bm_flux} we compare the Hamiltonian constraint
violations for both the BMEIN\_FLUX (``Einstein" system in BM formulation) and the BMRIC\_FLUX (``Ricci" system) schemes. 
As explained in section~\ref{sec:discrete} these systems represent
two different ways of casting the Einstein evolution equations in
hyperbolic form~\cite{Bona98b}.  We find that both systems converge to
second order, with the BMRIC system giving a slightly smaller value
for the Hamiltonian constraint.  The long term stability of the two systems
will be analyzed in forthcoming investigations.

\begin{table}
\begin{tabular}{|c||c|c|c|c|}
 & \# of points & & & \\
resolution & in each &$\Delta x$ & $c \frac {\Delta t}{\Delta x}$ &
\# of timesteps\\
                & direction    &       &  & \\ \hline
low             & 10           & 0.01  & 0.25 & 1 \\
medium      & 20               & 0.05  & 0.25 & 2 \\
high            & 40           & 0.025 & 0.25 & 4 \\
\end{tabular}
\caption{Different run parameters for the FRW $k=0$ convergence tests.}
\label{table:frwk0grid}
\end{table}

% Figure 

\begin{figure}
\psfig{figure=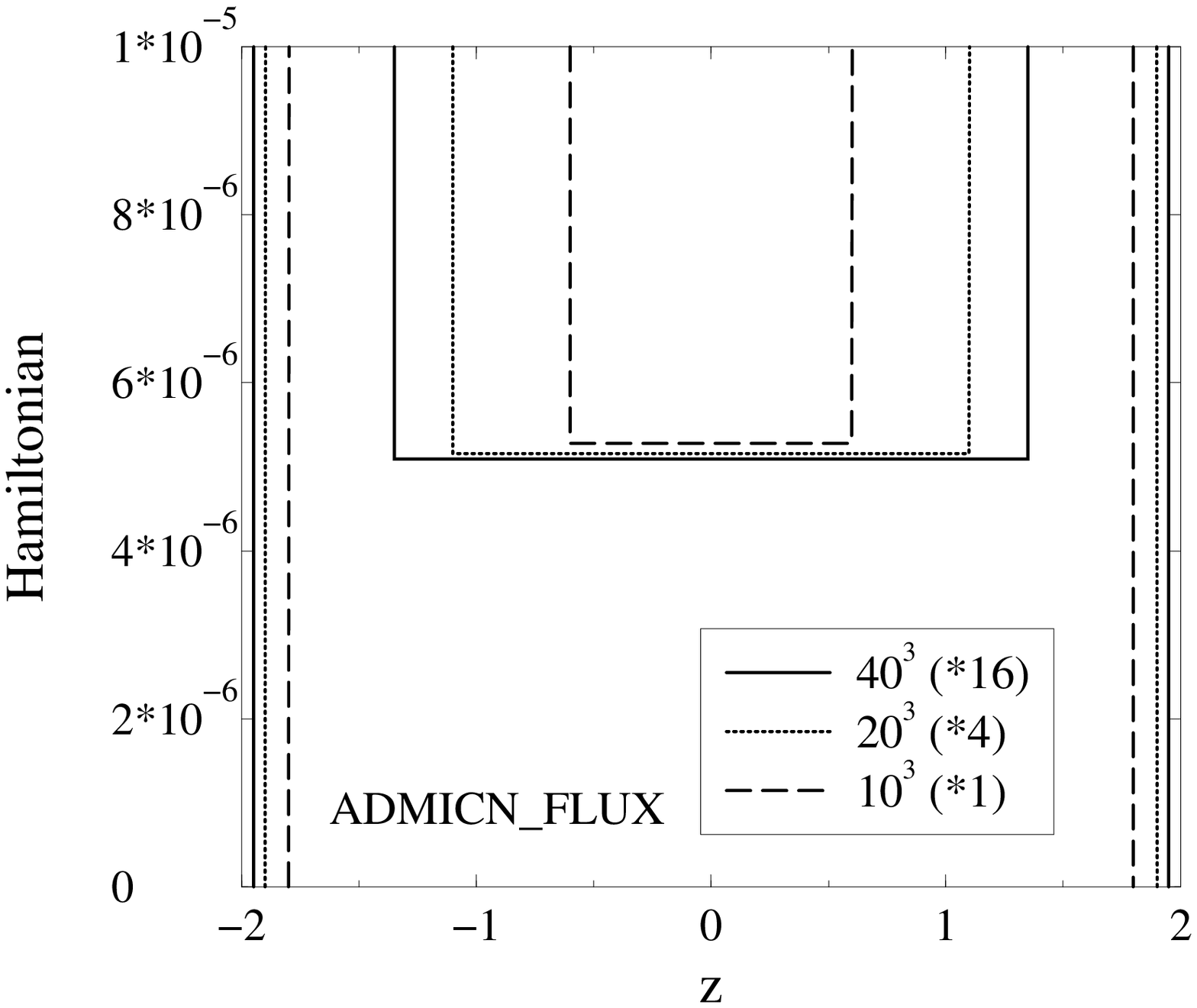,width=7.5cm}
\vspace{-6.4cm}
\hspace{7.5cm}
\psfig{figure=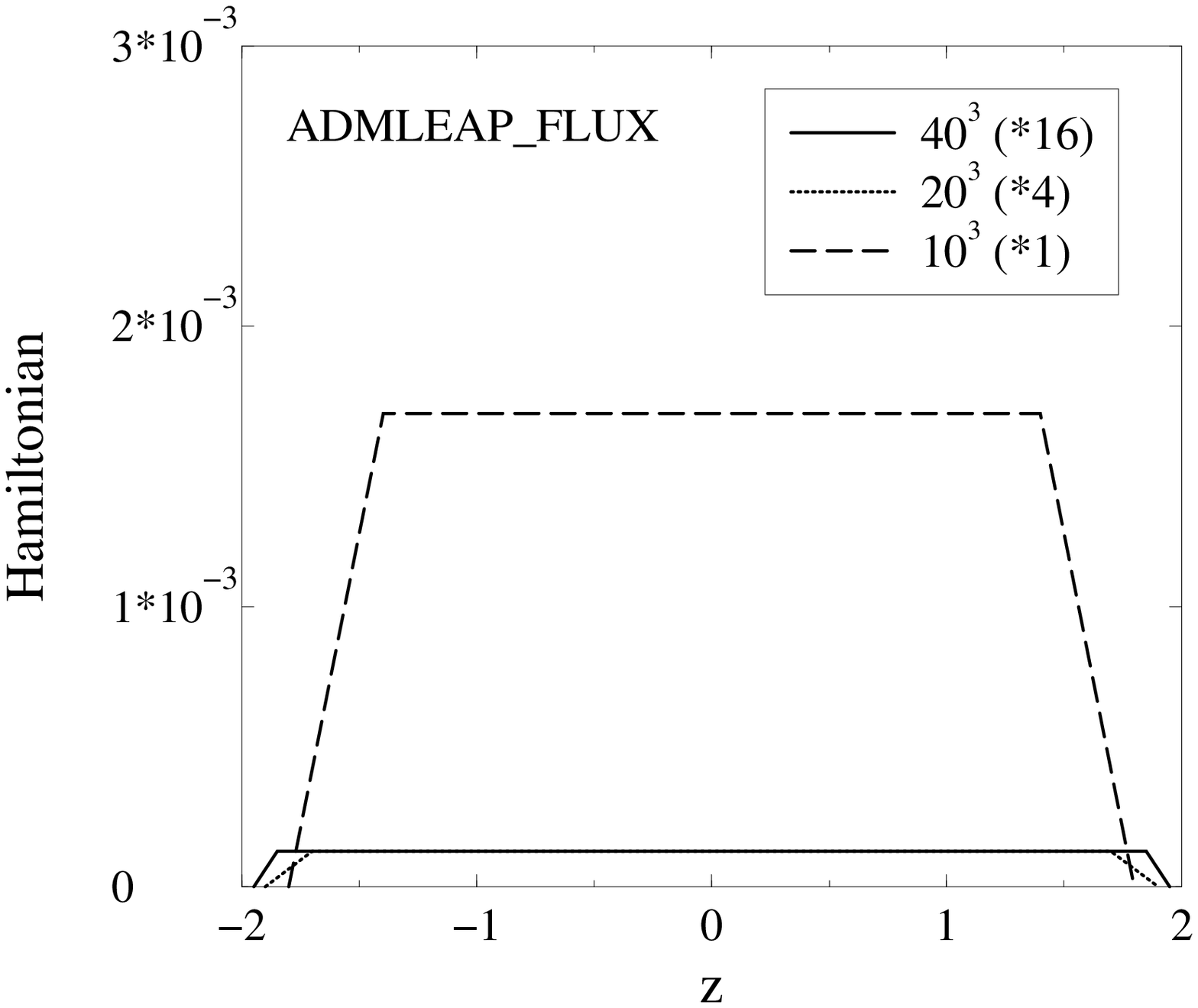,width=7.5cm}
\caption{Comparison of the Hamiltonian constraint values
for the ADMICN\_FLUX (left plot) and ADMLEAP\_FLUX systems. 
For the ADMICN\_FLUX system we obtain second order convergence apart from some 
numerical ``noise" at the boundaries.  
The ADMLEAP\_FLUX system does {\it not} appear to be converging.}
\label{fig:k0_adm_flux}
\end{figure}

We turn next to evolutions with ADM based evolution schemes.
In Fig.~\ref{fig:k0_adm_flux} we 
compare the Hamiltonian constraint for the ADMICN\_FLUX (iterative
Crank-Nicholson) and ADMLEAP\_FLUX (leapfrog) systems.  For the
ADMICN\_FLUX system we get second order convergence, with some ``noise''
at the boundaries.  This noise propagates into the grid more quickly
than with other methods, due to the iterative nature of the
ICN scheme.  

\begin{figure}
\psfig{figure=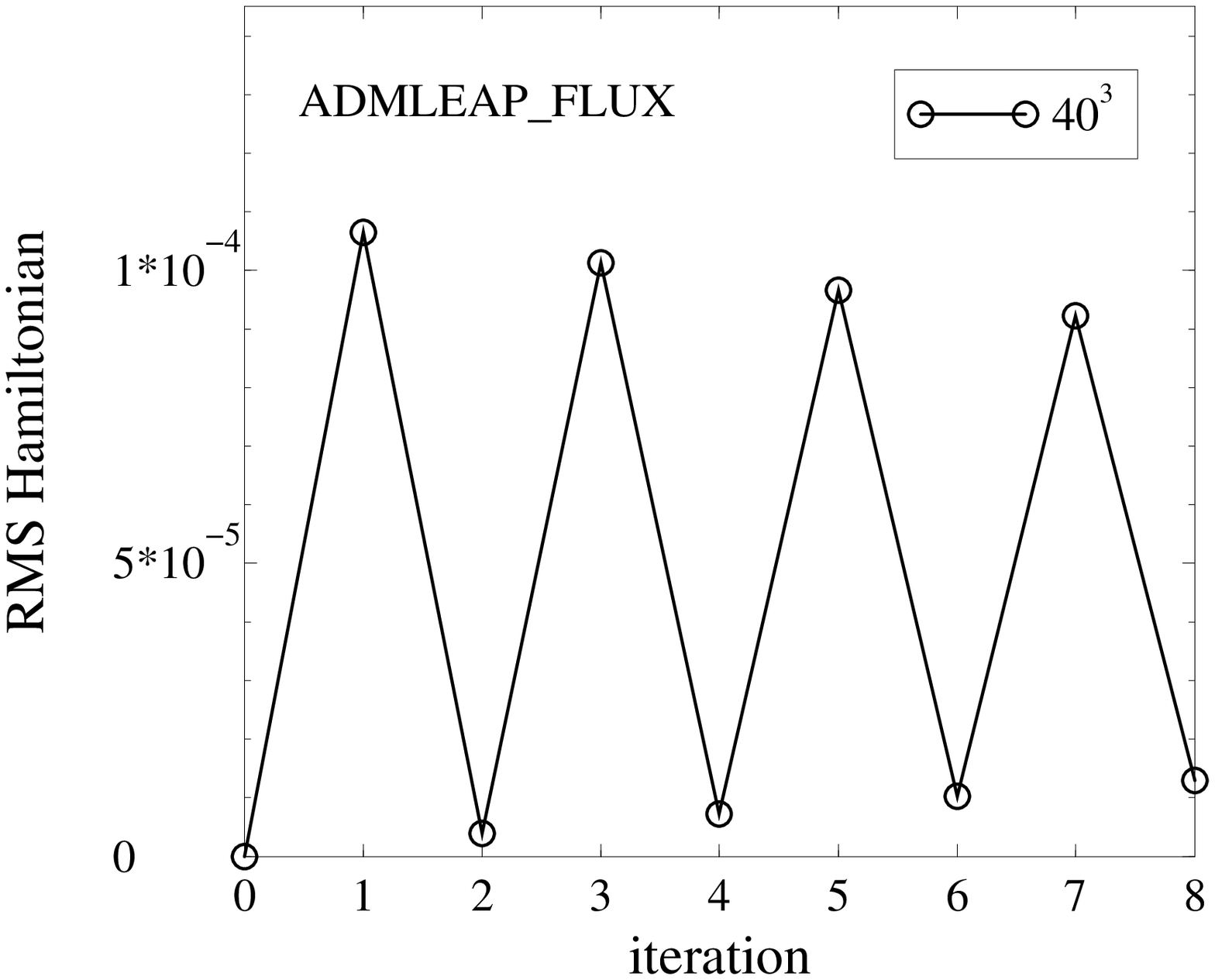,width=7.5cm}
\vspace{-6.4cm}
\hspace{7.5cm}
\psfig{figure=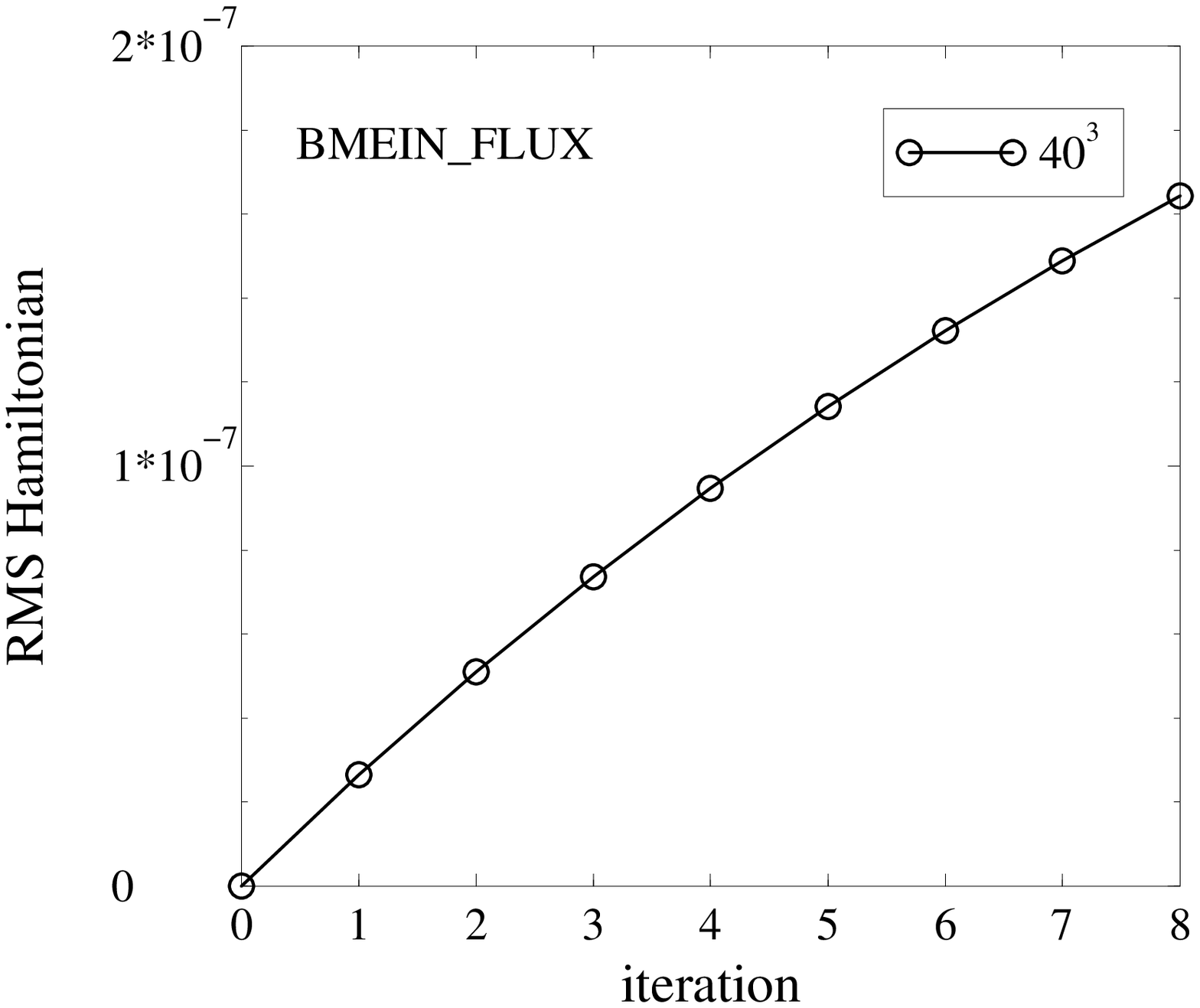,width=7.5cm}
\caption{Plot of the RMS Hamiltonian as a function of the time step
for the ADMLEAP\_FLUX (left) and BMEIN\_FLUX systems. We observe that the Hamiltonian is oscillating only for the ADMLEAP\_FLUX scheme.  
This has been reported in the literature as a non-physical ``loused"
solution. 
The degree to which the oscillations occur depends strongly on the 
initial data and gauge choice.}
\label{fig:loused}
\end{figure}

The ADMLEAP\_FLUX system does {\it not} 
appear to be converging at second order.
When we plot the root mean square (RMS) Hamiltonian 
as a function of time step for this system, 
Fig.~\ref{fig:loused}, we observe an oscillatory behavior. This appears
to be a ``loused'' solution to the
finite difference equations, as described by New et al.~\cite{New98},
occurring when non-staggered leapfrog methods are used to evolve
certain non-linear systems of equations.  The ``loused'' solution is a
non-physical solution characterized by oscillations from time step to
time step.  We have seen evidence of this solution only when
using the ADMLEAP system, both in vacuum evolutions, and when coupled to
hydrodynamics. It does not appear with any other system. For
comparison, see the RMS Hamiltonian for the BMEIN\_FLUX evolution in 
Fig.\ref{fig:loused}.  The degree to which this ``loused'' solution 
occurs depends strongly on the initial data and choice of gauge.

% Figure 

% Figure 

\begin{figure}
\psfig{figure=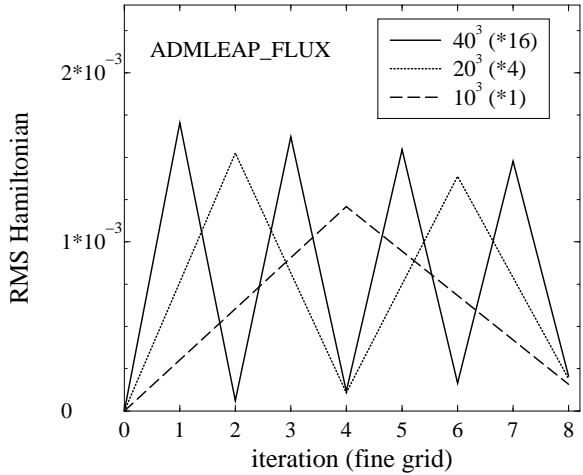,width=3.0in}
\caption{The convergence of the ``loused'' oscillations appearing 
for the ADMLEAP\_FLUX system is examined
by plotting the RMS Hamiltonian at each
iteration for three different resolutions: $40^3, 20^3,$ and $10^3$.
The value is scaled by a factor of sixteen
for the finest resolution grid, and a factor of four for the medium
resolution grid.  The iteration number refers to the finest grid.
As explained in the text, iteration eight is the first time when all
three grids coincide to a minimum value. 
At this point we see that the solution
converges at second order.}
\label{fig:loused_admleap_conv}
\end{figure}

It is important to note that this ``loused'' solution is a solution
to the finite difference equations, {\it not} to the 
differential equations.  Thus, it must
converge away with increased resolution.  In
Fig.~\ref{fig:loused_admleap_conv} we show the RMS Hamiltonian at each
iteration for three different resolutions: $40^3, 20^3,$ and $10^3$.
Its value has been scaled by a factor of
sixteen for the finest resolution grid, and a factor of four for the
medium resolution grid, so that if the solution is converging to
second order, the value graphed should remain constant.  The iteration
number refers to the finest grid.  Since
the ``loused'' solution oscillates between a maximum and minimum value
every time step, at iteration number four the fine and medium grid
correspond to a minimum (they have evolved an even number of time steps)
while the coarse grid is at a maximum (having evolved an odd number of
time steps).  Hence, iteration eight is the first time when all three 
grids correspond to a minimum. At this point 
we see that the solution is indeed converging at second order.

To summarize this subsection, we have verified that the coupling
between the spacetime and hydrodynamic methods, described in
Section ~\ref{sec:discrete}, yield second order convergence
in time.  We have seen evidence of a ``loused'' solution in the
ADMLEAP system.  This ``loused'' solution produces a non-physical
oscillation in time.  By comparing this oscillation at three
resolutions, and at the same stage of oscillation (when all three
resolutions are at a minimum), we see that it is converging away.

% Figure 

\subsection{k=-1 Convergence Tests}

Having confirmed the fidelity of the (second order
accurate) coupling in time between the spacetime and hydrodynamical
evolutions, we next test the spatial
derivatives by studying the non-spatially flat $k=-1$ FRW solution.
Since we now have to resolve spatial gradients, we increase the grid
resolution using $160^3, 80^3,$ and $40^3$ zones. The
different runs are summarized in Table
\ref{table:frwk-1grid}.  
For initial conditions we choose $R(t=0)=1$, and $\rho(t=0)=0.01$.

To monitor the correctness of the spacetime evolution we cut off our
grid at $x=y=z=\pm 1$ and ignore errors caused by boundary effects.
To analyze the spacetime, we look at the
Hamiltonian constraint, $H$, and the x-component of
the momentum constraint, $M^x$. Correspondingly, to address the accuracy of the
hydrodynamic evolution we look at the conservation of $\rho R^3$,
$\Delta (\rho R^3) = \rho R^3 |_{t} - \rho R^3 |_{t=t_0}$.  This
quantity should remain constant for a matter dominated FRW solution.

\begin{figure}
\psfig{figure=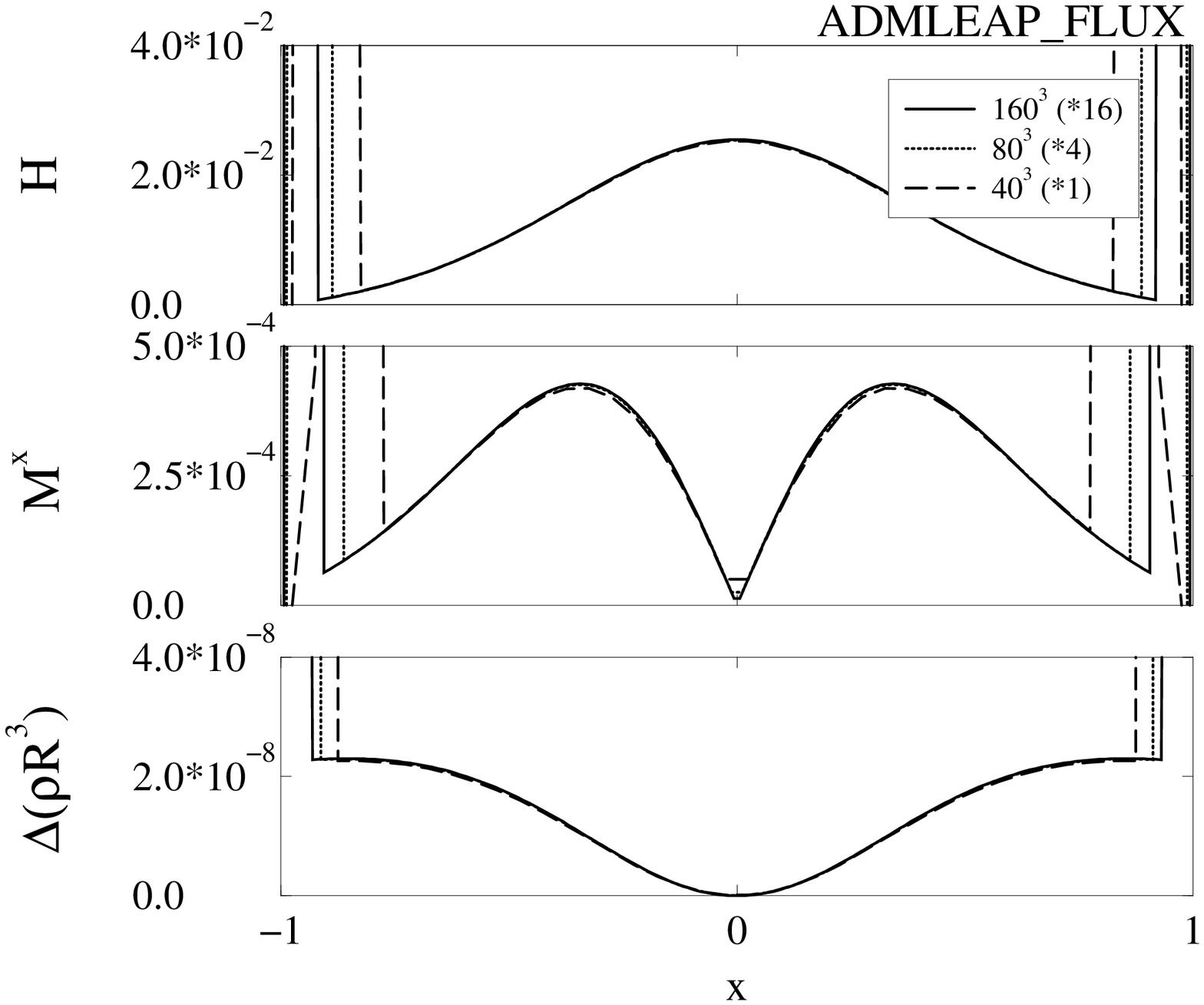,width=7.5cm}
\vspace{-6.5cm}
\hspace{7.5cm}
\psfig{figure=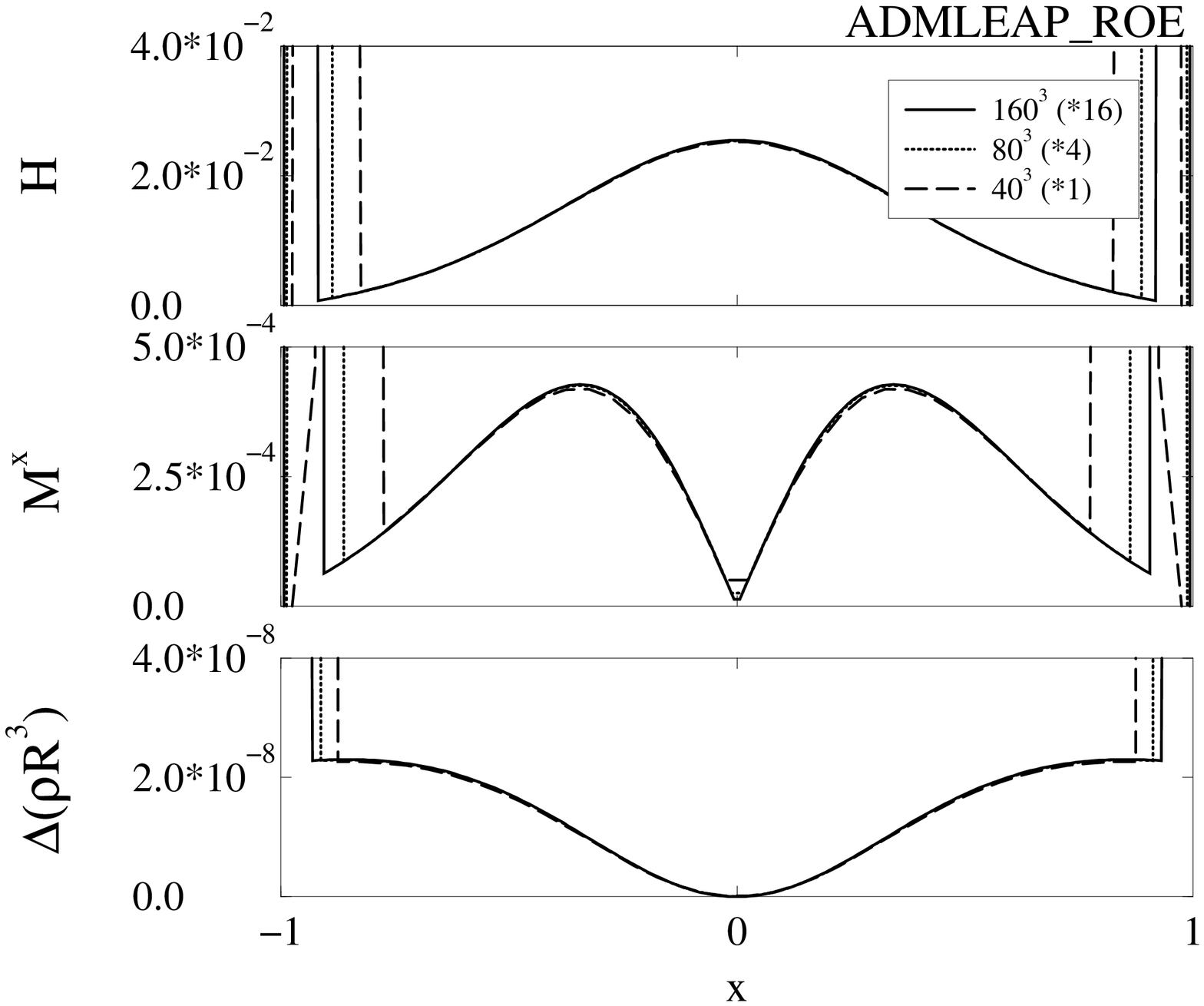,width=7.5cm}
\psfig{figure=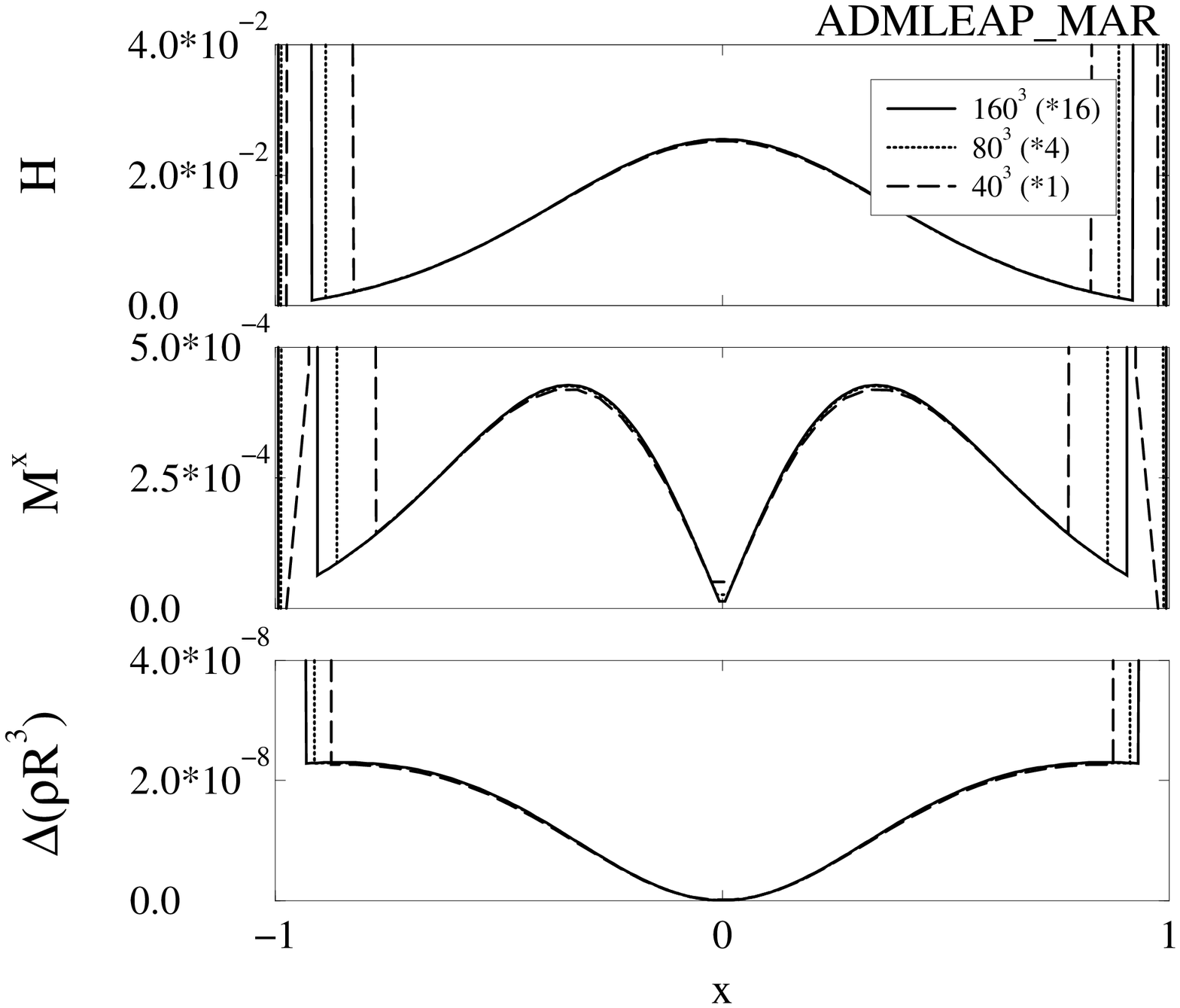,width=7.5cm}
\caption{Convergence plot of the
Hamiltonian constraint ($H$), the x-component of the momentum
constraint ($M^x$), and error plot for the conserved quantity $\rho R^3$
($\Delta (\rho R^3)$) for the ADMLEAP\_FLUX,
ADMLEAP\_ROE, and ADMLEAP\_MAR systems.  All three systems show 
second order convergence. 
No differences are noticeable between the three hydrodynamic 
methods for this initial data.}
\label{fig:k-1_adm_leap}
\end{figure}

In Fig.~\ref{fig:k-1_adm_leap} we plot $H$, $M^x$,
and the error in the conserved quantity $\rho R^3$
($\Delta (\rho R^3)$) for the ADMLEAP\_FLUX
ADMLEAP\_ROE, and the ADMLEAP\_MAR systems. 
All three systems
show second order convergence. We also find little difference between
the various hydrodynamic methods for this initial data.  Indeed, for
all the spacetime evolution systems studied, the 
different hydrodynamical methods give equivalent results.  
Because of this, we only show
results using Roe's method for the remainder of this subsection.

% Figure 

\begin{figure}
\psfig{figure=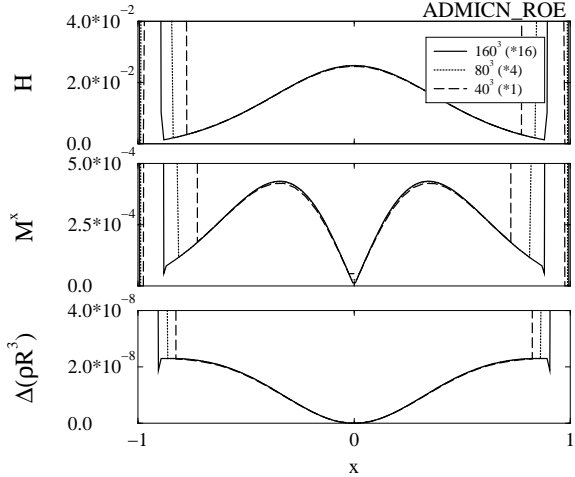,width=7.5cm}
\caption{Same as Fig.~\ref{fig:k-1_adm_leap} but using the ADMICN
evolution scheme for the spacetime.
Only results for Roe's method are presented, since
all three hydrodynamical methods give equivalent results for this
initial data. 
The ADMICN system gives a result similar to the ADMLEAP system.}
\label{fig:k-1_adm_icn}
\end{figure}

We now analyze the convergence of the same three quantities using the
ADMICN system for the spacetime evolution.
In Fig.~\ref{fig:k-1_adm_icn} we plot the results
for the ADMICN\_ROE system. Again we see second order convergence.
We also note that the ADMICN system
gives a similar evolution to the ADMLEAP system.

% Figure 

\begin{figure}
\psfig{figure=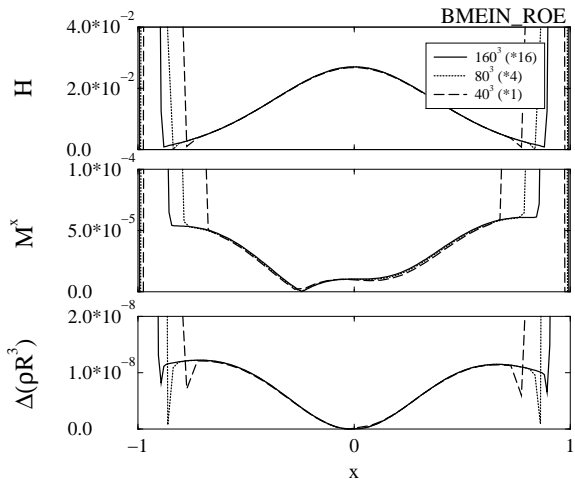,width=7.5cm}
\caption{Same as Fig.~\ref{fig:k-1_adm_leap} but using the BM equations
for the spacetime.
Only results for Roe's method are presented, since
all three hydrodynamical methods give equivalent results for this
initial data.  We note that the BMEIN system is slightly more
accurate than either of the ADM systems.}
\label{fig:k-1_bmein}
\end{figure}

We next repeat these convergence tests using the hyperbolic BM formulation.
In Fig.~\ref{fig:k-1_bmein} we plot $H$, $M^x$, and $\Delta (\rho R^3)$ 
for Roe's method
coupled to the BM ``Einstein" spacetime evolution.
As in the previous cases, we also obtain second order convergence, and
find little difference between the various hydrodynamic solvers.
Note the asymmetry present in $M^x$ which is a consequence of using an 
asymmetric method (MacCormack) to update the spacetime in the BM
system (contrary to the ADM systems where symmetric methods are employed).
This asymmetry is converging away with increased resolution, 
in the sense that $M^x$ is
converging to zero. We note that the BMEIN system is slightly more
accurate than either of the ADM systems.

\begin{table}
\begin{tabular}{|c||c|c|c|c|}
 & \# of points & & & \\
resolution & in each &$\Delta x$ & $c \frac {\Delta t}{\Delta x}$ &
\# of timesteps\\
                & direction    &       &  & \\ \hline
low             &  40          & 0.05  & 0.25 & 2 \\
medium          &  80          & 0.025  & 0.25 & 4 \\
high            & 160          & 0.0125 & 0.25 & 8 \\
\end{tabular}
\caption{Run parameters for the FRW $k=-1$
convergence tests.}
\label{table:frwk-1grid}
\end{table}

% Figure 

\begin{figure}
\psfig{figure=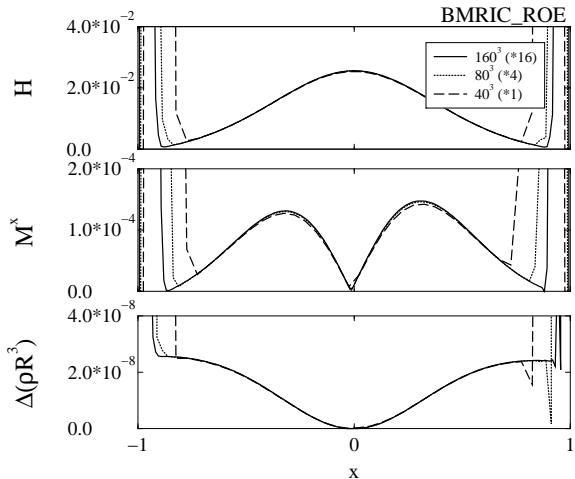,width=7.5cm}
\caption{Same as Fig.~\ref{fig:k-1_adm_leap} but using the BM equations
for the spacetime.
Only results for Roe's method are presented, since
all three hydrodynamical methods give equivalent results for this
initial data.  The BMRIC system has an accuracy similar to the BMEIN
system in this case.}
\label{fig:k-1_bmric}
\end{figure}

Finally, in 
Fig.~\ref{fig:k-1_bmric} we plot the convergence results obtained
when Roe's method is coupled to the BM 
``Ricci" system. As in all previous cases, we obtain 
second order convergence and no difference between the 
various hydrodynamical solvers.
The asymmetry in $M^x$, found previously using the BM ``Einstein" version of
the equations, is now somewhat smaller, and is also converging away ($M^x$ is
converging to zero).  The BMRIC system has an accuracy similar to the BMEIN
system in this case.

% Figure 

To summarize, we find second order convergence for all systems, even
when non-trivial spatial gradients are present.  The FRW $k=-1$ 
spacetime tests out many of the terms in the spacetime evolution,
however, the assumption that the matter be composed of dust
excludes many of the terms in the hydrodynamical equations.  
Due to this, we see little difference between the three
different hydrodynamical methods when evolving this initial data.

\section{Tolman-Oppenheimer-Volkoff Tests}
\label{sec:tov}

The FRW tests analyzed in the previous section assumed
the matter fields to be dust, i.e., $P=\epsilon=0$.
We now turn to a case where the pressure gradients
play a central role. We evolve a static star 
in the general relativistic setting, that is, 
a self-gravitating matter distribution
satisfying the (equilibrium structure) 
Tolman-Oppenheimer-Volkoff (TOV) equations.  

The ability to numerically evolve a compact, strongly gravitating
object is crucial for our program of developing a general 
purpose code for general relativistic astrophysics.  
Although the TOV solution is static, we evolve it with the 
full set of evolution equations for the 
hydrodynamics and the spacetime.  To maintain 
the static solution during the numerical evolution,
the pressure gradient
must exactly balance the gravitational force in the general
relativistic setting where both the energy density and pressure
are sources of gravity. 
Even though the extrinsic curvature and fluid velocities
are initially zero, 
finite differencing errors will allow these quantities 
to evolve away from zero.

Besides the complexity of the full GR-Hydro equations, there are two
other difficulties in numerically evolving this strongly
gravitating stellar configuration:  (1) the treatment of the surface
of the star, and (2) providing a coordinate condition
which maintains the long term stability
of the evolution.  We will discuss the first difficulty, including 
our numerical treatment, later in this section.  We 
will defer discussing the problem of
long term stability to the second paper in this series.

We begin with a discussion of the initial data.
The TOV equations \cite{Tolman39,Oppenheimer39b} are the Einstein
equations coupled to a perfect fluid stress-energy tensor under
the assumptions that the solution is static and spherically symmetric.
Specifically, the metric is given by
\begin{equation}
ds^2 = - {\alpha (r)}^2 \: dt^2 + \gamma_{rr}(r) \: dr^2 + r^2 \: {d\theta}^2 
+ r^2 {\sin}^2 \theta \: {d\phi}^2,
\end{equation}
where the metric functions, $\alpha (r)$ and $\gamma_{rr}(r)$, along with
the primitive hydrodynamic variables, $\rho(r)$ and $\epsilon(r)$,
are assumed to depend only on the circumferential radius $r$. (We
note that this form of the metric is only used in computing the initial
data, {\it not} in our dynamic evolution code, where a general
form of the metric in a Cartesian coordinate system is used.)
Under these assumptions,
the Einstein equations, along with the hydrodynamic
equations, reduce to
\begin{eqnarray}
\label{eq:tov}
\frac {\partial P(r)}{\partial r} & = & 
- \frac {(\rho + \rho \epsilon + P)(m + 4 \pi r^3 P)}
        {r(r - 2 m)},\\
\frac {\partial (\ln{\alpha(r)})}{\partial r} & = & 
  \frac {m + 4 \pi r^3 p}{r(r - 2 m)},\\
\frac {\partial m(r)}{\partial r} & = &
  4 \pi r^2 (\rho + \rho \epsilon),\\
\gamma_{rr}(r) & = &
{\left( 1 - \frac {2 m(r)}{r} \right) }^{-1},
\end{eqnarray}
\noindent where $m=m(r)$ is the mass energy contained
inside a sphere of circumferential radius $r$.
To determine the initial data, the above coupled ordinary differential
equations are integrated using a 4th order Runge-Kutta method, supplemented
by a polytropic equation of state (EOS)
\begin{equation}
P = {\cal K} \rho^{\Gamma}.
\end{equation}
We note that in the dynamical evolution we use the more generic 
EOS $P=(\Gamma -1) \rho
\epsilon $.  In the simulations shown in this section, we use
$\Gamma = \frac {5}{3}$,
${\cal K} = 5.380 \times 10^9 \frac {cm^4}{g^{2/3} {sec}^2}$, and a 
central mass density of 
${\rho}_c=5 \times 10^{14} \frac {g}{{cm}^3}$.
This configuration corresponds, roughly, to a neutron star
\cite{Shapiro86}.   Other initial parameters, showing 
essentially the same features as the one 
presented here, have also been tested.  
This choice of parameters leads to a TOV solution with
a total ADM mass of $0.566 M_{\odot}$ and a circumferential
coordinate radius of $14.9 km$.  The TOV solution is then matched at the
star's surface to an exterior
Schwarzschild spacetime with the appropriate
mass and coordinates\cite{Misner73}.  A
coordinate transformation from spherical to Cartesian
coordinates is then performed to obtain initial data for the 3D evolution
code.  The evolution is performed
with ``harmonic slicing'', that is 
$\dot{\alpha} = \alpha \sqrt{\gamma} \gamma^{ij}K_{ij}$.

We now present convergence tests for the evolution code.
These tests are performed with the
parameters described in Table \ref{table:tovgrid}.  As the 
initial data exhibits octant symmetry, only one octant is evolved
(with appropriate boundary conditions used at the inner
faces of the computational domain).  
This is an important capability of the code, in that it enables us
to achieve a higher resolution and make more efficient use
of available computational resources when allowed by the symmetry 
of the problem.
The initial data is then evolved up to $t = 0.986 \: \mu s$.  
We calculate $\Delta \rho = {\rho}_{\verb+num+} - {\rho}_{\verb+exact+}$,
the difference between the numerically 
computed mass density, ${\rho}_{\verb+num+}$,
and the exact solution, ${\rho}_{\verb+exact+}$.  We also
monitor the Hamiltonian constraint, $H$, and the $x$-momentum
constraint, $M^x$.  The order of convergence of the code, for 
various combinations of the spacetime and 
hydrodynamics integrations, is demonstrated as before.

For the TOV solution, we have tested the consistency 
of the flux-split, Roe, and 
Marquina methods both with and without the limiting functions.
We note that the limiters result in a truncation error that is first 
order in $\Delta x$ at points that attain minimum and maximum 
in the hydrodynamical
variables, while all other points have truncation errors which are second
order in $\Delta x$.  For the convergence tests 
in this section we will only present
results without the flux limiters.
Specifically, we
have built a switch in the code that will set $\psi = 1$ for the
limiting of the flux split method, Eq. (\ref{vanleer}), and turn off the
minmod function for the Roe and Marquina methods, 
so that we can perform tests with
and without limiters to ensure that all components of our code have the
expected convergence properties. 

\begin{figure}
\psfig{figure=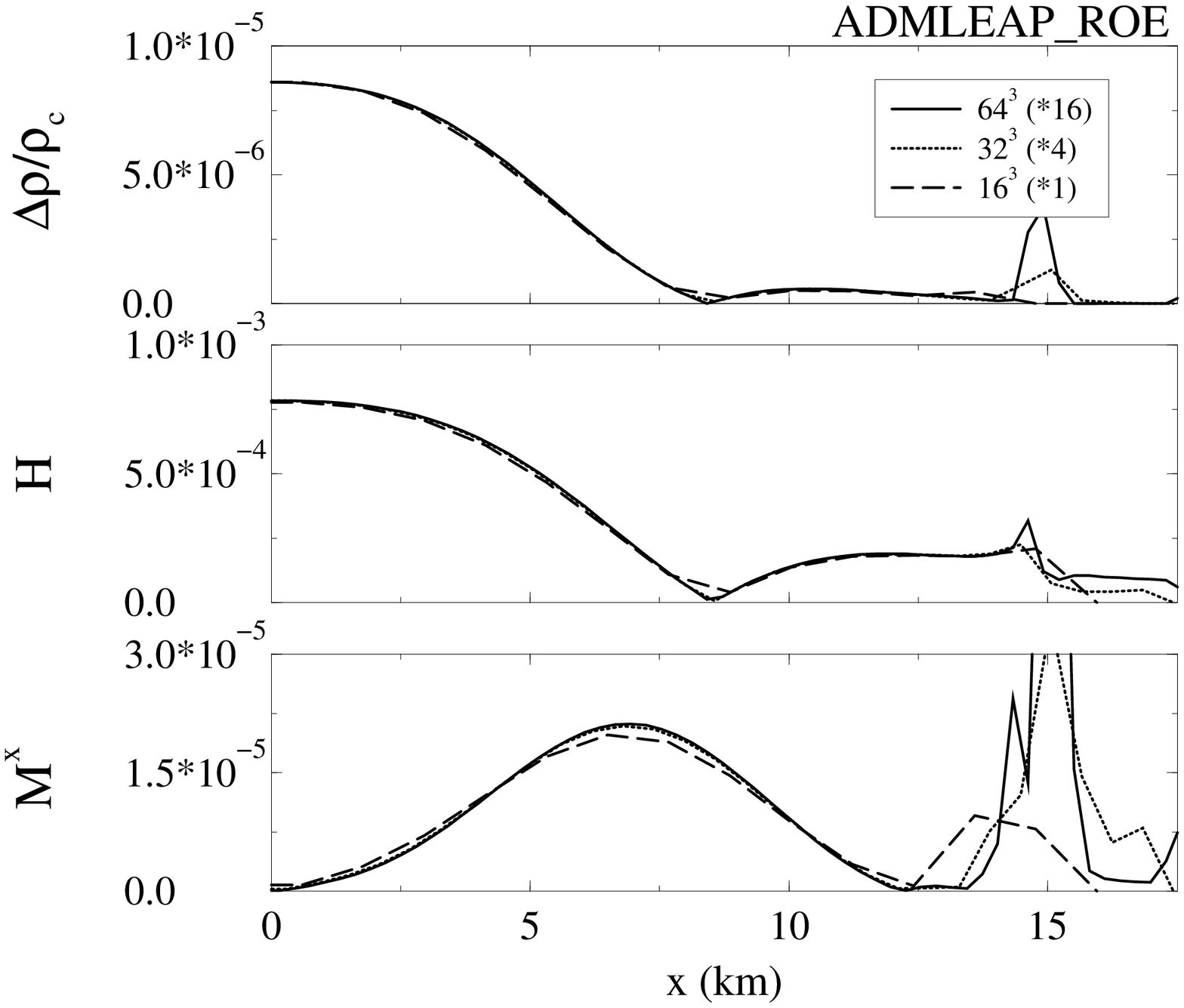,width=7.5cm}
\vspace{-6.4cm}
\hspace{7.5cm}
\psfig{figure=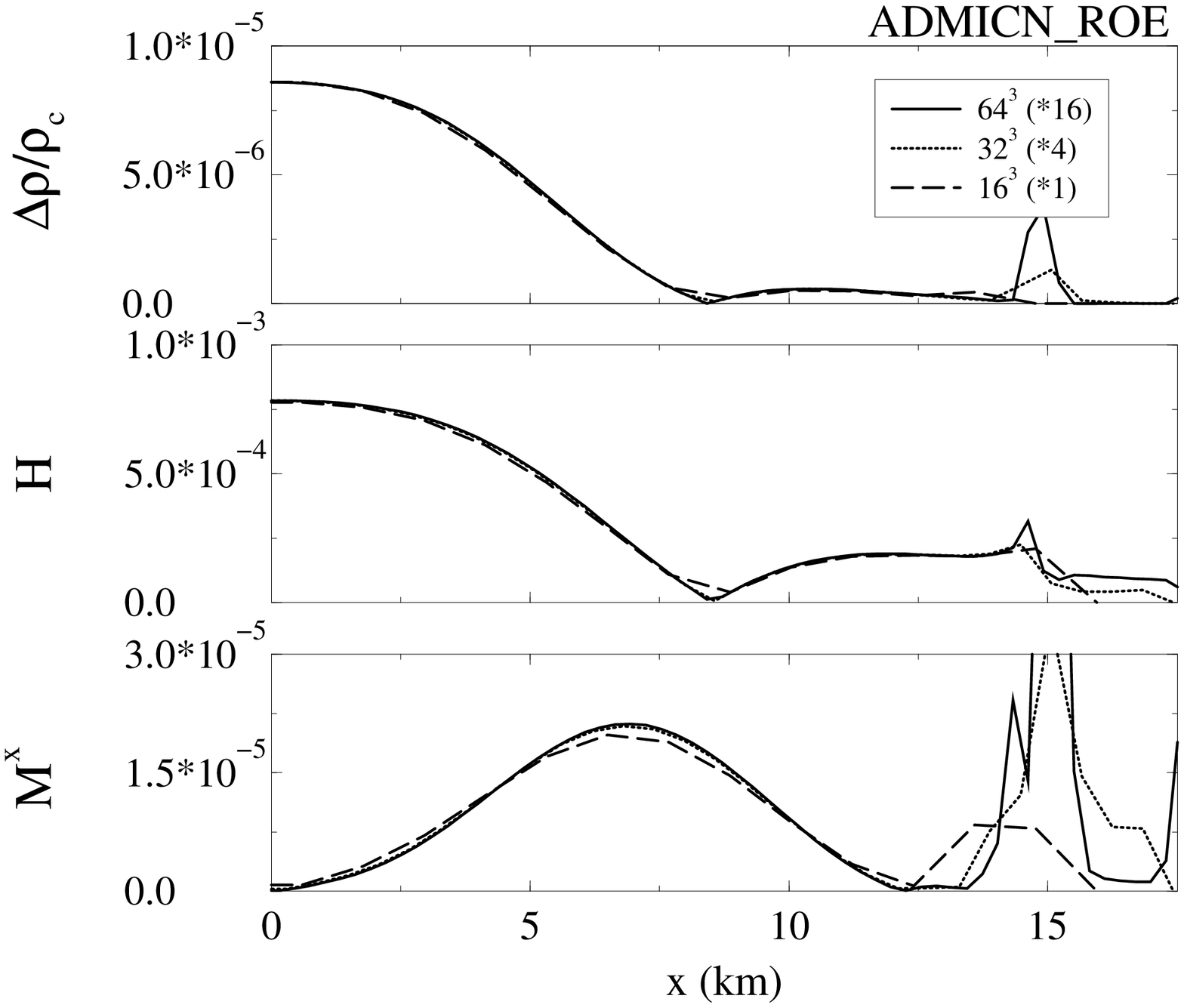,width=7.5cm}
\caption{We demonstrate the convergence of the ADMLEAP\_ROE and
ADMICN\_ROE evolution systems for
three different error functions:
the difference between the analytic and computed
rest mass density (normalized by the central rest mass density $\rho_c$)
$\Delta \rho / \rho_c$, the
Hamiltonian constraint $H$,
and the $x$-momentum constraint $M^x$.
In each of the three cases, we multiply the high resolution
result by sixteen and the medium resolution by four to show second order
convergence.  All results are shown at $t=0.986 \mu s$ which corresponds to
eight iterations at the highest resolution.  The graphs are taken along
the $x$ axis (results on the $y$ and $z$ axes are identical, and results
on the diagonal axis are similar).}
\label{fig:tov_adm_roe}
\end{figure}

\begin{figure}
\psfig{figure=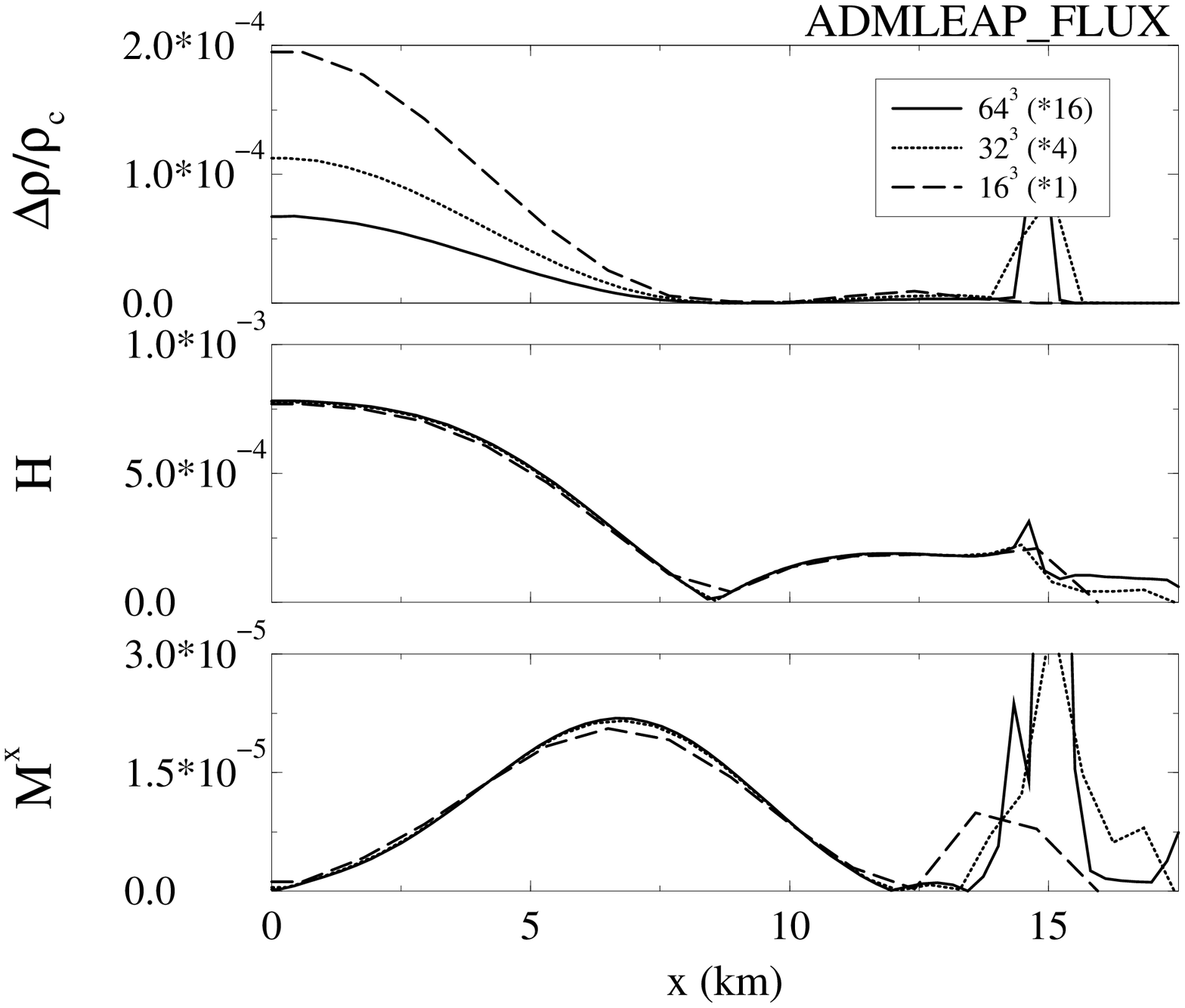,width=7.5cm}
\vspace{-6.4cm}
\hspace{7.5cm}
\psfig{figure=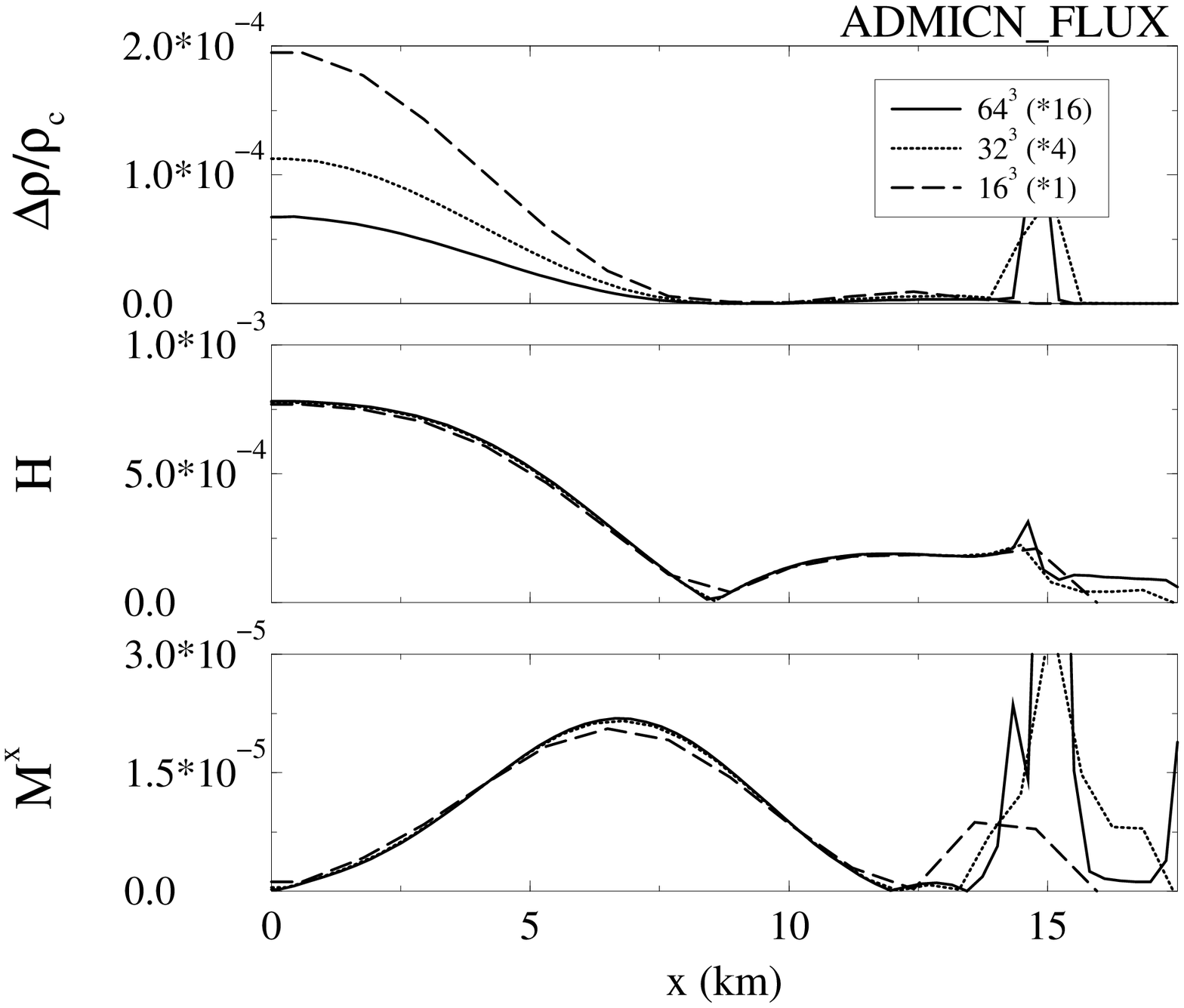,width=7.5cm}
\caption{We demonstrate the convergence of the ADMLEAP\_FLUX and
ADMICN\_FLUX evolution systems for
three different error functions:
the difference between the analytic and computed
rest mass density (normalized by the central rest mass density $\rho_c$)
$\Delta \rho / \rho_c$, the
Hamiltonian constraint $H$,
and the $x$-momentum constraint $M^x$.
In each of the three cases, we multiply the high resolution
result by sixteen and the medium resolution by four to show second order
convergence.  Note that the rest mass density (top frame) is converging
faster than second order in $\Delta x$ (see text for explanation).
All results are shown at $t=0.986 \mu s$ which corresponds to
eight iterations at the highest resolution.  The graphs are taken along
the $x$ axis (results on the $y$ and $z$ are identical, and
results on the diagonal axis are similar).}
\label{fig:tov_adm_flux}
\end{figure}

\begin{figure}
\psfig{figure=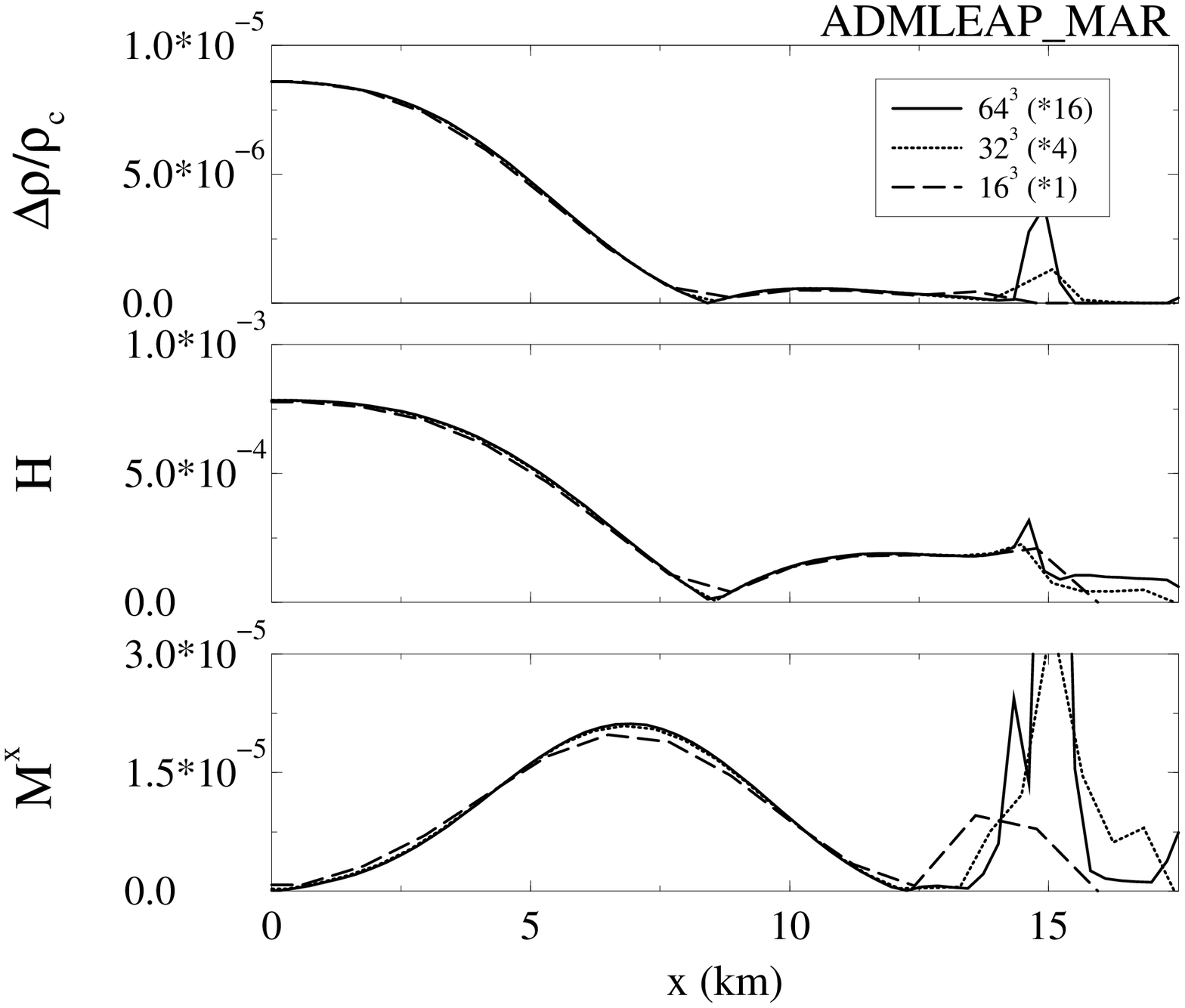,width=7.5cm}
\vspace{-6.4cm}
\hspace{7.5cm}
\psfig{figure=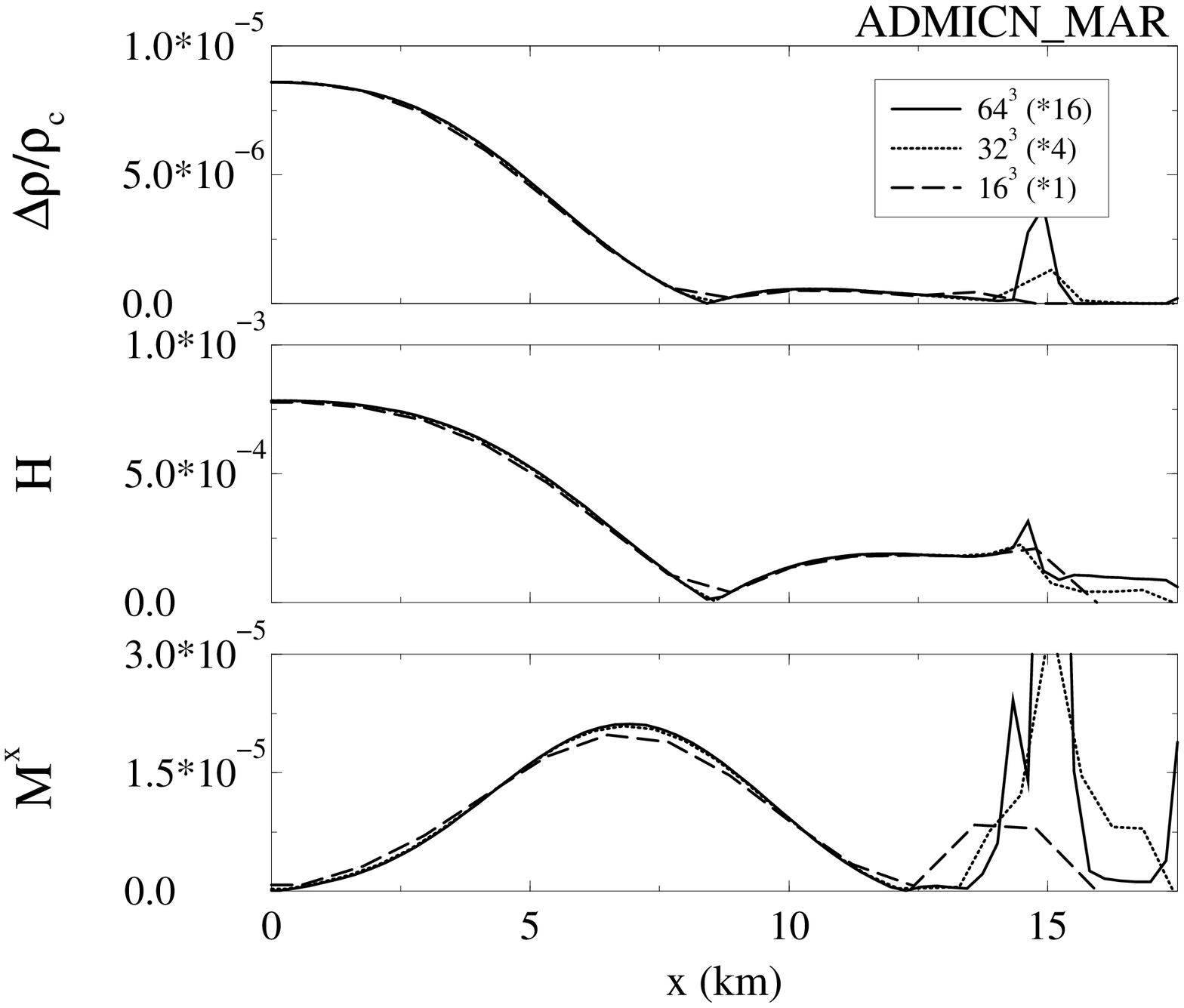,width=7.5cm}
\caption{We demonstrate the convergence of the ADMLEAP\_MAR and
ADMICN\_MAR evolution systems for
three different error functions:
the difference between the analytic and computed
rest mass density (normalized by the central rest mass density $\rho_c$)
$\Delta \rho / \rho_c$, the
Hamiltonian constraint $H$,
and the $x$-momentum constraint $M^x$.
In each of the three cases, we multiply the high resolution
result by sixteen and the medium resolution by four to show second order
convergence.  All results are shown at $t=0.986 \mu s$ which corresponds to
eight iterations at the highest resolution.  The graphs are taken along
the $x$ axis (results on the $y$ and $z$ axes are identical, and results
on the diagonal axis are similar).}
\label{fig:tov_adm_mar}
\end{figure}

\begin{figure}
\psfig{figure=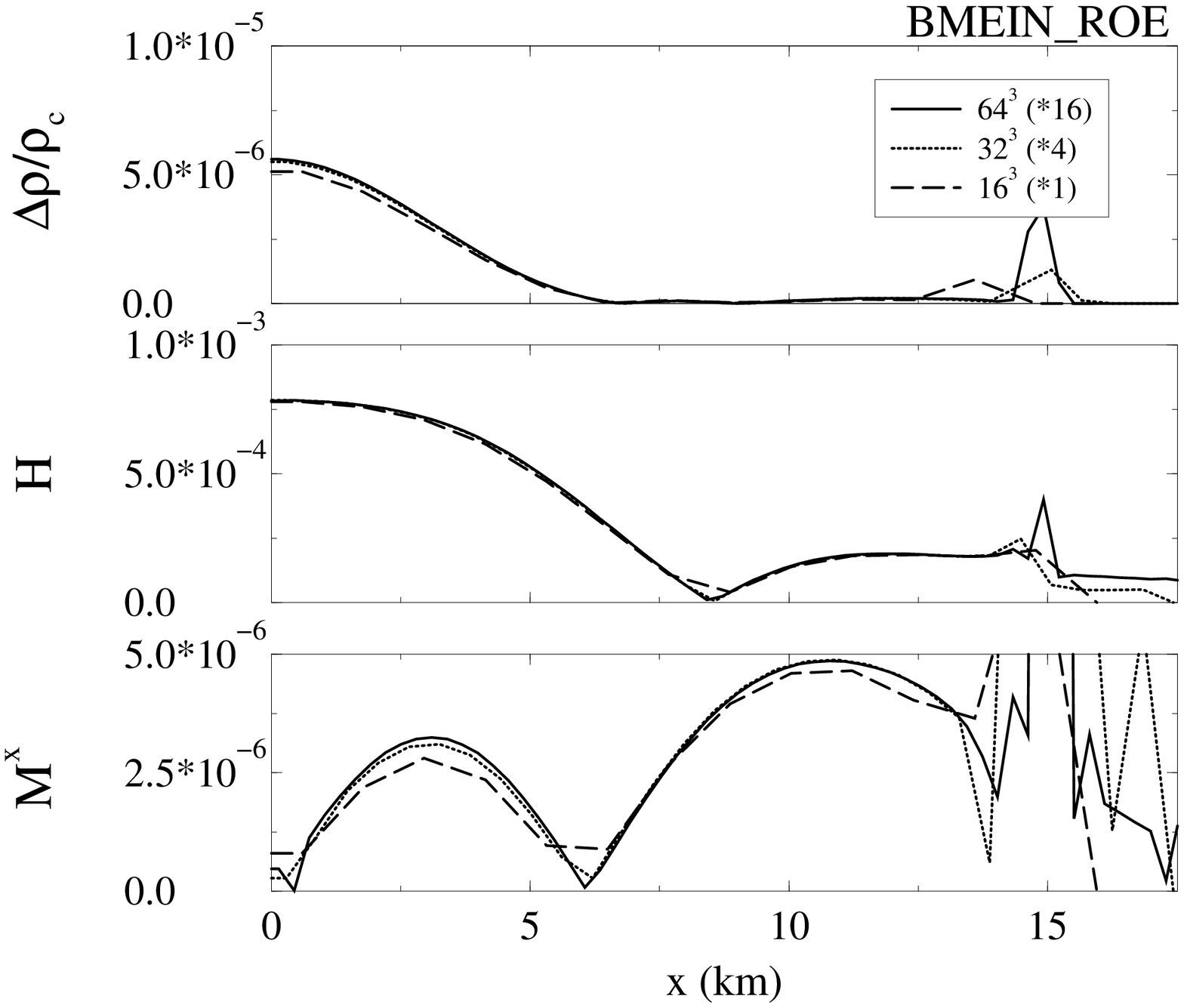,width=7.5cm}
\vspace{-6.4cm}
\hspace{7.5cm}
\psfig{figure=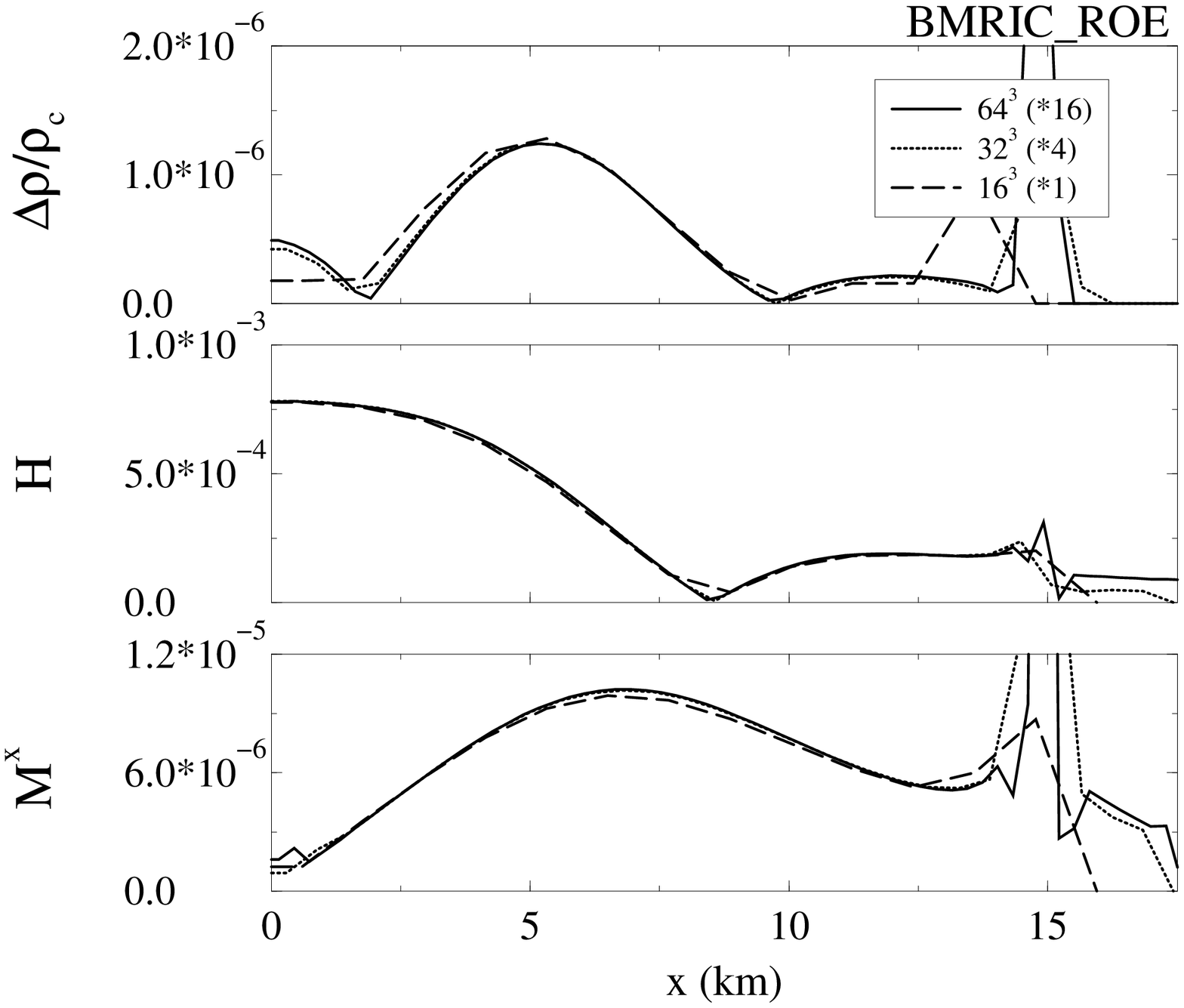,width=7.5cm}
\caption{We demonstrate the convergence of the BMEIN\_ROE and
BMRIC\_ROE evolution systems for
three different error functions:
the difference between the analytic and computed
rest mass density (normalized by the central rest mass density $\rho_c$)
$\Delta \rho / \rho_c$, the
Hamiltonian constraint $H$,
and the $x$-momentum constraint $M^x$.
In each of the three cases, we multiply the high resolution
result by sixteen and the medium resolution by four to show second order
convergence.  All results are shown at $t=0.986 \mu s$ which corresponds to
eight iterations at the highest resolution.  The graphs are taken along
the $x$ axis (results on the $y$ and $z$ axes are identical, and results
on the diagonal axis are similar).}
\label{fig:tov_bm_roe}
\end{figure}

\begin{figure}
\psfig{figure=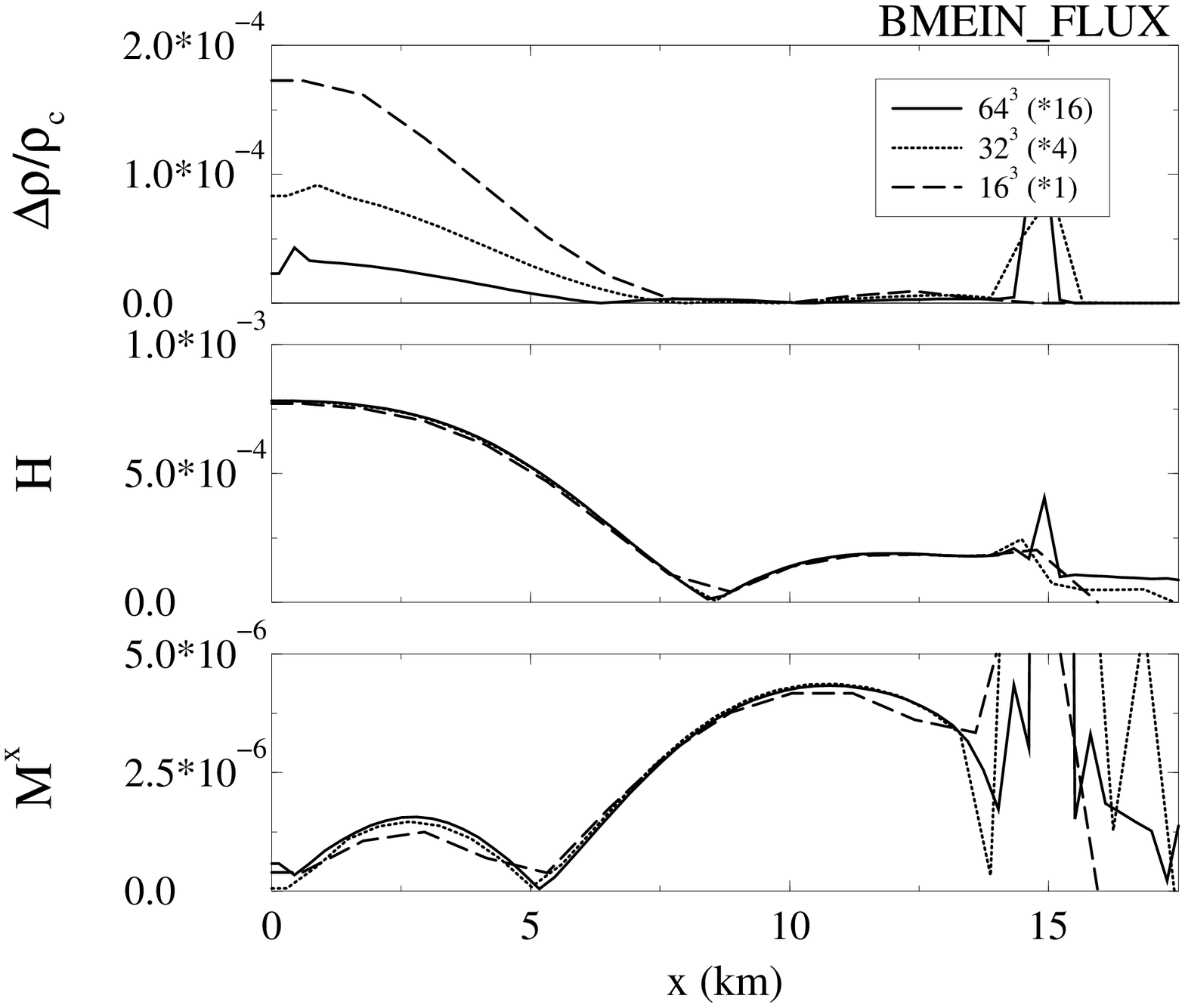,width=7.5cm}
\vspace{-6.4cm}
\hspace{7.5cm}
\psfig{figure=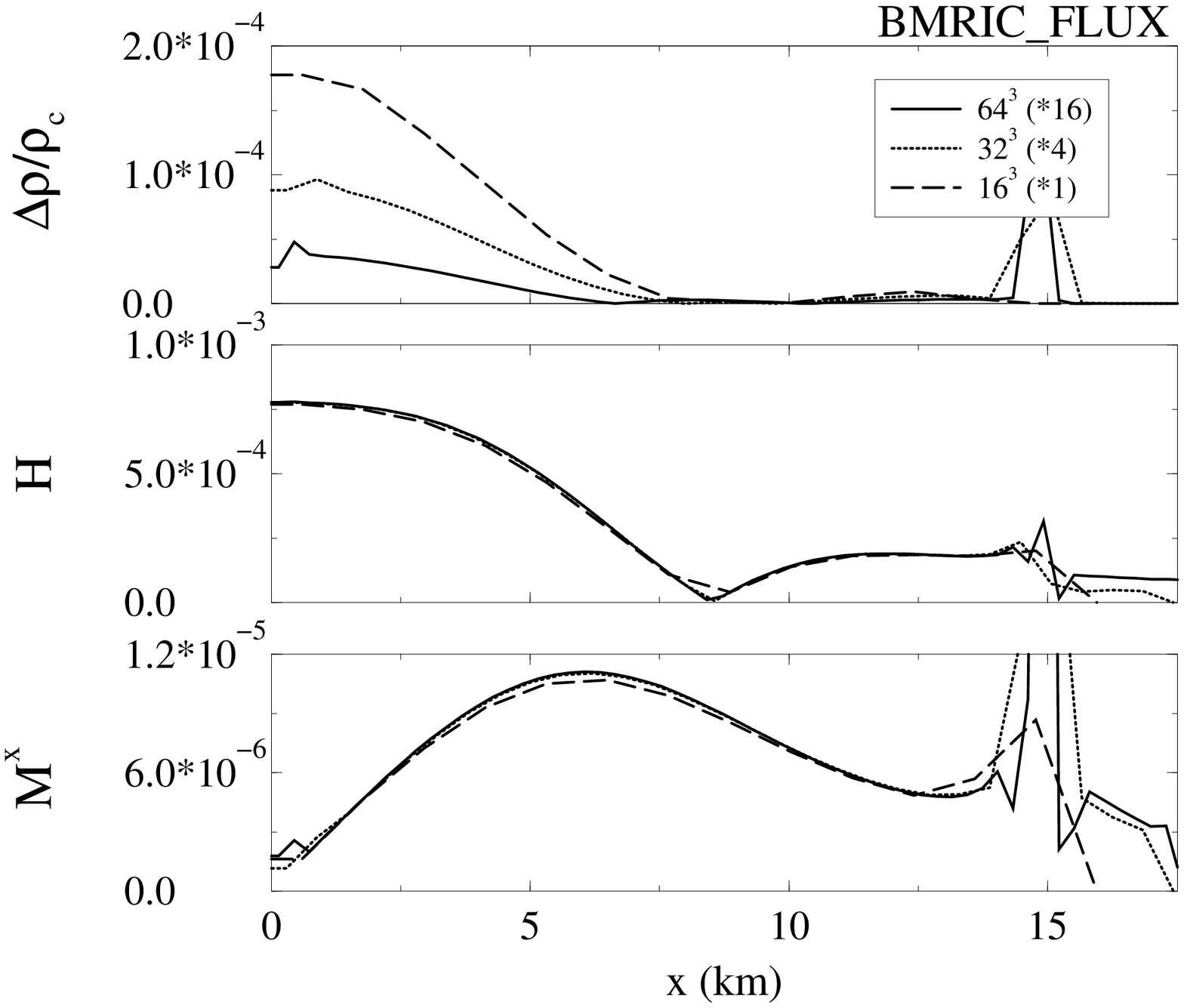,width=7.5cm}
\caption{We demonstrate the convergence of the BMEIN\_FLUX and
BMRIC\_FLUX evolution systems for
three different error functions:
the difference between the analytic and computed
rest mass density (normalized by the central rest mass density $\rho_c$)
$\Delta \rho / \rho_c$, the
Hamiltonian constraint $H$,
and the $x$-momentum constraint $M^x$.
In each of the three cases, we multiply the high resolution
result by sixteen and the medium resolution by four to show second order
convergence.  Note that the rest mass density (top frame) is converging
faster than second order in $\Delta x$ (see text for explanation).
All results are shown at $t=0.986 \mu s$ which corresponds to
eight iterations at the highest resolution.  The graphs are taken along
the $x$ axis (results on the $y$ and $z$ axes are identical, and
results on the diagonal axis are similar).}
\label{fig:tov_bm_flux}
\end{figure}

\begin{figure}
\psfig{figure=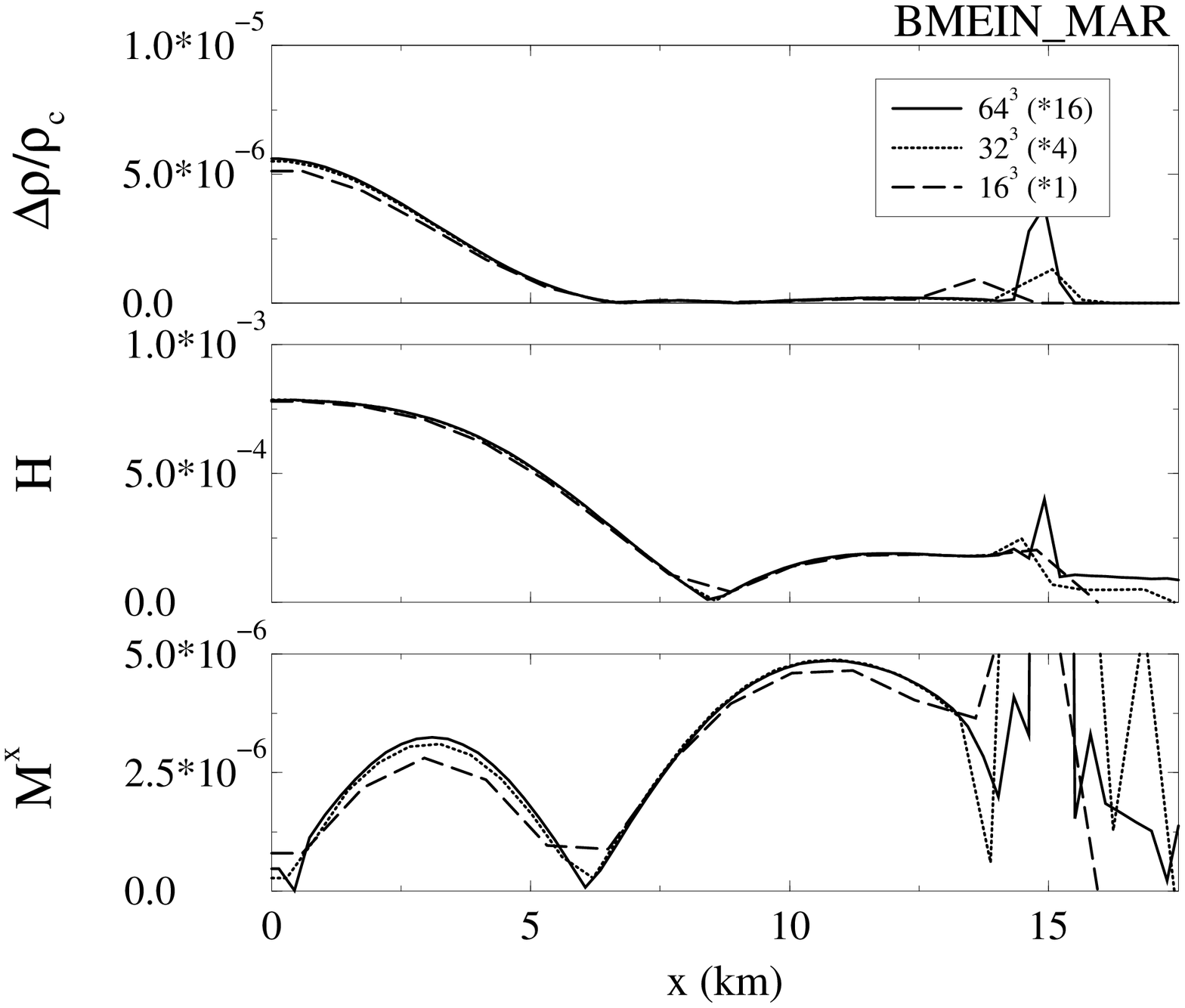,width=7.5cm}
\vspace{-6.4cm}
\hspace{7.5cm}
\psfig{figure=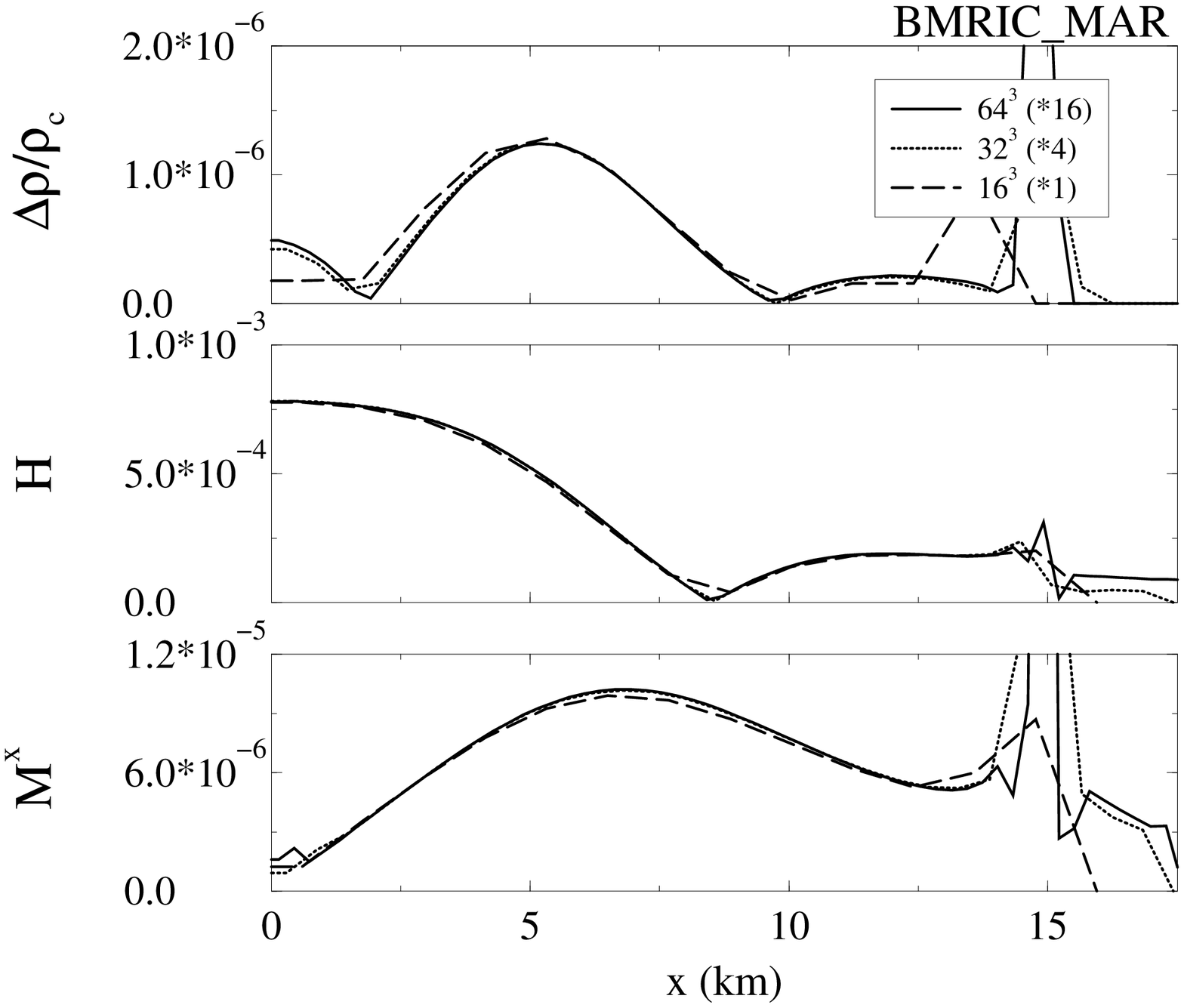,width=7.5cm}
\caption{We demonstrate the convergence of the BMEIN\_MAR and
BMRIC\_MAR evolution systems for
three different error functions:
the difference between the analytic and computed
rest mass density (normalized by the central rest mass density $\rho_c$)
$\Delta \rho / \rho_c$, the
Hamiltonian constraint $H$,
and the $x$-momentum constraint $M^x$.
In each of the three cases, we multiply the high resolution
result by sixteen and the medium resolution by four to show second order
convergence.  All results are shown at $t=0.986 \mu s$ which corresponds to
eight iterations at the highest resolution.  The graphs are taken along
the $x$ axis (results on the $y$ and $z$ axes are identical, and results
on the diagonal axis are similar).}
\label{fig:tov_bm_mar}
\end{figure}

Figs.~\ref{fig:tov_adm_roe}~--~\ref{fig:tov_bm_mar} 
show convergence plots for all 12 system combinations
resulting from three hydrodynamical evolution schemes (ROE, FLUX, MAR)
and four spacetime evolution schemes (ADMLEAP,ICN,BMEIN,BMRIC),
see Table~\ref{table:discrete_names}.
Plots of the difference between the
numerically evolved rest mass density and the analytic solution
normalized by the central density $(\Delta \rho /
{\rho}_c)$, the Hamiltonian constraint $(H)$, and the $x$-component of
the momentum constraint $(M^x)$ are shown.  
In  Figs.~\ref{fig:tov_adm_roe}~--~\ref{fig:tov_bm_mar},
we focus on the interior of the star, $r < 14.9 \: km$.
The treatment of the surface of the star will be discussed in detail later.

% Figure

\begin{table}
\begin{tabular}{|c||c|c|c|c|c|}
 & \# of points & &  & & total  \\
resolution & in each & $\Delta x$ (km) & $c \frac {\Delta t}{\Delta x}$ &
\# of timesteps & evolved time \\
 & coordinate direction & & & & $(\mu s)$ \\ \hline
low & 16 & 1.182 & 0.125 & 2 & 0.986 \\
medium & 32 & 0.591 & 0.125 & 4 & 0.986 \\
high & 64 & 0.2955 & 0.125 & 8 & 0.986 \\
\end{tabular}
\caption{Computational grid parameters for the TOV tests.}
\label{table:tovgrid}
\end{table}

% Figure

% Figure

% Figure

% Figure

% Figure

\begin{figure}
\psfig{figure=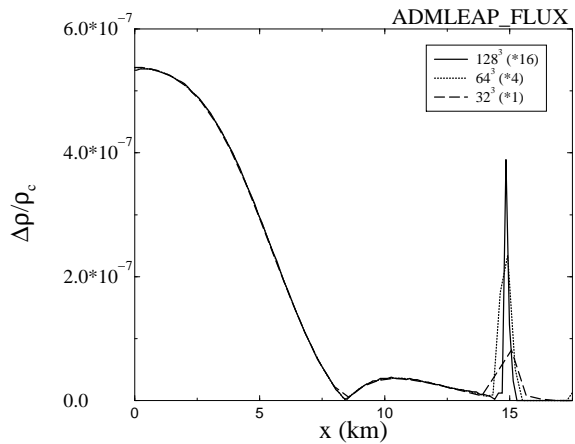,width=7.5cm}
\caption{We demonstrate the convergence of the 
evolved rest mass density with the ADMLEAP\_FLUX
evolution system for a set of grids with twice the resolution
as that displayed in table~\ref{table:tovgrid}.  
All results are shown at $t=0.493 \mu s$ which corresponds to
eight iterations at the highest resolution.  The graphs are taken along
the $x$ axis (results on the $y$ and $z$ axes are identical, and
results on the diagonal axis are similar).}
\label{fig:tov_admleap_flux_highres}
\end{figure}

Notice that the rest mass density
for all systems using the flux-split hydrodynamical evolution
scheme (FLUX)
is converging at a rate that is
higher than second order for the ${\Delta x}$ used here.
This is due to the fact that, for the resolutions used here,
the truncation error terms that are proportional to ${\Delta
x}^3$ have a magnitude comparable to the truncation error terms
that are proportional to ${\Delta x}^2$.  This causes the
appearance of ``hyper-convergence'' as seen in
Figs.~\ref{fig:tov_adm_flux}~and~\ref{fig:tov_bm_flux}.  
To confirm this point, we plot in Fig.~\ref{fig:tov_admleap_flux_highres}
the results of a convergence test for the ADMLEAP\_FLUX
system with two times the spatial resolution as described in
table~\ref{table:tovgrid}.  A clear indication of 
second order convergence is observed at this resolution.

% Figure

We also note that for the same resolution, the BMRIC spacetime
evolution scheme is slightly more accurate than the other
spacetime evolution schemes (from a standpoint of the absolute
value of the error functions plotted).  Also, the Roe and Marquina
schemes for evolving the hydrodynamics are more accurate (by
an order of magnitude) than the flux-split method.

\begin{figure}
\psfig{figure=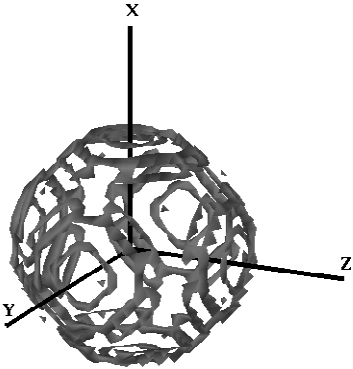,width=8.5cm}
\caption{A 3D isosurface of constant internal specific energy density 
at time $t=0.04 \: ms$ is shown.
This value of $\epsilon = 4.1 \times 10^{-4} \: \epsilon_c$ 
corresponds to $87\%$ of the atmosphere specific internal energy
density. Notice that the troublesome regions are not along the coordinate
or diagonal axes and hence would not be observable in 1D plots along
these axes.}
\label{fig:atmos1}
\end{figure}

\begin{figure}
\psfig{figure=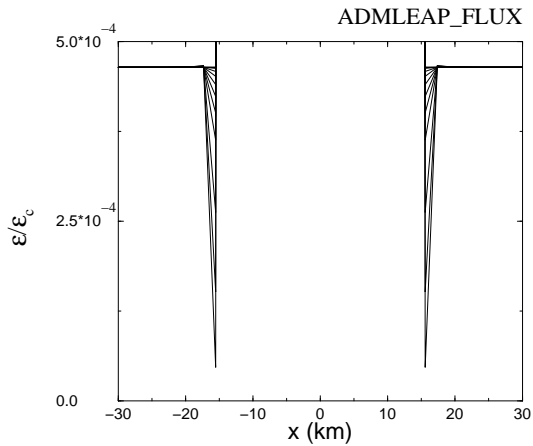,width=7.0cm}
\vspace{0.0cm}
\hspace{7.0cm}
\caption{1D plot of  $\epsilon / {\epsilon}_c$ for the
ADMLEAP\_FLUX system evolving a TOV configuration. The 
lines are plotted in the $x$-direction at time intervals of $0.002 \: ms$, and 
correspond to coordinate values $z=4.6 \: km$, 
$y=-6.4 \: km$.  The final time is $t=0.042 \: ms$. The code crashes
shortly afterwards.}
\label{fig:atmos2}
\end{figure}

Next we turn to the star's surface treatment.  There are three
related issues in the numerical evolution of the surface region of a
compact self-gravitating object in general relativity.  (1) At the
surface of the star, the second normal derivatives of some of the
hydrodynamic quantities, e.g., the density, are discontinuous for most
equations of state.  This discontinuity is also present in the TOV
solution with $\Gamma = 5/3 $ studied in this section.  The Einstein
equations imply that the curvature tensor (which contains second
derivatives of the metric functions) has a kink at the stellar
surface.  The curvature tensor enters explicitly in the evolution
equation of the extrinsic curvature.  This makes the numerical 
treatment of the stellar surface considerably more difficult in the
relativistic framework than in the Newtonian case. Accordingly,
the numerical evolutions are less stable in the relativistic case.
(2) In the exterior of the star there is vacuum and hence, the density 
must drop to zero. As the density approaches zero, the transformation
from the evolved variables $(\tilde{D},\tilde{S_i},\tilde{\tau})$
to the primitive variables $(\rho,v^i,\epsilon)$ becomes singular.
The ``standard'' treatment of this problem is to add an ad hoc
``atmosphere'', with some choice of thermal properties in the exterior
region. This atmosphere typically has a density 
orders of magnitude smaller than that of the interior of the star and 
should have a negligible effect on the dynamics of the system.  In our
simulations, we typically pick the atmosphere to be $10 ^ {-4}$ to $10
^ {-5}$ of the central density ${\rho}_c$ of the TOV star and with the
same EOS as the star.  This is sufficient to ensure that the GR-Hydro
equations are neither singular nor degenerate in our treatment, while
having negligible effect on the actual dynamical evolution of the star.
(3) In regions of low density, especially in the atmosphere near the surface
of the star, there are two related difficulties: (i) it is
difficult to accurately recover the pressure (which is a power of the density) 
from the evolved variables $(\tilde{D},\tilde{S_i},\tilde{\tau})$, and
(ii) it is easy to develop high velocity flows due to the
strong gravitational field there.  In particular, the
atmosphere ($r > 14.9 \: km$) is not part of the
equilibrium TOV initial data, and the gravitational field is
driving it to collapse onto the surface of the star. Numerically, it is
problematic to have the atmosphere colliding with the surface of the
star, creating a shock, and leaving the specific internal
energy density $\epsilon$ in the atmosphere
behind plunging to zero. These difficulties eventually cause the code to crash.  
An explicit demonstration of these difficulties
is exhibited in Figs.~\ref{fig:atmos1}~and~\ref{fig:atmos2}.  
For these simple tests,
we evolve a TOV configuration as described above.  We implement an
atmosphere with density ${{\rho}_{}}_{atmos} = 10^{-5} \: {\rho}_c$.
The resulting atmosphere has a specific internal energy of
${{\epsilon}_{}}_{atmos} \approx 4.6 \times 10^{-4} \: {\epsilon}_c$.
We then evolve the configuration with the ADMLEAP\_FLUX evolution
scheme.  We first evolve this configuration {\it without} implementing
the surface treatment described below 
(which {\it is} implemented for all other
runs in this paper).  The code crashes after only $0.043 \: ms$, due
to the specific internal energy dropping to zero near the surface
of the star.
Fig.~\ref{fig:atmos1} shows a 3D isosurface plot of $\epsilon$
corresponding to a value of $\epsilon = 0.87 {{\epsilon}_{}}_{atmos}$ at
time $t=0.04 \: ms$.  These regions indicate where the specific
internal energy is dropping significantly in this
short time interval.  Fig.~\ref{fig:atmos2}
shows a 1D plot of $\epsilon / {\epsilon}_c$.  Clearly, the 
specific internal energy is dropping to zero rapidly in an 
unstable fashion near the surface
of the star $r=14.9 \: km$.  

% Figure

% Figure

To circumvent these problems, we have found a simple, yet effective
treatment for the stable numerical evolution of 
low density regions.  Again, since this scheme is enacted only
for very low density flow, it has a negligible effect on the 
dynamics of the system.  It is important to note that the first
indication of problems is in the recovery of the primitive
variables $(\rho,v^i,\epsilon)$ from the evolved variables 
$(\tilde{D},\tilde{S_i},\tilde{\tau})$, given by Eq.~(\ref{eq:evolvedvar}).
In regions where the rest mass density $\rho$ is less
than some specified minimum density ${\rho}_{_{\verb+min+}}$ (typically
some fraction of the atmosphere rest mass density), if the 
recovery of the primitive variables $(\rho,v^i,\epsilon)$ from 
the evolved variables $(\tilde{D},\tilde{S_i},\tilde{\tau})$ results
in a negative specific energy density $\epsilon$, 
then the primitive variables are solved again, with the
condition for adiabatic flow
\begin{equation}
P = {\cal K} {\rho}^{\Gamma},
\end{equation}
replacing the definition of $\tilde{\tau}$ (the fifth component of
Eq.~(\ref{eq:evolvedvar})).  

\begin{figure}
\psfig{figure=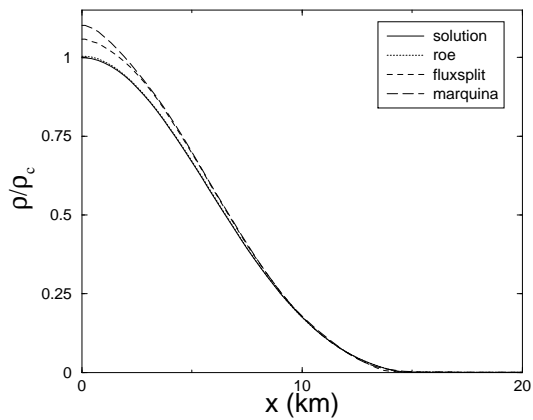,width=7.0cm}
\nopagebreak
\vspace{0.0cm}
\nopagebreak
\hspace{7.0cm}
\nopagebreak
\samepage
\caption{Long term ($1.0 ms$) evolution of TOV initial data. The final
rest mass density is plotted for the ADMLEAP\_ROE, ADMLEAP\_FLUX,
and ADMLEAP\_MAR systems, along with the static analytic solution. The
evolution with the ADMLEAP\_ROE system is indistinguishable from the
analytic solution.  The resolution used for these runs correspond
to $\Delta x = 0.2954 km$, with $\Delta t / \Delta x = 0.125/c$.
$8000$ timesteps were required to evolve to $1.0 ms$.}
\label{fig:tov_long}
\end{figure}

To demonstrate the stability of this scheme, we show in 
Fig.~\ref{fig:tov_long} the final rest mass density profile
for numerical evolutions of the TOV configuration described above.  
The three different hydrodynamics evolutions schemes (flux-split,Roe, and
Marquina) are used to evolve the same TOV configuration to $1.0 \: ms$
(approximately $8000$ time steps).  
Note that for Roe's method, the final mass density configuration
is indistinguishable from the initial configuration profile.  

% Figure

\section{Boosted TOV Tests}
\label{sec:boost}

The final test case we present is the most stringent.
The boosted TOV solution effectively tests many of
the features one requires of a general relativistic hydrodynamical
spacetime code:
relativistic fluid motion, strong gravitational
fields, and a non-trivial coordinate condition for both 
the lapse and shift. 
In addition, this test has an analytic solution to compare against.  
We simulate a
neutron star moving with a high velocity along the 
$\hat{x} + \hat{y} + \hat{z}$ diagonal of the computational domain. 
We find that in such
a simulation, every term in both the Einstein equations and the
GR-Hydro equations is activated; each evolved variable is a nontrivial
function of space and time.
We obtain an analytic solution of a boosted neutron star
by applying a coordinate transformation that corresponds to a Lorentz
boost at spatial infinity on a solution to the 
TOV equations.  Specifically, if the solution to the TOV equations
(see section~\ref{sec:tov}) is expressed in Cartesian $(t,x,y,z)$ coordinates,
we transform to another set of coordinates
$(t^\prime,x^\prime,y^\prime,z^\prime)$ via the transformation
\begin{equation}
\left [ \begin{array}{c} t^\prime \\ x^\prime \\ y^\prime \\ z^\prime
        \end{array} \right ] = 
\left [ \begin{array}{cccc}
\gamma_b & \xi_x \gamma_b & \xi_y \gamma_b & \xi_z \gamma_b  \\
\xi_x \gamma_b & \left ( 1 + \frac {(\gamma_b - 1) {\xi_x}^2}{\xi^2} \right ) &
               \left ( \frac {(\gamma_b - 1) \xi_x \xi_y}{\xi^2}   \right ) &
               \left ( \frac {(\gamma_b - 1) \xi_x \xi_z}{\xi^2}   \right )  \\
\xi_y \gamma_b & \left ( \frac {(\gamma_b - 1) \xi_x \xi_y}{\xi^2}   \right ) &
               \left ( 1 + \frac {(\gamma_b - 1) {\xi_y}^2}{\xi^2} \right ) &
               \left ( \frac {(\gamma_b - 1) \xi_y \xi_z}{\xi^2}   \right )  \\
\xi_z \gamma_b & \left ( \frac {(\gamma_b - 1) \xi_x \xi_z}{\xi^2}   \right ) &
               \left ( \frac {(\gamma_b - 1) \xi_y \xi_z}{\xi^2}   \right ) &
               \left ( 1 + \frac {(\gamma_b - 1) {\xi_z}^2}{\xi^2} \right ) 
        \end{array} \right ]
\left [ \begin{array}{c} t \\ x \\ y \\ z
        \end{array} \right ],
\end{equation}
where $\xi^2 = {\xi_x}^2 + {\xi_y}^2 + {\xi_z}^2$ and
$\gamma_b = {(1 - \xi^2)}^{-1/2}$.  The resulting metric and stress-energy
tensor in the primed frame, 
\begin{eqnarray}
g_{ {\mu}^\prime {\nu}^\prime} (t^\prime,x^\prime,y^\prime,z^\prime) & 
\equiv &
\frac {\partial x^\alpha}{\partial x^{ {\mu}^\prime}} 
\frac {\partial x^\beta}{\partial x^{ {\nu}^\prime}}
g_{\alpha \beta}, \\
T_{ {\mu}^\prime {\nu}^\prime} (t^\prime,x^\prime,y^\prime,z^\prime) & 
\equiv &
\frac {\partial x^\alpha}{\partial x^{ {\mu}^\prime}} 
\frac {\partial x^\beta}{\partial x^{ {\nu}^\prime}}
T_{\alpha \beta} ,
\end{eqnarray}
are also solutions to the Einstein equations coupled to the GR-Hydro
equations.  Notice that the shift in the boosted
(primed) coordinates is non-zero.  For example, the $x$-component
of the shift vector is given by
\begin{eqnarray}
{\beta}_{x^\prime}(t^\prime,x^\prime,y^\prime,z^\prime) = 
g_{t^\prime x^\prime}(t^\prime,x^\prime,y^\prime,z^\prime) & = & 
{\gamma_b}^2 \xi_x \alpha ^2 -
\gamma_b \left ( 1 + \frac {(\gamma_b - 1) {\xi_x}^2}{\xi^2} \right )
\left ( \xi_x \gamma_{xx} + \xi_y \gamma_{xy} + \xi_z \gamma_{xz} \right ) - \\
 & & - \gamma_b  \frac {(\gamma_b - 1) \xi_x \xi_y}{\xi^2} 
\left ( \xi_x \gamma_{xy} + \xi_y \gamma_{yy} + \xi_z \gamma_{yz} \right ) - \nonumber \\
 & & - \gamma_b  \frac {(\gamma_b - 1) \xi_x \xi_z}{\xi^2}         
\left ( \xi_x \gamma_{xz} + \xi_y \gamma_{yz} + \xi_z \gamma_{zz} \right ) \nonumber ,
\end{eqnarray}
where $\alpha$ and $\gamma_{ij}$ are the lapse and 3-metric,
respectively, computed in the rest
frame (unprimed coordinates) evaluated at coordinates $(t,x,y,z)$ that
correspond to the primed coordinates
$(t^\prime,x^\prime,y^\prime,z^\prime)$.  In the tests performed in
this section, we specify the lapse and shift to be given by these
values and check that all 
evolved variables converge to
the analytic solution.  For the pre-boosted TOV solution, we use
the same configuration used in the previous section.

\begin{figure}
\psfig{figure=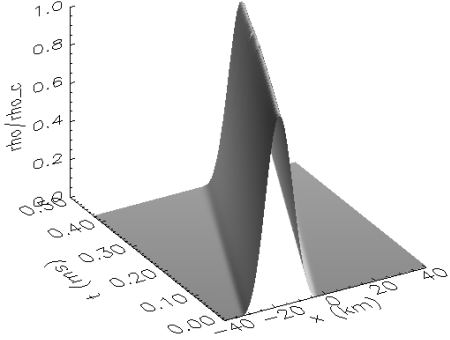,width=7.5cm}
\caption{A plot of the evolution of the
rest mass density $\rho$, scaled
by the central rest mass density $\rho_c$ along the diagonal as
a function of time.  The initial data corresponds to the TOV 
configuration from Sec.~\ref{sec:tov} boosted in the 
$\hat{x}+\hat{y}+\hat{z}$ diagonal direction with velocity
$v/c = 0.3$.  The star is evolved for $0.4 \: ms$.  The spatial
resolution corresponds to approximately 30 points across the star.  }
\label{fig:boost_tov_long}
\end{figure}

In Fig.~\ref{fig:boost_tov_long} we plot the evolution of
the rest mass density along the direction 
of boost ($\hat{x}+\hat{y}+\hat{z}$ diagonal) as a function of time.
The boosted star is evolved for $0.4 \: ms$, and is boosted
with a velocity of
$v/c = 0.3$.  As can be seen, the neutron
star has traversed approximately $36 \: km$ during the $0.4 ms$,
maintaining its original profile.

% Figure

For completeness, we also present convergence
tests for the boosted star.  The grid parameters for the tests
are given in table~\ref{table:boostgrid}. 
The initial data is evolved with the three different resolutions.
To test the code in the highly relativistic regime, 
we use the boost parameters 
$\xi_x = \xi_y = \xi_z = 0.5$ giving a lorentz factor $\gamma_b = 2$.
These boost parameters correspond to a neutron star 
moving in the $\hat{x}+\hat{y}+\hat{z}$ diagonal direction with
a velocity of $v/c = 0.87$.

\begin{figure}
\psfig{figure=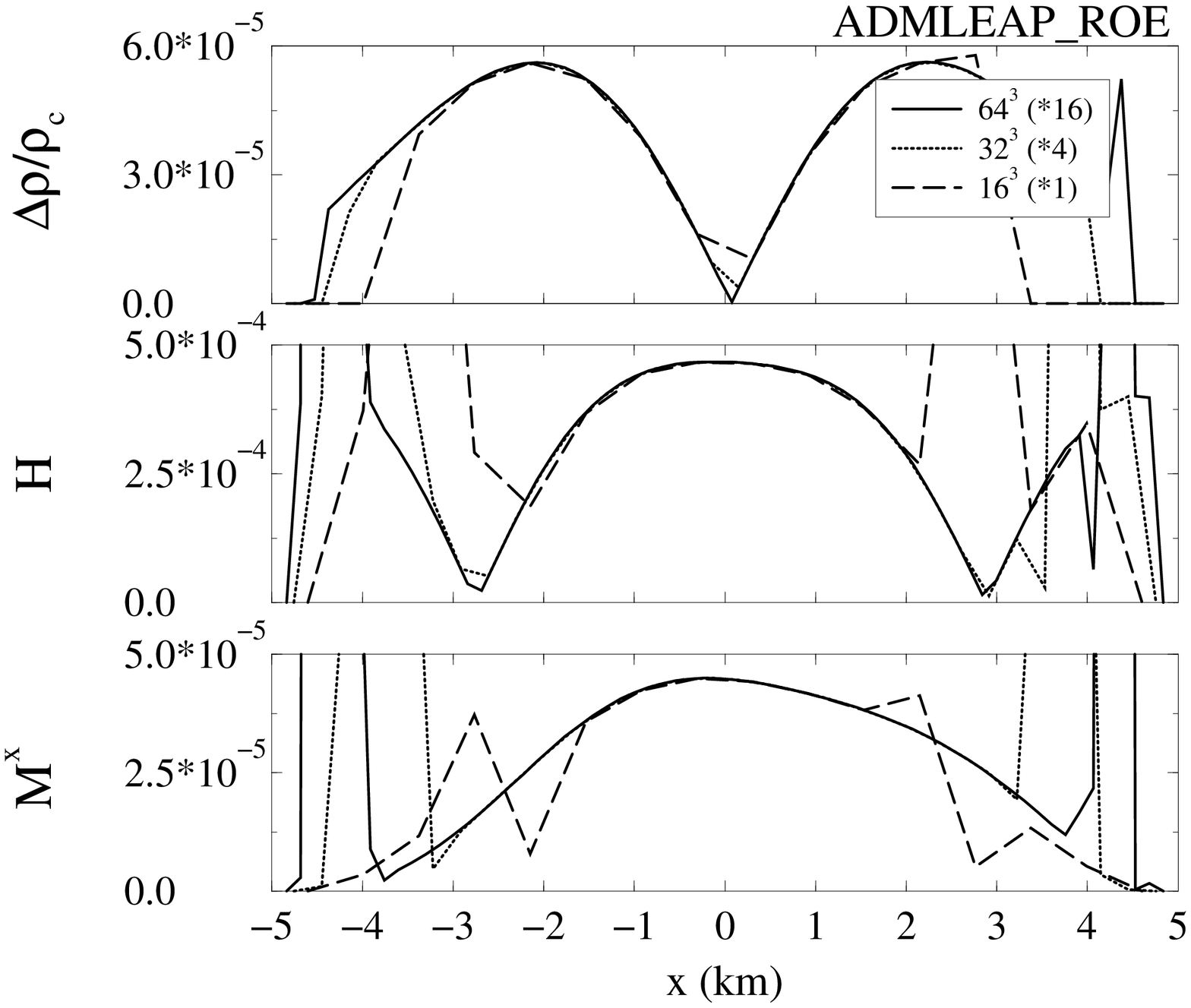,width=7.5cm}
\vspace{-6.4cm}
\hspace{7.5cm}
\psfig{figure=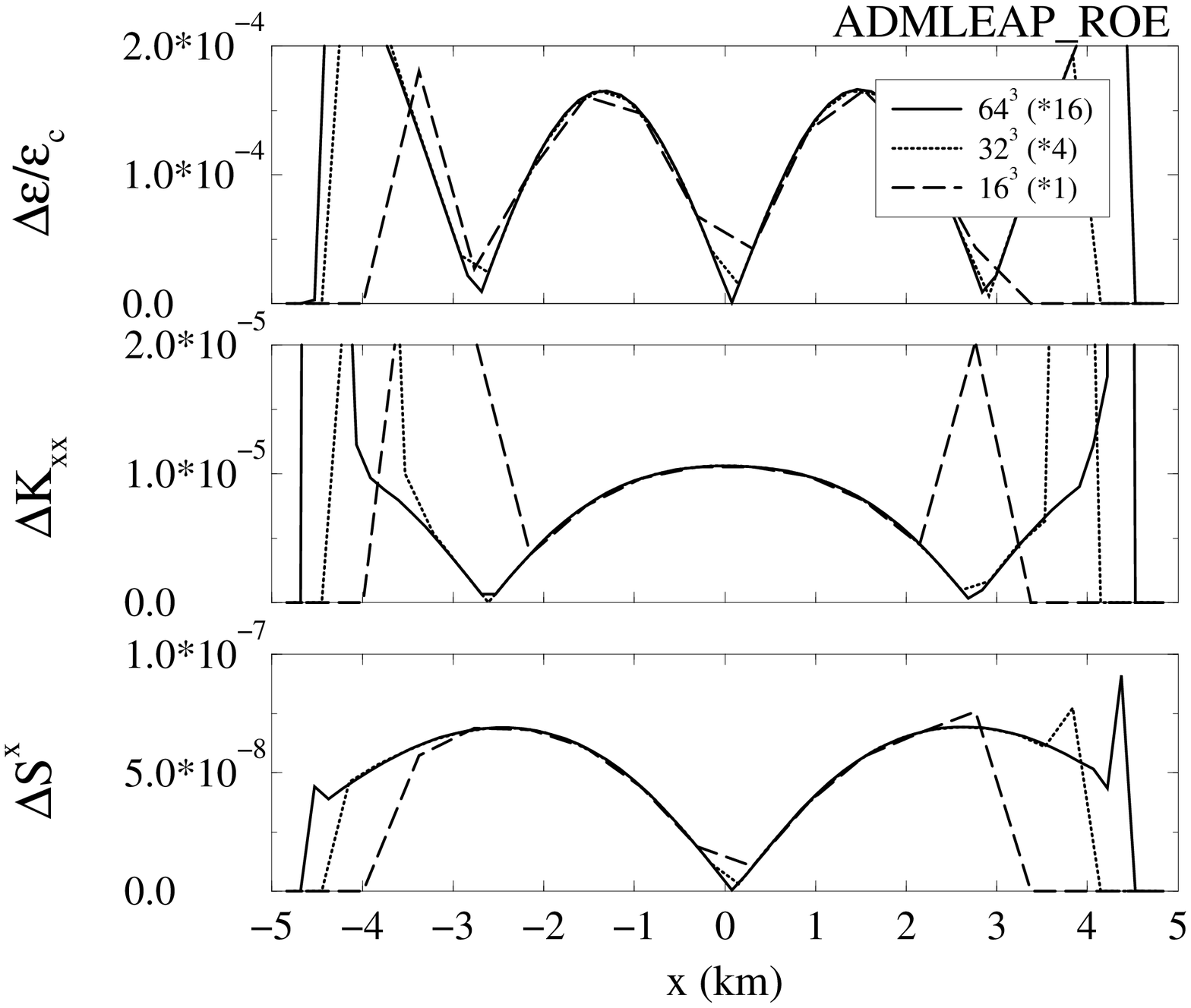,width=7.5cm}
\caption{We demonstrate the convergence of the ADMLEAP\_ROE 
evolution system for six different error functions.  On the left
panel, we plot 
the difference between the analytic and computed
rest mass density (normalized by the central rest mass density $\rho_c$)
$\Delta \rho / \rho_c$, the
Hamiltonian constraint, $H$,
and the $x$-momentum constraint, $M^x$.
On the right panel, we plot the 
difference between the analytic and computed
specific energy density
(normalized by the central specific energy density), $\Delta \epsilon /
{\epsilon}_c$, the difference between the $xx$ component of the
extrinsic curvature and the analytic solution,
$\Delta K_{xx}$, and the difference between the
$x$ component of the momentum and the analytic solution,$\Delta S^x$.
In each case, we multiply the high resolution
result by sixteen and the medium resolution by four to show second order
convergence.  All results are shown at $t=0.296 \mu s$ which corresponds to
eight iterations at the highest resolution.  The graphs are taken along
the $\hat{x}+\hat{y}+\hat{z}$ diagonal axis (results on the 
coordinate axis ($x$,$y$,$z$) are similar).}
\label{fig:boost_leap_roe}
\end{figure}

\begin{figure}
\psfig{figure=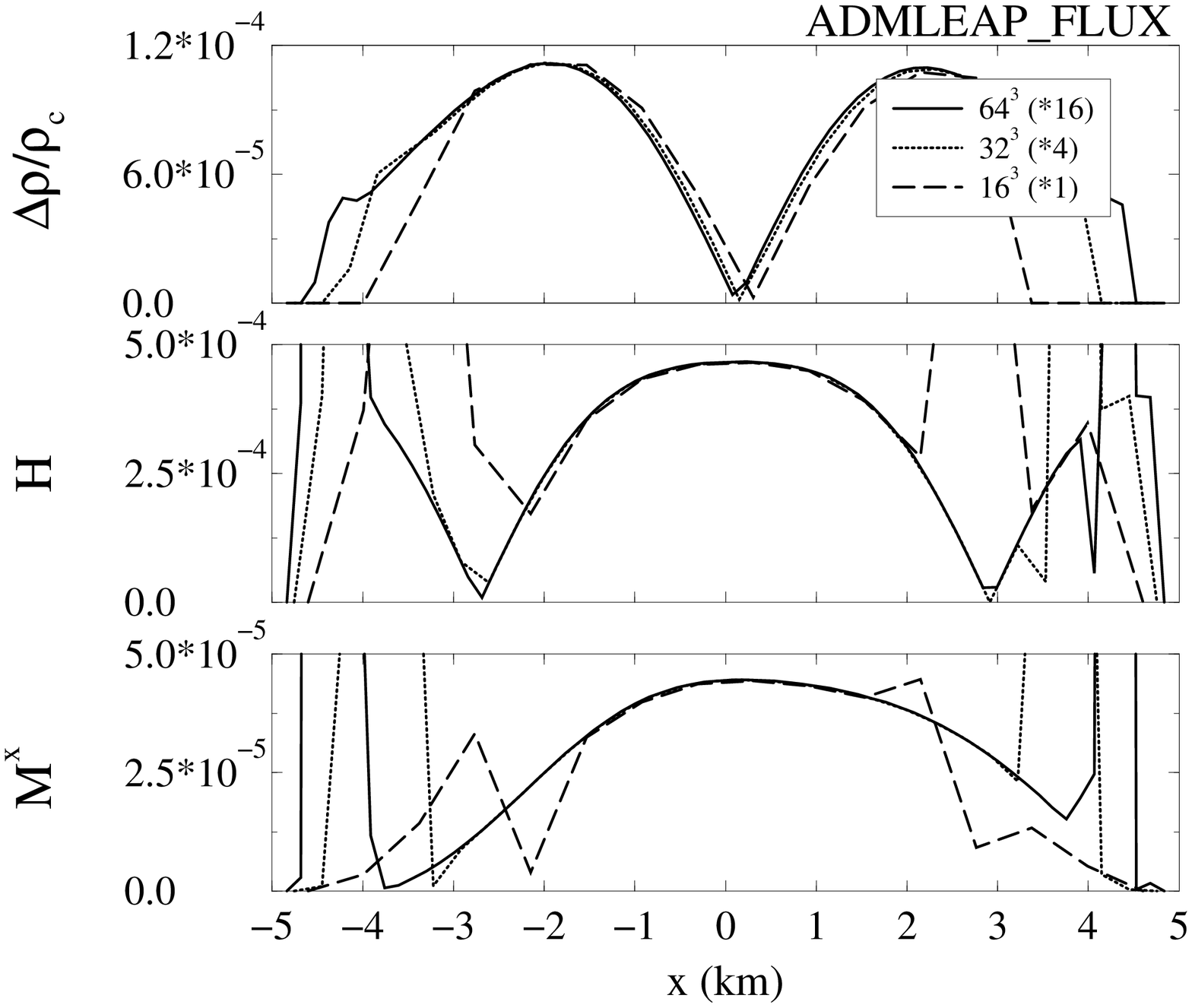,width=7.5cm}
\vspace{-6.6cm}
\hspace{7.5cm}
\psfig{figure=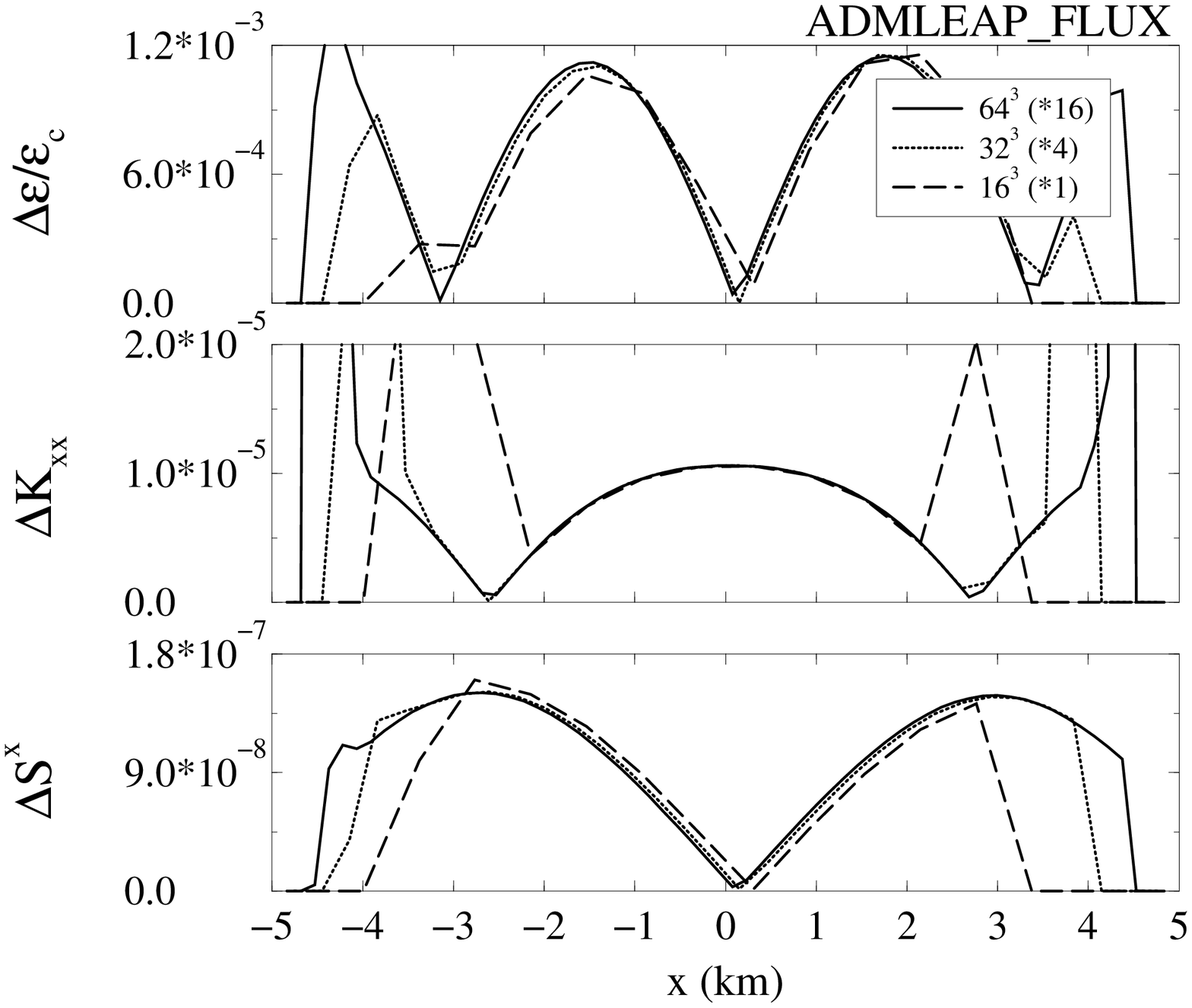,width=7.5cm}
\caption{We demonstrate the convergence of the ADMLEAP\_FLUX
evolution system for six different error functions.  On the left
panel, we plot
the difference between the analytic and computed
rest mass density (normalized by the central rest mass density $\rho_c$)
$\Delta \rho / \rho_c$, the
Hamiltonian constraint, $H$,
and the $x$-momentum constraint, $M^x$.
On the right panel, we plot the
difference between the analytic and computed
specific energy density
(normalized by the central specific energy density), $\Delta \epsilon /
{\epsilon}_c$, the difference between the $xx$ component of the
extrinsic curvature and the analytic solution,
$\Delta K_{xx}$, and the difference between the
$x$ component of the momentum and the analytic solution,$\Delta S^x$.
In each case, we multiply the high resolution
result by sixteen and the medium resolution by four to show second order
convergence.  All results are shown at $t=0.296 \mu s$ which corresponds to
eight iterations at the highest resolution.  The graphs are taken along
the $\hat{x}+\hat{y}+\hat{z}$ diagonal axis (results on the
coordinate axis ($x$,$y$,$z$) are similar).}
\label{fig:boost_leap_flux}
\end{figure}

\begin{figure}
\psfig{figure=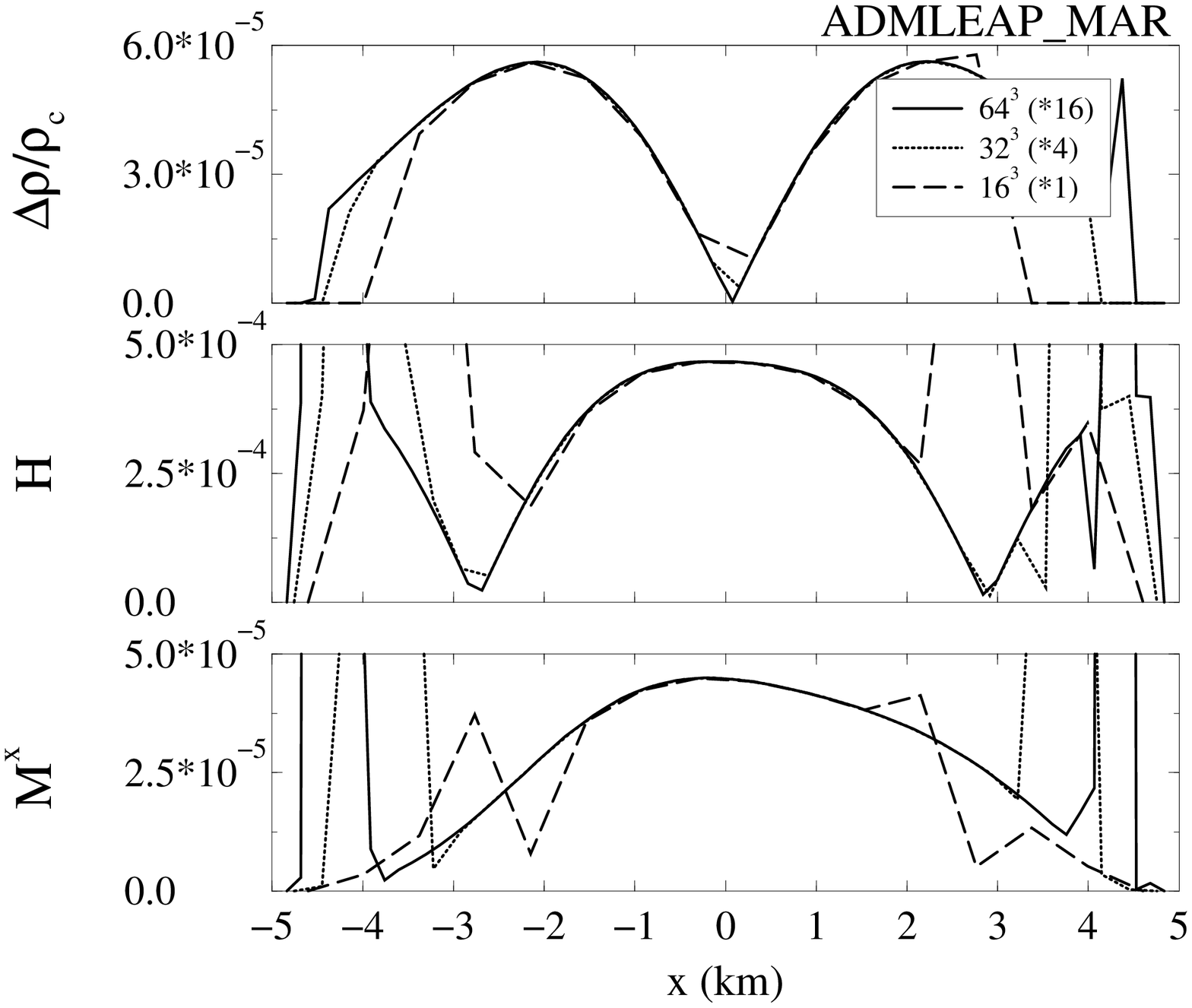,width=7.5cm}
\vspace{-6.4cm}
\hspace{7.5cm}
\psfig{figure=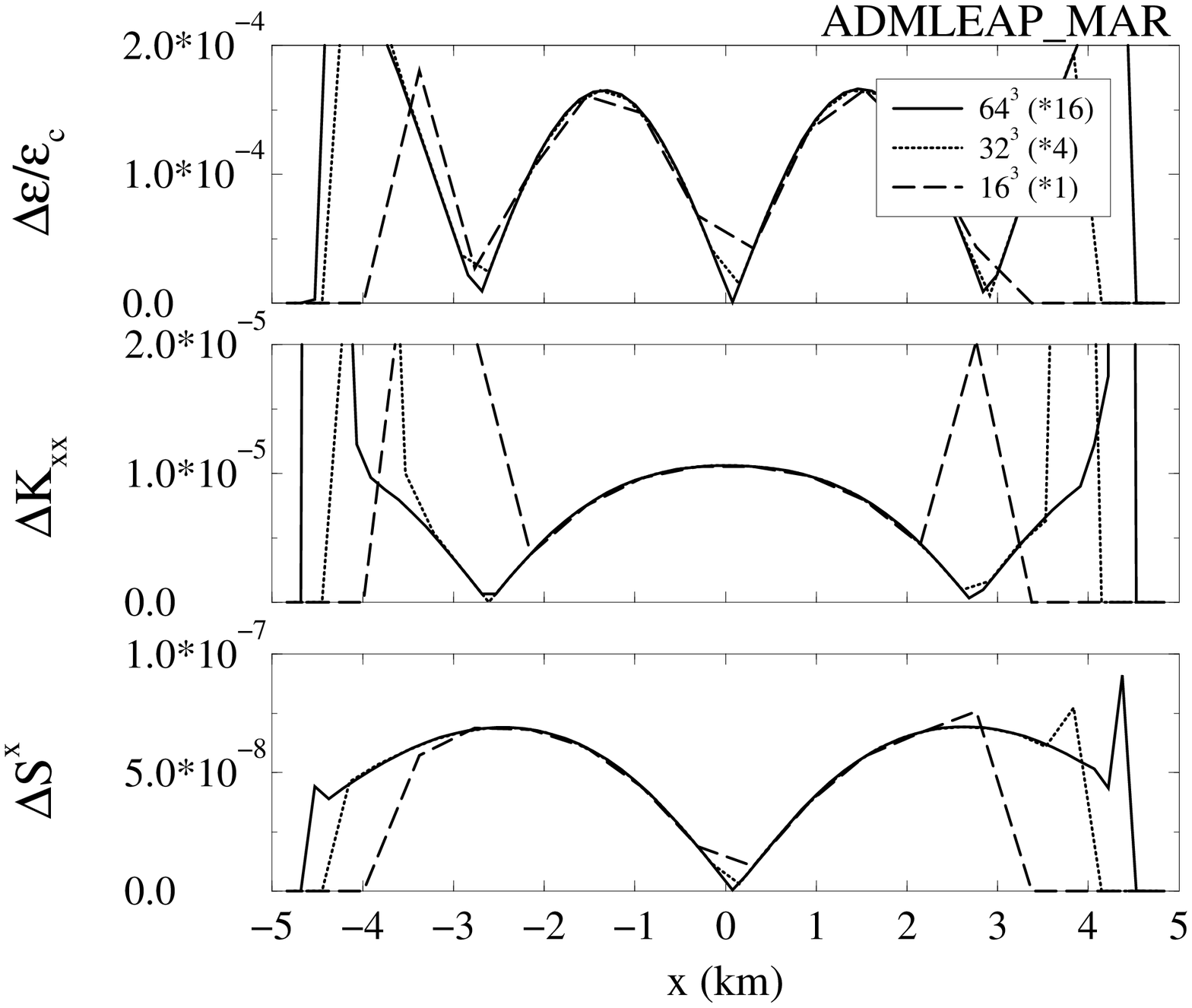,width=7.5cm}
\caption{We demonstrate the convergence of the ADMLEAP\_MAR
evolution system for six different error functions.  On the left
panel, we plot
the difference between the analytic and computed
rest mass density (normalized by the central rest mass density $\rho_c$)
$\Delta \rho / \rho_c$, the
Hamiltonian constraint, $H$,
and the $x$-momentum constraint, $M^x$.
On the right panel, we plot the
difference between the analytic and computed
specific energy density
(normalized by the central specific energy density), $\Delta \epsilon /
{\epsilon}_c$, the difference between the $xx$ component of the
extrinsic curvature and the analytic solution,
$\Delta K_{xx}$, and the difference between the
$x$ component of the momentum and the analytic solution,$\Delta S^x$.
In each case, we multiply the high resolution
result by sixteen and the medium resolution by four to show second order
convergence.  All results are shown at $t=0.296 \mu s$ which corresponds to
eight iterations at the highest resolution.  The graphs are taken along
the $\hat{x}+\hat{y}+\hat{z}$ diagonal axis (results on the
coordinate axis ($x$,$y$,$z$) are similar).}
\label{fig:boost_leap_mar}
\end{figure}

\begin{figure}
\psfig{figure=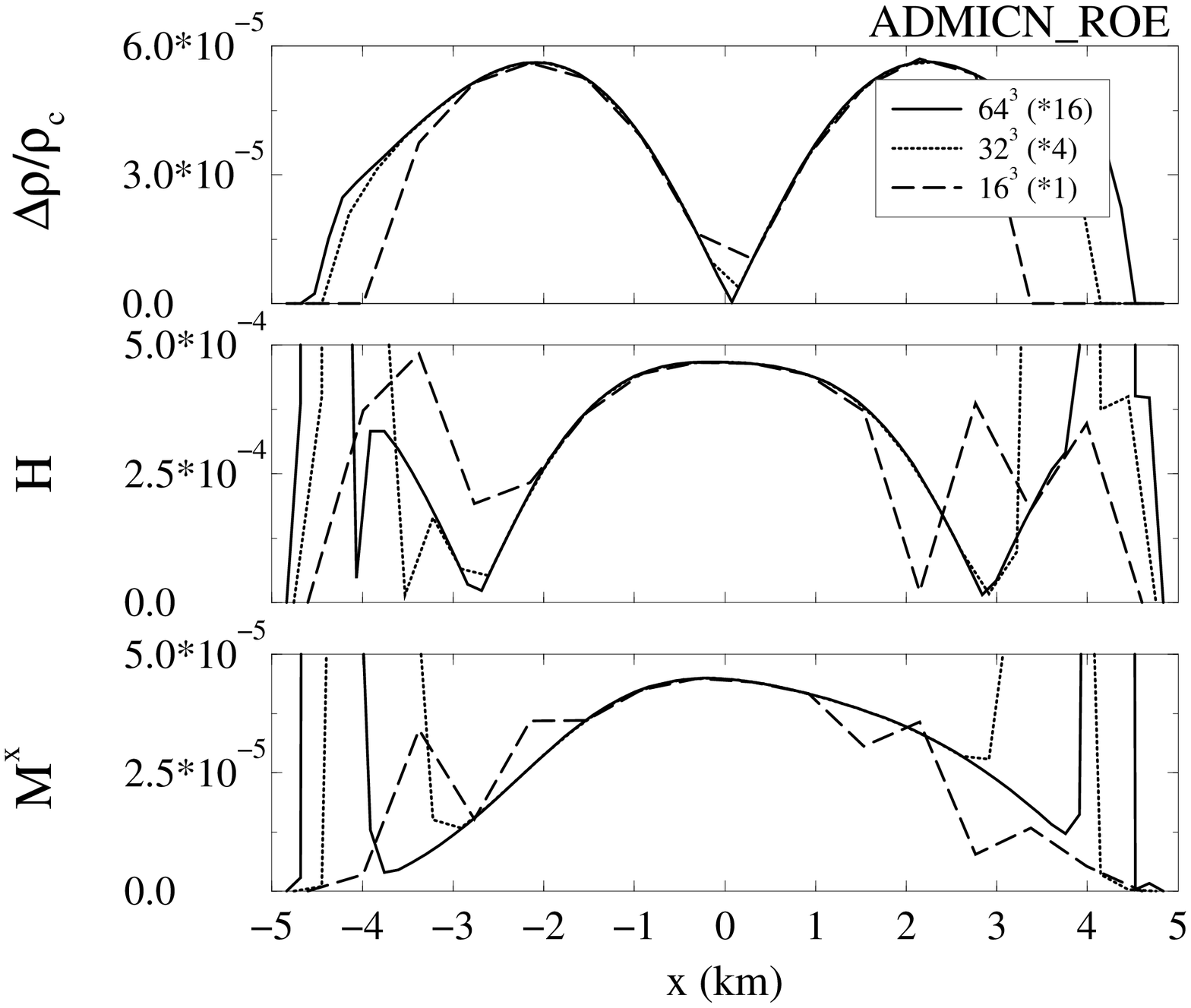,width=7.5cm}
\vspace{-6.4cm}
\hspace{7.5cm}
\psfig{figure=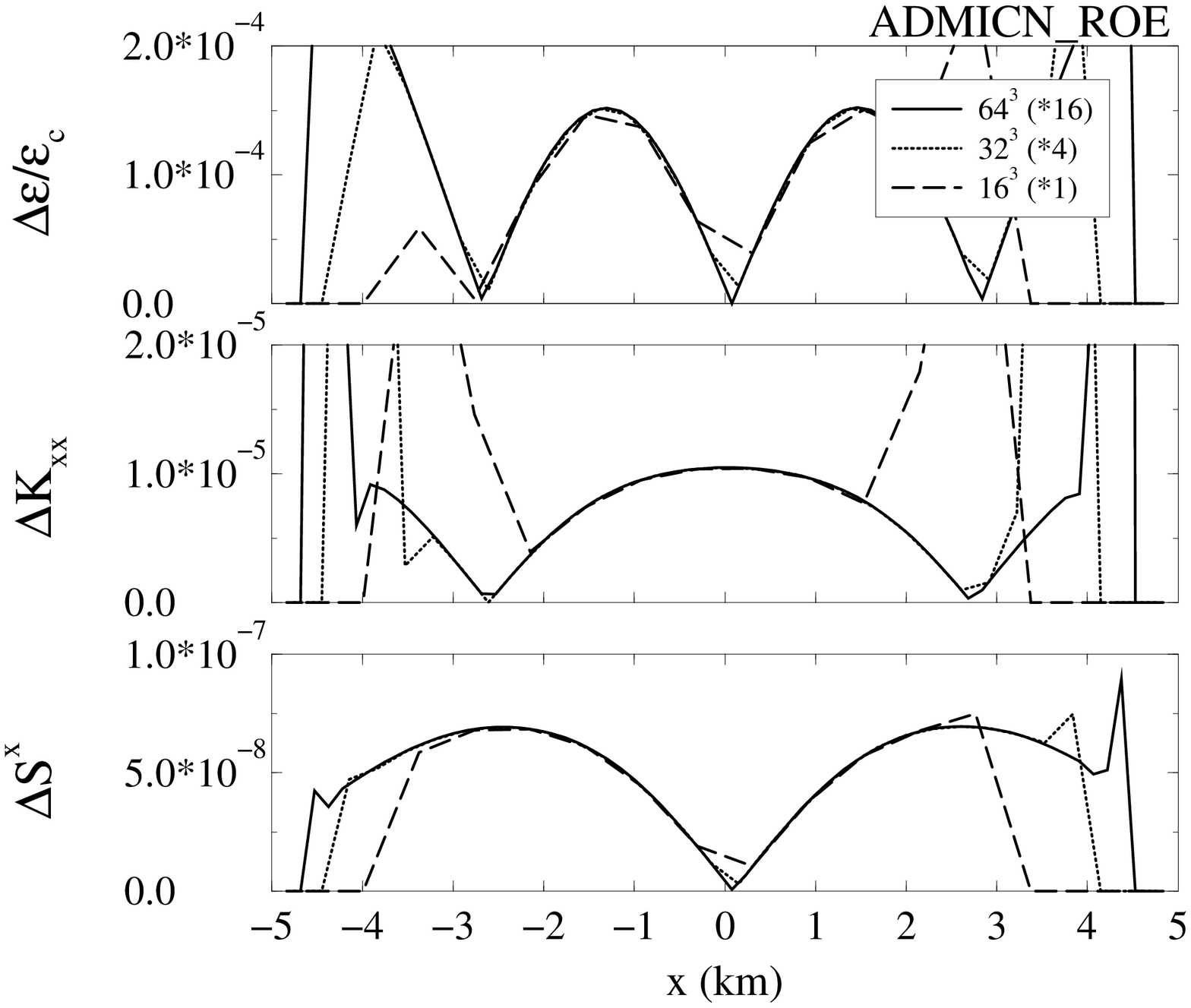,width=7.5cm}
\caption{We demonstrate the convergence of the ADMICN\_ROE
evolution system for six different error functions.  On the left
panel, we plot
the difference between the analytic and computed
rest mass density (normalized by the central rest mass density $\rho_c$)
$\Delta \rho / \rho_c$, the
Hamiltonian constraint, $H$,
and the $x$-momentum constraint, $M^x$.
On the right panel, we plot the
difference between the analytic and computed
specific energy density
(normalized by the central specific energy density), $\Delta \epsilon /
{\epsilon}_c$, the difference between the $xx$ component of the
extrinsic curvature and the analytic solution,
$\Delta K_{xx}$, and the difference between the
$x$ component of the momentum and the analytic solution,$\Delta S^x$.
In each case, we multiply the high resolution
result by sixteen and the medium resolution by four to show second order
convergence.  All results are shown at $t=0.296 \mu s$ which corresponds to
eight iterations at the highest resolution.  The graphs are taken along
the $\hat{x}+\hat{y}+\hat{z}$ diagonal axis (results on the
coordinate axis ($x$,$y$,$z$) are similar).}
\label{fig:boost_icn_roe}
\end{figure}

\begin{figure}
\psfig{figure=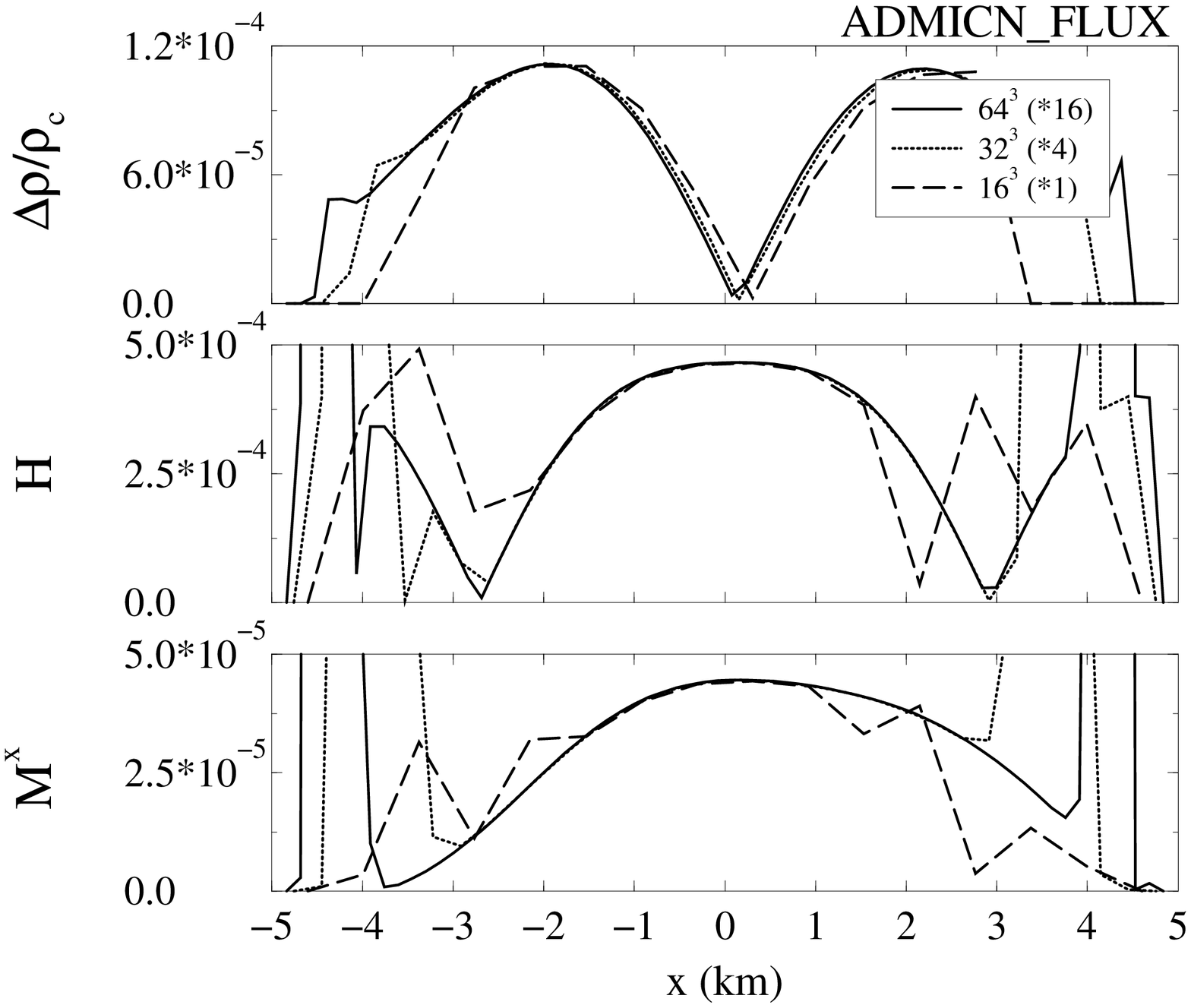,width=7.5cm}
\vspace{-6.4cm}
\hspace{7.5cm}
\psfig{figure=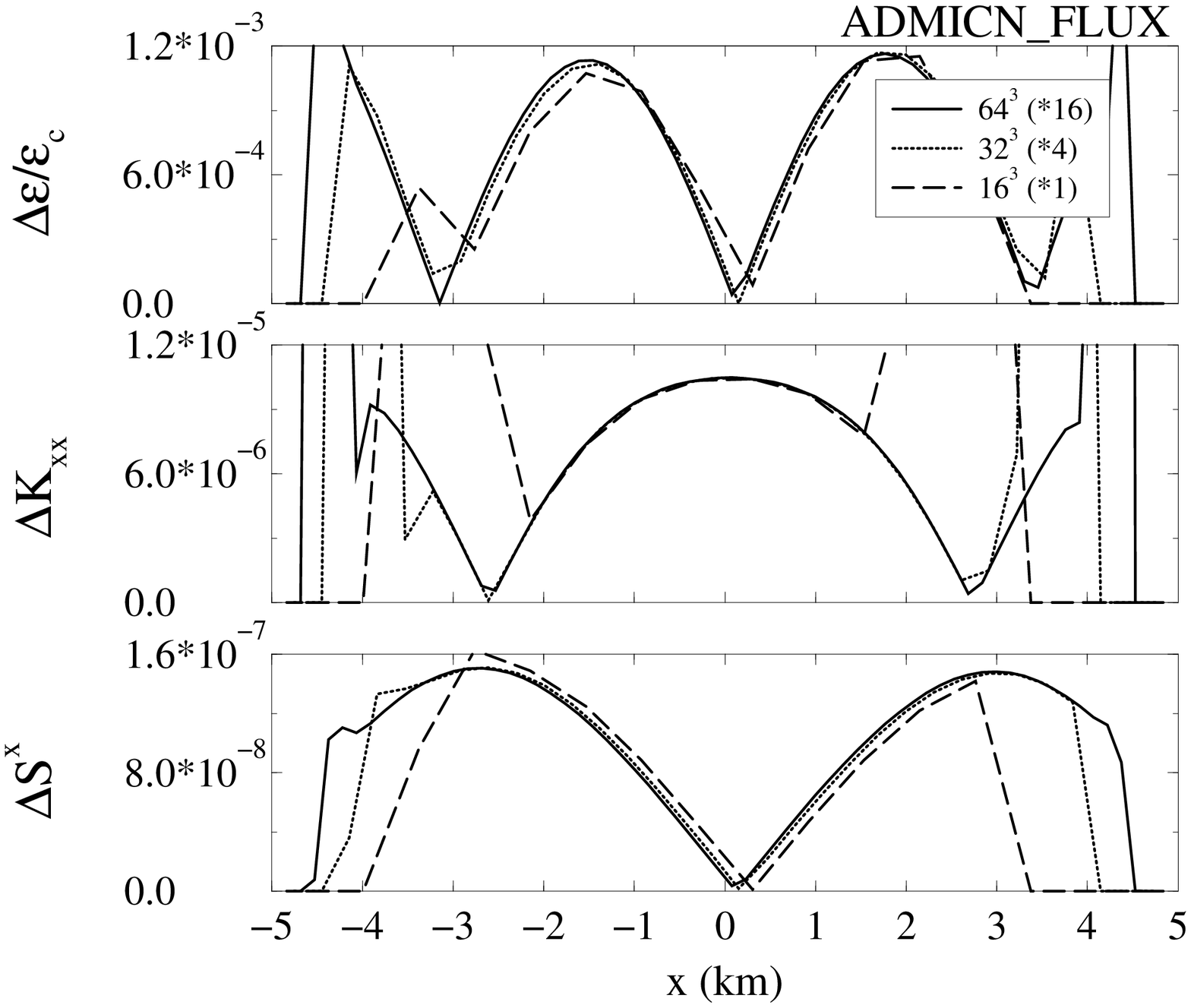,width=7.5cm}
\caption{We demonstrate the convergence of the ADMICN\_FLUX
evolution system for six different error functions.  On the left
panel, we plot
the difference between the analytic and computed
rest mass density (normalized by the central rest mass density $\rho_c$)
$\Delta \rho / \rho_c$, the
Hamiltonian constraint, $H$,
and the $x$-momentum constraint, $M^x$.
On the right panel, we plot the
difference between the analytic and computed
specific energy density
(normalized by the central specific energy density), $\Delta \epsilon /
{\epsilon}_c$, the difference between the $xx$ component of the
extrinsic curvature and the analytic solution,
$\Delta K_{xx}$, and the difference between the
$x$ component of the momentum and the analytic solution,$\Delta S^x$.
In each case, we multiply the high resolution
result by sixteen and the medium resolution by four to show second order
convergence.  All results are shown at $t=0.296 \mu s$ which corresponds to
eight iterations at the highest resolution.  The graphs are taken along
the $\hat{x}+\hat{y}+\hat{z}$ diagonal axis (results on the
coordinate axis ($x$,$y$,$z$) are similar).}
\label{fig:boost_icn_flux}
\end{figure}

\begin{figure}
\psfig{figure=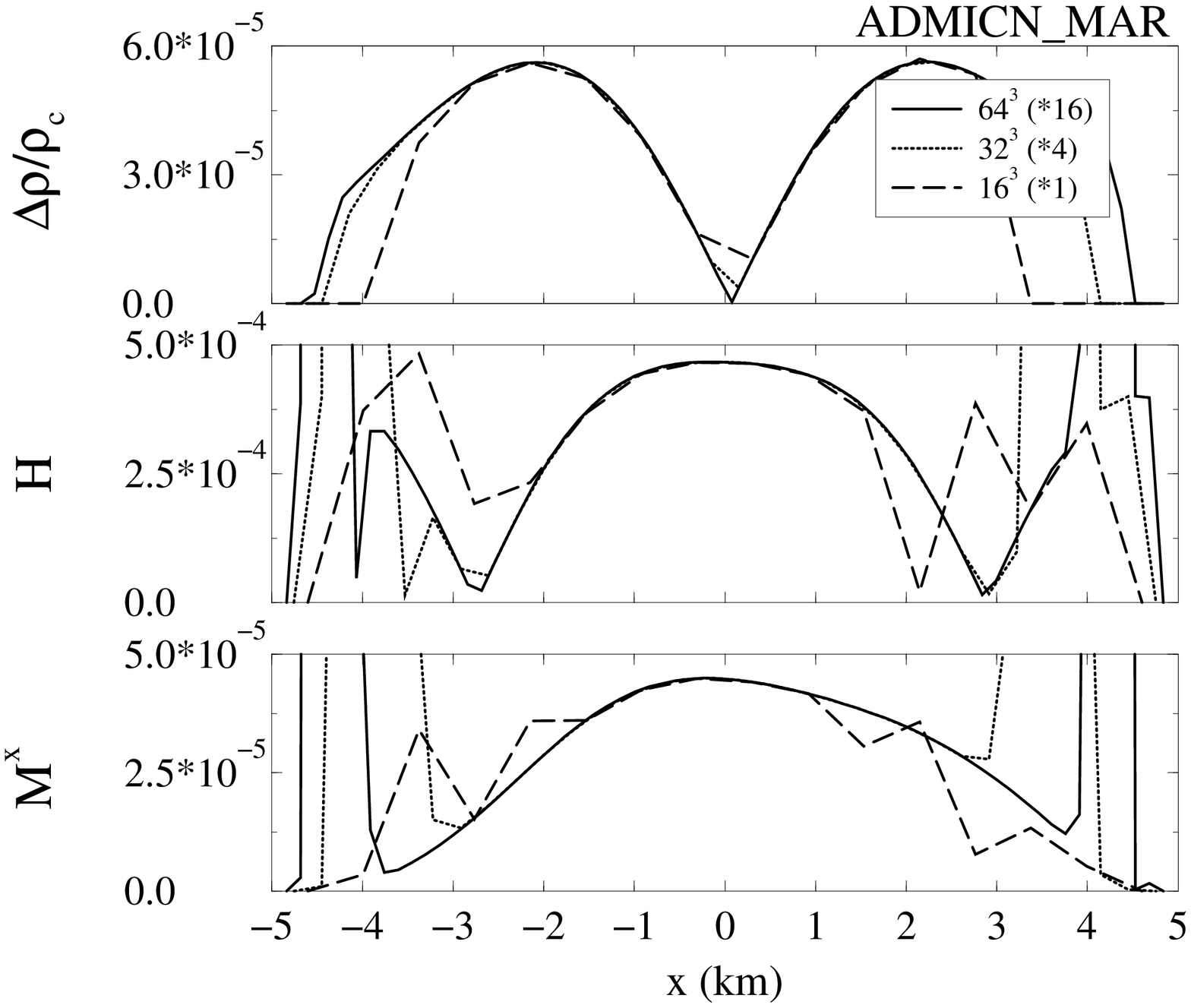,width=7.5cm}
\vspace{-6.4cm}
\hspace{7.5cm}
\psfig{figure=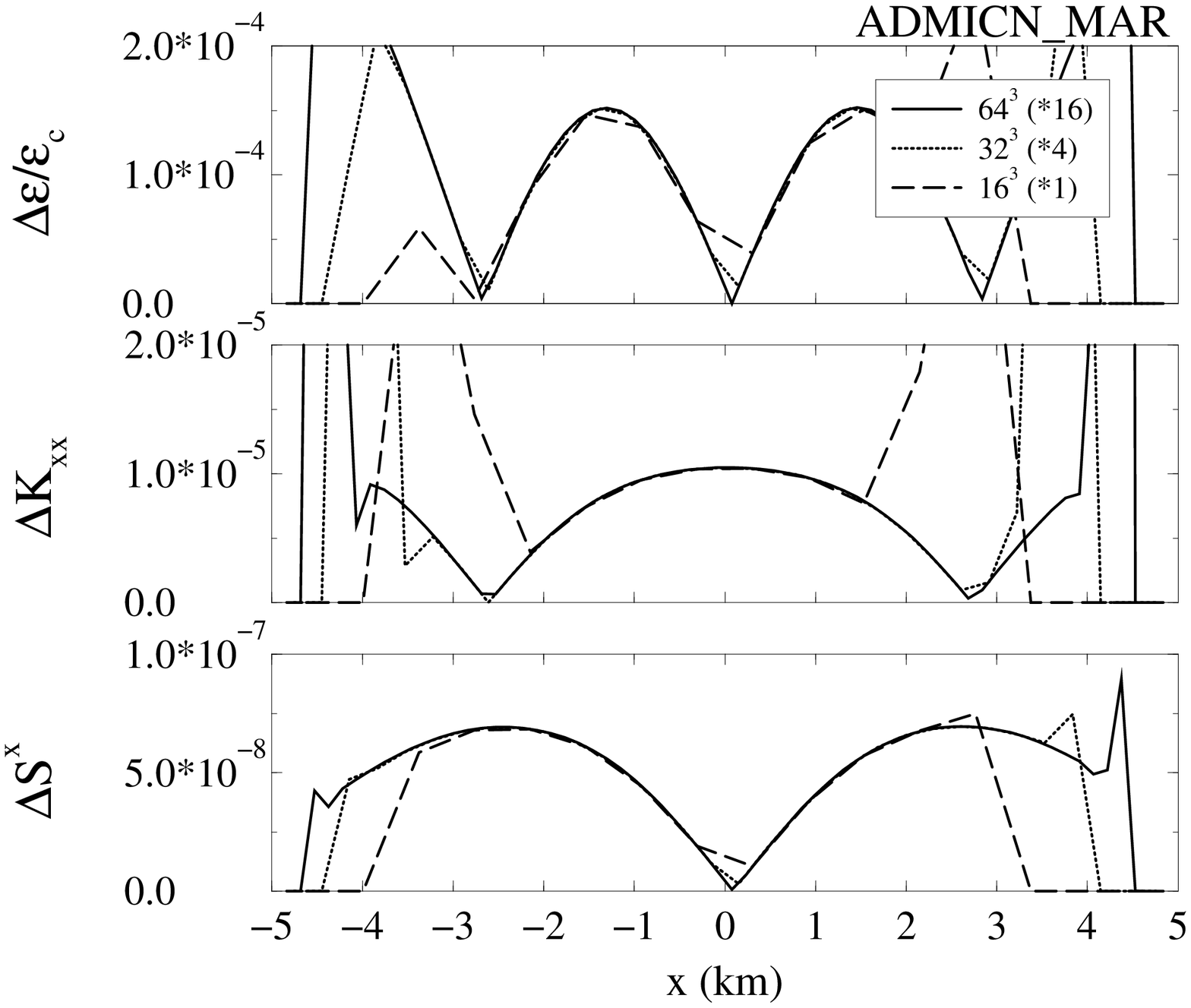,width=7.5cm}
\caption{We demonstrate the convergence of the ADMICN\_MAR
evolution system for six different error functions.  On the left
panel, we plot
the difference between the analytic and computed
rest mass density (normalized by the central rest mass density $\rho_c$)
$\Delta \rho / \rho_c$, the
Hamiltonian constraint, $H$,
and the $x$-momentum constraint, $M^x$.
On the right panel, we plot the
difference between the analytic and computed
specific energy density
(normalized by the central specific energy density), $\Delta \epsilon /
{\epsilon}_c$, the difference between the $xx$ component of the
extrinsic curvature and the analytic solution,
$\Delta K_{xx}$, and the difference between the
$x$ component of the momentum and the analytic solution,$\Delta S^x$.
In each case, we multiply the high resolution
result by sixteen and the medium resolution by four to show second order
convergence.  All results are shown at $t=0.296 \mu s$ which corresponds to
eight iterations at the highest resolution.  The graphs are taken along
the $\hat{x}+\hat{y}+\hat{z}$ diagonal axis (results on the
coordinate axis ($x$,$y$,$z$) are similar).}
\label{fig:boost_icn_mar}
\end{figure}

Figs.~\ref{fig:boost_leap_roe}~--~\ref{fig:boost_icn_mar}
show convergence plots for six system combinations
resulting from three hydrodynamical evolution schemes (ROE, FLUX, MAR)
and two spacetime evolution schemes (ADMLEAP,ICN),
see Table~\ref{table:discrete_names}.
The left panel of each figure contains plots of the 
difference between the
numerically evolved rest mass density and the analytic solution,
(normalized by the central density $(\Delta \rho /
{\rho}_c)$, the Hamiltonian constraint $(H)$, and the $x$-component of
the momentum constraint $(M^x)$.
The right panel of each figure contains plots of the 
difference between the
numerically evolved specific energy density and the analytic solution
(normalized by the central specific energy density), $\Delta \epsilon /
{\epsilon}_c$, the difference between the $xx$ component of the 
extrinsic curvature and the analytic solution
,$\Delta K_{xx}$, and the difference between the
$x$ component of the momentum and the analytic solution,$\Delta S^x$.

To provide adequate resolution for the convergence tests, we move
the boundaries of our computational domain inside of the star.
We ignore errors caused by the boundary, and focus on the convergence
properties of the interior solution.

In comparing the absolute value of the errors for the different schemes,
we notice no significant difference between the two spacetime
evolution schemes (ADMLEAP, ADMICN).  However, there is
clearly a difference between the hydrodynamical evolution schemes,
where the Roe and Marquina methods (ROE,MAR) are equally more accurate
than the flux-split method (FLUX) for the resolutions used.

\begin{table}
\begin{tabular}{|c||c|c|c|c|c|} \hline
\label{table:boostgrid}
 & \# of points & &  & & total  \\
resolution & in each & $\Delta x$ (km) & $c \frac {\Delta t}{\Delta x}$ &
\# of timesteps & evolved time \\
 & coordinate direction & & & & $(\mu s)$ \\ \hline
low & 16 & 0.3545 & 0.125 & 2 & 0.296 \\
medium & 32 & 0.1772 & 0.125 & 4 & 0.296 \\
high & 64 & 0.0886 & 0.125 & 8 & 0.296 \\ \hline
\end{tabular}
\caption{Computational grid parameters for boosted TOV tests.}
\end{table}

% Figure

% Figure

% Figure

% Figure

% Figure

% Figure

\section{Conclusions}

In this paper we present a new three-dimensional, Eulerian, general
relativistic hydrodynamical code constructed for general relativistic
astrophysics.  This code is capable of evolving the coupled system of
the Einstein and hydrodynamic equations.  The code is constructed for a
completely general spacetime metric based on a Cartesian
coordinate system, with arbitrarily specifiable lapse and shift
conditions.  This paper discussed the general relativistic hydrodynamics
part of the code, and its coupling to the spacetime code, in parallel to
the presentation of the spacetime (vacuum) part of the code in ~\cite{Bona98b}.

We have derived a spectral decomposition for the GR-Hydro equations
valid for general spatial metrics, generalizing the results of
~\cite{Banyuls97} which were only valid for the case of a diagonal metric.  
Based on this spectral
decomposition, three different approximate (linearized) Riemann
solvers, flux-split, Roe and Marquina, were used to integrate the
relativistic hydrodynamic equations.  We tested these methods
individually and compared the results against one another.  While we
found all methods converging to second order in the discretization
parameter, we also compared the absolute values of errors of the
different methods.  

Which method produced the smallest absolute error, and whether the
spacetime or hydrodynamical evolution was the dominant source of
error, depends on the initial data being evolved.
For the shocktube problem, only the hydrodynamical evolution
was relevant since the evolution took place on a flat background
metric.  For an evolution along a coordinate axis, the Roe and
Marquina methods were superior to the flux split method.
For an evolution where the shockfront is along the diagonal, 
the flux split method was slightly more accurate than both
the Roe and Marquina method.
For the FRW evolutions, the spacetime
evolution is the main source of error.  The BM system tends to
be more accurate than the ADM system.  For the TOV tests, we
find that the Roe and Marquina methods are more accurate than
the flux split method, and the BM system is more accurate than
the ADM system.  For the boosted TOV test, the Roe and Marquina
methods are again superior to flux split.
We caution that these statements 
could depend on the resolution used and the duration of evolution.

The hydrodynamic evolution is coupled to the spacetime
evolution in a manner which is second order accurate in {\it
both} space and time.  The coupled code was subjected to a series of
convergence tests, with different combinations of the spacetime and
hydrodynamics finite differencing schemes, demonstrating the
consistency of the discrete equations with the differential
equations \cite{Lax56}.  The extensive convergence tests performed
are important not only for the validation of the code, but have also
been important debugging tools during the code development process.
We consider the tests presented to be a minimal set that
any 3D GR-Hydro code should pass before actual applications.
The test-beds that we report on in
this paper include: special relativistic shock tubes,
Friedmann-Robertson-Walker cosmology tests, evolution of equilibrium
configurations of compact stars (solutions to the
Tolman-Oppenheimer-Volkoff equations), and the evolution of 
relativistically boosted
TOV stars transversing diagonally across the computational
domain.  The degree of complexity presented in these tests
increases from purely special relativistic flows in flat backgrounds
to fully general relativistic flows in dynamical spacetimes.  In
particular, the last test-bed (the boosted star) involves {\it all}
possible terms in the coupled set of GR-Hydro evolution equations and
were carried out with a non-trivial lapse and shift vector.

We found a simple, yet effective treatment for handling the surface
region of a general relativistic self-gravitating compact object.  The
key idea is to replace the energy equation update by the condition
of adaiabatic flow in regions of low density.
While the
surface region is not changing the overall dynamics of the star,
numerical instabilities there could halt the numerical evolution if
uncontrolled.  The capability to handle the surface region in
a stable fashion is important for the application of the
code to the study of neutron star astrophysics.  We have demonstrated
this capability in the equilibrium and boosted star
test-beds.  Refinement of this treatment for long term stability is
presently being investigated.

Additional code calibrations that are underway include long-term
stability analysis of single neutron stars, comparisons of waveforms
from perturbed neutron stars, and comparisons with one-dimensional and
axisymmetric (2D) independent GR-Hydro codes that we (together with
our collaborators) constructed ~\cite{Font97,Brandt98}.
Those will be reported in later papers in this series.

The formulation of the coupled set of equations and the numerical code
reported in this paper were used for the construction of the milestone
code ``GR3D'' for the NASA Neutron Star Grand Challenge project (for a
description of the project, see
http://wugrav.wustl.edu/Relativ/nsgc.html).  The goal of this project
is to develop a code for general relativistic astrophysics, and in
particular, one that is capable of simulating the inspiral coalescence
of a neutron star binary system.  The
coalescences of neutron star binaries are expected to be important
sources of gravitational waves for interferometric detectors.  The
strongest signal will come from the highly dynamic ``plunge'' during
the final phase of the inspiral; a fully general relativistic code
provides the only way to calculate this portion of the waveform.  A
version of the code which passed the milestone requirement of the
NASA Grand Challenge project, has recently been released to the
community~\cite{GR3D}.
This code has been benchmarked at over 140 GFlop/sec on
a $1024$ node Cray T3E with a scaling efficiency of over 95\%,
showing the potential for large scale 3D simulations
of realistic astrophysical systems.
Further development of our general relativistic code, and its
application to the specific study of the neutron star coalescence
scenario, will be described in later papers in this series.

To summarize, this paper presents the first (and necessary)
steps towards constructing an accurate and reliable tool for the numerical 
study of astrophysical phenomena involving matter at relativistic speeds 
and strong gravitational fields.

\section{Acknowledgments}

The general relativistic hydrodynamical module ``MAHC"
presented and studied in this paper is coupled to the ``Cactus"
code for the spacetime evolution. The
Cactus code is being developed by an international collaboration, with a
major contribution coming from the Albert Einstein Institute in Potsdam
(Germany), and with significant contribution from the relativity group 
(WUGRAV) at Washington University in St. Louis, Missouri, and from colleagues 
at the National Center for Supercomputing Applications in Urbana, Illinois.
Further development has been carried out at the University of the Balearic
Islands in Mallorca (Spain), the University of Valencia (Spain), and elsewhere. 
The hydrodynamical module was developed mainly at WUGRAV, with significant 
contributions from the Potsdam group, and has benefited 
from interactions with the hydro group at the University of Valencia.

We would like to thank Miguel Alcubierre, Gabrielle Allen, Pete
Anninos, Toni Arbona, Carles Bona, Steve Brandt,
Bernd Br\"ugmann, Dan Bullok, Tom Clune, Teepanis
Chachiyo, Ming C. Chu, Greg Comer, Thomas Dramlitsch, Comer Duncan, Ed
Evans, Ian Foster, Tom Goodale, Carsten Gundlach, Philip Gressman,
Philip Hughes, Jos\'e Mar\'{\i}a Ib{\'a}{\~n}ez, Sai Iyer, 
Gerd Lanferman, Joan Mass\'o, Peter
Miller, Philippos Papadopoulos, Manish Parashar, Bo Qin,
K.V. Rao, Paul Saylor, Bernard Schutz, Edward Seidel, John Shalf, Hisa-aki
Shinkai, Joan Stela, Doug Swesty, Ryoji Takahashi, Robert
Young, Paul Walker, Ed Wang, and William Wu for useful discussions and various
help with the code development.

This work is supported by NSF grants Phy 96-00507 and 96-00049, NSF NRAC
Allocation Grant no. MCA93S025, NASA grant
NASA-NCCS5-153, the Albert Einstein Institute, and the Institutes
of Mathematical Sciences, Chinese University of Hong Kong.
One of us (J.A.F) acknowledges financial support from the TMR
program of the European Union (contract number ERBFMBICT971902).

% Bibliography
\bibliographystyle{prsty}

\end{document}